\documentclass[prd,onecolumn,superscriptaddress,amsfonts,amssymb,amsmath,floatfix]{revtex4-2}
\usepackage{bm}
\usepackage{amsfonts}
\usepackage{latexsym}
\usepackage{graphicx}
\usepackage{amsmath}
\usepackage{palatino}
\usepackage{bigints}
\usepackage{mathpazo}
\usepackage{soul}
\usepackage{graphicx,amssymb,amsmath,amsthm,amsfonts,amscd, mathrsfs,epsfig,epsf}
\usepackage{textcomp}
\usepackage{float}
\usepackage{booktabs}
\usepackage{dcolumn}
\usepackage{booktabs}
\usepackage{multirow}
\usepackage{hyperref}
\hypersetup{colorlinks,citecolor=blue}
\usepackage{amsmath}
\usepackage[mathcal]{euscript}
\usepackage{xcolor}
\usepackage{orcidlink}
\usepackage{commath}
\usepackage{subcaption}
\usepackage{caption}

\def\be{\begin{eqnarray}}
\def\ee{\end{eqnarray}}

\def\jnl@style{\it}
\def\aaref@jnl#1{{\jnl@style#1}}

\def\aaref@jnl#1{{\jnl@style#1}}

\def\aj{\aaref@jnl{AJ}}                   
\def\apj{\aaref@jnl{ApJ}}                 
\def\apjl{\aaref@jnl{ApJ}}                
\def\apjs{\aaref@jnl{ApJS}}               
\def\apss{\aaref@jnl{Ap\&SS}}             
\def\aap{\aaref@jnl{A\&A}}                
\def\aapr{\aaref@jnl{A\&A~Rev.}}          
\def\aaps{\aaref@jnl{A\&AS}}              
\def\mnras{\aaref@jnl{Mon.~Not.~Roy.~Astron.~Soc.}}             
\def\prd{\aaref@jnl{Phys.~Rev.~D}}        
\def\prc{\aaref@jnl{Phys.~Rev.~C}}  
\def\prl{\aaref@jnl{Phys.~Rev.~Lett.}}    
\def\qjras{\aaref@jnl{QJRAS}}             
\def\skytel{\aaref@jnl{S\&T}}             
\def\ssr{\aaref@jnl{Space~Sci.~Rev.}}     
\def\zap{\aaref@jnl{ZAp}}                 
\def\nat{\aaref@jnl{Nature}}              
\def\aplett{\aaref@jnl{Astrophys.~Lett.}} 
\def\apspr{\aaref@jnl{Astrophys.~Space~Phys.~Res.}} 
\def\physrep{\aaref@jnl{Phys.~Rep.}}      
\def\physscr{\aaref@jnl{Phys.~Scr}}       
\def\commat{\aaref@jnl{Comm.~Math.~Phys.}}              
\def\science{\aaref@jnl{Science}}               
\def\cqg{\aaref@jnl{Classical Quant.~Grav.}}            
\def\jpcs{\aaref@jnl{JPCS}}                                     
\def\ijmpd{\aaref@jnl{Int.~J.~Mod.~Phys.~D}}                    
\def\grg{\aaref@jnl{Gen.~Relat.~Gravit.}}               
\def\rpp{\aaref@jnl{Rep.~Prog.~Phys.}}          
\def\npa{\aaref@jnl{Nucl.~Phys.~A}}        
\def\lrr{\aaref@jnl{Living Rev.~Rel.}}                   
\def\jcap{\aaref@jnl{J.~Cosmology Astropart.~Phys.}}    
\def\rmp{\aaref@jnl{Rev.~Mod.~Phys.}}   
\def\epjc{\aaref@jnl{Eur.~Phys.~J.~C}}

\allowdisplaybreaks[1]

\newcommand{\ben}{\begin{equation}}
\newcommand{\en}{\end{equation}}
\newcommand{\bea}{\begin{eqnarray}}
\newcommand{\ena}{\end{eqnarray}}

\newcommand{\bff}{\begin{figure}}
\newcommand{\eff}{\end{figure}}

\newcommand{\refb}[1]{(\ref{#1})}

\begin{document}

\color{black}       

\title{Quasinormal modes and absorption cross-section of a Bardeen black hole surrounded by perfect fluid dark matter in four dimensions}

\author{Angel Rincon \orcidlink{0000-0001-8069-9162}}
\email[Email: ]{angel.rincon@ua.es}
\affiliation{Departamento de Física Aplicada, Universidad de Alicante, Campus de San Vicente del Raspeig, E-03690 Alicante, Spain}
\author{Sharmanthie Fernando \orcidlink{0000-0003-4198-255X}}
\email[Email: ]{fernando@nku.edu}
\affiliation{Department of Physics, Geology $\&$ Engineering Technology Northern Kentucky University
Highland Heights, Kentucky 41099, U.S.A.}
\author{Grigoris Panotopoulos \orcidlink{0000-0002-7647-4072
}}
\email[Email: ]{grigorios.panotopoulos@ufrontera.cl}
\affiliation{Departamento de Ciencias Físicas,
Facultad de Ingeniería y Ciencias
Universidad de La Frontera, Casilla 54-D
Temuco, Chile}
\author{Leonardo Balart \orcidlink{0000-0002-2425-4369}}
\email[Email: ]{leonardo.balart@ufrontera.cl}
\affiliation{Departamento de Ciencias Físicas,
Facultad de Ingeniería y Ciencias
Universidad de La Frontera, Casilla 54-D
Temuco, Chile}
%


\begin{abstract}
In this paper we study quasinormal modes and absorption cross sections for the  $(1+3)$-dimensional Bardeen black hole surrounded by perfect fluid dark matter. Studies of the massless scalar field is already done in \cite{Sun:2023slzl}. Hence, in this paper we will focus on the massive scalar field perturbations and massless Dirac field perturbations. To compute the quasinormal modes we use the semi-analytical 3rd-order WKB method, which has been shown to be one of the best approaches when the effective potential is adequate and when $n < \ell$ and $n < \lambda$. We have also utilized the Pöschl-Teller method to compare the valus obtained using the WKB approach. We have computed quasinormal frequencies by varying various parameters of the theory such as the mass of the scalar field $\mu$, dark matter parameter $\alpha$ and the magnetic charge $g$. We have summarized our solutions in tables and figures for clarity. As for the absorption cross section, we used third order WKB approach to compute reflection, transmission coefficients and partial absorption cross sections. Graphs are presented to demonstrate the behavior of the above quantities when the dark matter parameter and mass of the massive scalar field are varied.
\end{abstract}

\maketitle

\section{Introduction}

General Relativity (GR in what follows) has a special place over the rest of theories of gravity given that it provides a relatively simple and accurate description of how gravity works ~\cite{Einstein:1915ca}: it justifies the Universe at large scale, and also fits well for moderate length scales: however, it has failed to describe physics at a small scale.
More precisely, GR is considered the privileged theory because a significant amount of evidence supports it. In particular, two facts play a crucial role and are too important to be ignored. They are: i) the existence of gravitational waves (hereafter GWs), and ii) the existence of black holes (hereafter BHs).
They are of critical significance and are undeniable proofs of the validity of Einstein's general theory of relativity.

After the first direct detection of GWs with the emission of signals from a binary BH merger (about a decade ago) and the subsequent ringdown of the single resulting BH~\cite{LIGOScientific:2016aoc}, a completely new era for gravitational wave physics began.
In addition, the detection of X-rays from intensely heated material swirling around an unknown object provided strong evidence for a central BH such as Cygnus X-1 in the Milky Way.
The massive amount of data coming from gravitational wave astronomy gives us the opportunity to test GR as well as alternative theories of gravity. Regardless of the theory, the study of BH physics can be used to clarify the mysteries of the cosmos at a fundamental level. For instance, BH physics could provide insights into the unification of GR, quantum theory, and statistical mechanics~\cite{Hawking:1975vcx,Bekenstein:1973ur,Bardeen:1973gs,Hawking:1982dh}, which could represent a significant advance in the understanding of quantum gravity.

In the context of BH physics, it becomes relevant to study how BHs are affected when external perturbations are applied. External perturbations are interesting because realistic BHs do not exist in isolation in nature; rather, they are continuously interacting with their surrounding environment. Thus, depending on how the BH responds to perturbations, it will be stable (if the amplitude decays with time) or unstable (if the amplitude grows with time). More precisely, the so-called Quasi-Normal Modes (hereafter QNMs) are complex frequencies that encode how a BH responds to small perturbations (see \cite{Regge:1957td,Zerilli:1970se,Zerilli:1970wzz,Zerilli:1974ai,Moncrief:1975sb,Teukolsky:1972my} and also the seminal monograph \cite{Chandrasekhar:1985kt}).

The BH perturbations can be broken down as follows:
i) the generation of radiation in response to initial conditions, 
ii) damped oscillations described by complex frequencies, and finally
iii) a power-law decay of the fields involved.
The QNM frequencies characterizing phase (ii) and are defined as 
$\omega \equiv \omega_R + i \omega_I$. One of the most remarkable facts about QNMs is that they depend on only a few parameters of the BH, such as i) mass, ii) electric charge, and iii) rotation.
It should be mentioned that both the real and imaginary parts of the QNM frequencies have a physical meaning: the real part $\omega_R$ governs the period of oscillation, expressed as $T = 2\pi/\omega_R$, and the imaginary part $\omega_{I}$ encodes the fluctuation decay with a time scale of $t_D = 1/\omega_I$.

%
%
QNMs of BHs have been studied extensively over the years, using classical and novel methods to compute the exact and numerical spectrum, not only for static BHs but also for rotating ones. There are many works in the literature on computing QNMs for a variety of BHs and methods of computing QNMs. Due to the large volume of work,we will mention only a few here: Li et al.\cite{Li:2015ly} calculated QNM frequencies for a regular BH with a magnetic charge. Scalar QNMs of accelerating Kerr-Newman-AdS BHs were studied in \cite{Amado:2024ag}. There has been indication that QNMs play a role in phase transitions of BHs: in \cite{Priya:2024sp}, dyonic BHs in Einstein-Maxwell-scalar gravity were studied to understand the interplay between QNMs and phase transitions. One of the models suggested for Quantum Gravity is Loop Quantum Gravity; in\cite{Zhang:2024zw} \cite{Livine:2024lm}, QNMs have been studied for BHs in Loop Quantum Gravity. QNM frequencies of BHs in modified gravity are studied in \cite{Chung:2024cy} \cite{Chung:2024cy1}. From the experimental side, BH spectroscopy with nonlinear QNMs has been studied in \cite{Lagos:2025la}. 
Other contributions related to the computation of QNMs can be easily found in the bibliography of the aforementioned papers, but other useful papers are \cite{Rincon:2018sgd,Panotopoulos:2017hns,Panotopoulos:2019qjk,Panotopoulos:2020mii,Rincon:2020cos,Rincon:2024won,Balart:2023swp,Skvortsova:2024eqi}.

Greybody factor (GBF) is the probability of a wave tunneling through an effecive potential due to a perturbation by a field: it is also given as $|T(\omega)|^2$ with $T(\omega)$ being the transmission coefficient. One can sum the partial cross sections, $\sigma_l = \frac{ \pi (2 l +1) |T(\omega)|^2}{\omega^2}$ to obtain the total absorption cross section for that particular field. There are many works on GBFs of BHs in the literature: we will mention only a few here. From an observational standpoint, GBFs have been proposed as stable gravitational wave observable \cite{Rosato:2024rdp}. Konoplya and Zinhailo \cite{Konoplya:2019kz} studied GBFs of non-Schwarzschild BHs in higher derivative gravity. GBF, along with QNMs and thermodynamics of AdS BHs, were studied in \cite{Lin:2024lbz}. Few more related to GBFs of BHs are \cite{li:2024lzls, Al-Badawi:2023al, Baruah:2023bod}. There is a nice review on BH GBFs in \cite{Sakalli:2022sk}.
Additional related paper can be consulted in \cite{Ovgun:2023ego,Panotopoulos:2018pvu,Panotopoulos:2016wuu,Panotopoulos:2017yoe,Lambiase:2023zeo,Javed:2022rrs,Kim:2007gj,Lutfuoglu:2025ljm,Lutfuoglu:2025hjy,Hamil:2024njs}, for instance.

Rubin et. al \cite{Rubin:1970rf} studied rotation curves of Andromeda galaxy and observed that the stars at the edge were moving faster than expected: Since then, much work has been focused on understanding what causes this anomaly in the motion. Now it is accepted that there are nonbaryonic matter which does not emit light but exert gravitational attraction; these matter is called ``dark matter" (DM). It is the general acceptance that the current Universe consists of 
\noindent
26.8 $\%$ DM with the rest made of 4.9 $\%$ baryonic matter and 68.3 $\%$ dark energy. Recent observations have led to the belief that supermassive BHs at the center of almost all elliptical and spiral galaxies are surrounded by DM halos \cite{akiyama:2019eh}. Hence in the quest to understand the Universe, studying BHs with DM takes a central role.

There are several works that have focused on studying models with BHs immersed in DM: we will mention a few here. Mannheim and Kazanas \cite{Mannheim:1989mk} found Schwarzschild-like solutions with DM term from pure Weyl gravity without the Einstein term. QNMs of such BHs were studied recently by Konoplya et.al \cite{Konoplya:2025kk}. Al-Badawi and 
Shaymatov studied BHs with Dehnen DM halo \cite{Al-Badawi:2024bs}. This BH also has a string cloud around it. The authors studied QNMs and the shadow of the aforementioned BH. Thermodynamics and shadow bound of a BH surrounded by a DM halo was studied by Myung \cite{Myung:2025m}. Charged BHs surrounded by Hernquist DM were studied by Mollicone and Destounis \cite{Mollicone:2025md}. Schwarzschild-like BH surrounded by pseudo-isothermal DM halo was studied in \cite{Liu:2025lmtw}.

A complete landscape of BH physics is still incomplete because, among other things,  many BHs still have singularities: such a fact is a problem, although they are expected to occur under certain conditions \cite{Hawking:1973uf}. Note that the classical solution of Einstein's field equations presents both future singularities \cite{Penrose:1964wq} and past singularities \cite{Hawking:1965mf,Hawking:1966sx,Hawking:1966jv,Hawking:1967ju},and these singularities are usually hidden behind an event horizon \cite{Israel:1967za}. There are some static BHs which have an event horizon, combined with finite values for the curvature invariants (e.g., $R$, $R_{\mu\nu}R^{\mu\nu}$, $R_{\kappa\lambda\mu\nu}R^{\kappa\lambda\mu\nu}$) over the whole range of the radial coordinate $r$. Usually, these types of BHs are collectively called "regular" BHs, but to be precise, we should call them non-singular BHs. Note that in order to ensure that a BH is regular, it is necessary to study: (i) the divergence of the curvature invariants and (ii) the incompleteness of the geodesics (see \cite{Lan:2023cvz} for more details).

Kiselev \cite{Kiselev:2002dx} suggested that Perfect Fluid Dark Matter (hereafter PFDM) could be a realistic model, considered as a quintessence scalar field. Zhang and collaborators \cite{Zhang:2020mxi} considered a Bardeen BH (assuming non-linear electrodynamics) immersed in PFDM in four dimensions and found new analytical BH solutions (static and rotating solutions); they also studied thermodynamics and energy conditions of these BHs. There has been several work focusing on BHs with PFDM. We will mention a few here: The motion of particles around the Schwarzschild-de Sitter BH surrounded by PFDM was studied in \cite{Rayimbaev:2021rs}. Phase transitions of a rotating BH immersed in PFDM were studied in \cite{Hendi:2020hnlj}. Haroon et al. \cite{Haroon:2019hjjlm} studied the shadows and light deflection from a rotating BH with a cosmological constant immersed in PDFM. A method of distinguishing a Kerr BH from a naked singularity when both are immersed in PFDM was discussed in \cite{Rizwan:2019rjj}. Some additional works on BHs with PFDM are \cite{Narzilloev:2020nrsaab} \cite{Shaymatov:2021saj} \cite{Shaymatov:2021sma}.

Our goal in this paper is to study the BH solutions with PFDM derived by Zhang et al. \cite{Zhang:2020mxi} in the context of QNMs and absorption cross sections: here we will focus on massive scalar perturbations and Dirac massless perturbations.

The manuscript is structured as follows. In Sec.\eqref{sec2} we will present the main ingredients necessary to understand the background of the BH to be investigated. Then, in Sec.\eqref{sec3}, we will derive the wave equations, discuss the methods to obtain the QNM frequencies and compute the QNM frequencies. In Sec.\eqref{sec4} we compute the absorption cross section and finally in Sec.~\eqref{sec5} we summarize our main findings, discuss our results, and propose possible ideas for future work.

\section{Background: Bardeen black hole immersed in PFDM} \label{sec2}

In this section, we will present the relevant equations useful to generate the BH background of interest.
The corresponding Einstein's field equations for the nonlinear electromagnetic field and the PFDM are given by,
\begin{align}
G_{\mu}^{\,\nu} &= 8\pi \bigl( {T_{\mu}^{\nu}}^{\text{NLE}} +  {T_{\mu}^{\nu}}^{\text{M}} \bigl),\label{eq:EM1}
\end{align}
Here,  the electromagnetic tensor ${T_{\mu}^{\nu}}^{\text{NLE}}$ is defined as follows:
\begin{align}
 {T_{\mu}^{\nu}}^{\text{NLE}} &\equiv \frac{1}{4\pi} \bigg( \mathcal{L}_FF_{\mu\lambda}F^{\nu\lambda} - \delta_{\mu}^{\,\nu}\mathcal{L} \bigg) ,
\end{align}
where $ \frac{\partial\mathcal{L}\left(F\right)}{\partial F} \equiv \mathcal{L}_F$. In addition, the  equations of motion for the non-linear electromagnetic field are:
\begin{align}
\nabla_{\mu}\bigl( \mathcal{L}_F F^{\nu\mu} \bigl) &= 0 ,\label{eq:EM2}
\\
\nabla_{\mu}\left(*F^{\nu\mu}\right) & = 0.\label{eq:EM3}
\end{align}
Notice that $F_{\mu\nu}=2\nabla_{[\mu}A_{\nu]}$ and $\mathcal{L}$ is a function of $F\equiv\frac{1}{4}F_{\mu\nu}F^{\mu\nu}$ defined by \cite{Ayon-Beato:2000mjt}
\begin{equation}
\mathcal{L}\left(F\right)=\frac{3M}{\vert g\vert^{3}}\left(\frac{\sqrt{2g^{2}F}}{1+\sqrt{2g^{2}F}}\right)^{\frac{5}{2}}.
\end{equation}
Here, $g$ corresponds to the magnetic charge, and $M$ represents the BH mass. 
For a BH surrounded by PFDM, the energy momentum tensor can be expressed as ${T_{\nu}^{\mu}}^{\text{M}}=\mathrm{diag}(-\epsilon,p_r,p_\theta,p_\phi)$ with the density, radial and tangential pressures of DM given by \cite{Kiselev:2002dx,Li:2012zx}
\begin{equation}
-\epsilon=p_r=\frac{\alpha}{8\pi r^3}\qquad \text{and} \qquad p_\theta=p_\phi=-\frac{\alpha}{16\pi r^3}\ .
\end{equation}
Now, let us consider the line element for a static and spherically symmetric background. 
Assuming the Schwarzschild ansatz for the metric components, we  can assume the following:
\begin{equation}
 \label{eq:spherically metric}
 {ds}^{2}=-f(r) {d} t^{2}+f(r)^{-1} {~d} r^{2}+r^{2} {~d} \Omega^{2},
\end{equation}
where we can consider the metric potential defined as,
\begin{align}
    f(r) &= 1 - \frac{2 \mathcal{M}(r)}{r},
\end{align}
being $\mathcal{M}(r)$ defined as
\begin{align}
    \mathcal{M}(r) & = \frac{M}{\left(1 + \left(\frac{g}{r}\right)^{2}\right)^{\frac{3}{2}}} - \frac{1}{2}\alpha\ln{ \left(\frac{r}{\vert\alpha\vert} \right)},
\end{align}
Now, we can write $f(r)$ as,
\begin{equation}\label{eq:f}
  f\left(r\right)=1-\frac{2Mr^{2}}{\left(r^{2}+g^{2}\right)^{\frac{3}{2}}} + \frac{\alpha}{r}\ln{ \left( \frac{r}{\vert\alpha\vert} \right)}.
\end{equation}
At this point we should mention that $\alpha$ is a parameter that encodes the strength of the dark matter content, and is also directly related to the density and pressure of the PFDM (conventionally called the dark matter parameter). 
In principle, the parameter $\alpha$ is free, but it has been shown that the weak energy condition is satisfied when $\alpha<0$ (see for example \cite{Zhang:2020mxi}), so we will only consider negative values of the parameter $\alpha$.
Finally, it should be mentioned that we obtain the Schwarzschild BH when both $\{\alpha, g\} \rightarrow \{0,0\}$. If we consider the dark matter parameter $\alpha$ to be zero, it reduces to the Bardeen BH, and it reduces to the Schwarzschild BH immersed by PFDM when $g=0$.

\section{Quasinormal modes} \label{sec3}

In the following, we will present the relevant equations and the basic theory needed to introduce the reader to the key points involved in the computation of the corresponding QNM frequencies. In particular, we will study a Bardeen BH immersed in perfect fluid dark matter: we will compute QNM frequencies for perturbations of the BH by massive scalar field and massless Dirac field. We will use two methods: the first is the well-known WKB method, while the second approach takes advantage of the Pöschl-Teller potential.

Although we will focus on only two methods, it becomes essential to highlight the state-of-the-art about the plethora of methods used to obtain the corresponding QNMs, both analytical and numerical.
The computation of QNMs is treated differently depending on the complexity of the background and the fields we are considering, however, by ignoring unnecessary details, we can assure that exact analytical solutions for quasinormal spectra of BHs are available only in a limited number of cases. To name a few, we can point out:
i)  {\bf{Pöschl-Teller Potential}}: when the effective potential barrier is represented by the Pöschl-Teller potential, as explored in studies like \cite{Poschl:1933zz,Ferrari:1984zz,Cardoso:2001hn,Cardoso:2003sw,Molina:2003ff,Panotopoulos:2018hua,Hamil:2024rsg} the QNMs frequencies can be obtained using a simple relation (see subsequent paragraphs).
ii)  {\bf{Hypergeometric Functions}}: when the differential equation of the radial wave function can be transformed into Gauss's hypergeometric function, as discussed in \cite{Birmingham:2001hc,Fernando:2003ai,Fernando:2008hb,Gonzalez:2010vv,Destounis:2018utr,Ovgun:2018gwt,Rincon:2018ktz} QNM frequencies can be computed. 
iii) {\bf{Heun functions}}: If we consider Teukolsky's equations for certain cases (including Kerr-de Sitter BHs), it is possible to transform the resulting equations into Heun's equations, giving us the possibility to compute their QNM frequencies exactly by the Heun function \cite{Hatsuda:2020sbn,Fiziev:2011mm,Naderi:2024dhh}.
Due to the complexity and nontrivial nature of the differential equations involved, numerical or semi-analytical methods are often required to compute QNM frequencies. Several techniques have been developed for this purpose, including:
i) {\bf{Frobenius Method and its generalizations}} \cite{Destounis:2020pjk,Fontana:2022whx,Hatsuda:2021gtn},
ii) {\bf{Continued Fraction Method}}, along with its refinements \cite{Leaver:1985ax,Nollert:1993zz,Daghigh:2022uws}, 
iii) {\bf{Asymptotic Iteration Method}} \cite{Cho:2011sf,2003JPhA...3611807C,Ciftci:2005xn}, etc.
For a more comprehensive overview, see \cite{Konoplya:2011qq}.

\subsection{Summarize of Methods to compute QNMs}
\subsubsection{WKB method}

The WKB method is a semi-analytic technique used for compute the QNMs of BHs \cite{Schutz:1985km,Iyer:1986np,Iyer:1986nq,Kokkotas:1988fm,Seidel:1989bp}. It was first introduced by Schutz and Will \cite{Schutz:1985km}, it was later refined by Iyer and Will \cite{Iyer:1986np} to include second and third-order corrections. It is also efficient for determining the lower overtones of oscillating Schwarzschild BHs. Its accuracy improves with higher angular harmonic index $\ell$ but decreases for higher overtone indices.
Subsequently, R.A. Konoplya  generalized to the 6th order \cite{Konoplya:2003ii}, while J. Matyjasek and M. Opala derived the formulas for orders 7 through 13 \cite{Matyjasek:2017psv}.
The WKB formula is based on matching asymptotic solutions involving a combination of incoming and outgoing waves, with a Taylor expansion around the peak of the potential barrier at $x = x_0$.
This expansion includes the region between the two turning points (the roots of the effective potential $U(x,\omega) \equiv V(x) - \omega^2$).
The WKB formula to compute the QNM spectra at any order (up to 13th) is obtained by using the following expression
\begin{equation}
\omega_n^2 = V_0+(-2V_0'')^{1/2} \Lambda(n) - i \nu (-2V_0'')^{1/2} [1+\Omega(n)]\,,
\end{equation}
where 
i) $V_0''$ is the second derivative of the potential (evaluated at the maximum), 
ii) $V_0$ represents the maximum of the effective barrier,  
iii) $\nu = n+1/2$, and
iv) $n=0,1,2...$ is the overtone number.
The two functions: $\Lambda(n)$ and $\Omega(n)$ are extremely long expressions of $\nu$ (and derivatives of the potential evaluated at the maximum). In light of this, we decided to avoid showing the concrete form of them (the expression can be consulted in \cite{Kokkotas:1988fm}). 
Thus, to perform our computations, we have used here a Wolfram Mathematica \cite{wolfram} notebook utilizing the WKB method at any order from one to 13 \cite{Konoplya:2019hlu}.
In addition, for a given angular harmonic index, $\ell$, we will consider values $n < \ell$ (or $n < \lambda$) only. For higher order WKB corrections (and recipes for simple, quick, efficient, and accurate computations) see \cite{Konoplya:2019hlu,Hatsuda:2019eoj}. 
%

\subsubsection{Pöschl-Teller method}

Mashhoon and Ferrari proposed to use Pöschl-Teller (PT) potential for calculating the corresponding QNMs \cite{Ferrari:1984zz}. The idea is extremely simple, i.e., replace the effective potential by the PT potential, namely:
\begin{align}
V(r^*) \sim V_{\text{PT}}(r^*) = \frac{V_0}{\cosh^2 \beta(r^*-r^*_0)},
\end{align}
and, using that,  obtain directly the modes with the help of  the expression:
\begin{align}
    \omega_n &= \pm \beta \sqrt{\frac{V_0}{\beta^2} - \frac{1}{4}} - i \beta \bigg( n + \frac{1}{2} \bigg),
\end{align}
where $\beta$ is defined according to:
\begin{align}
    \beta &= \sqrt{-\frac{1}{2V_0} \frac{\mathrm{d}^2 V(r^*)}{\mathrm{d}{r^*}^2}} \Bigg{|}_{r^* = r^{*}_0}.
\end{align}
In the last expression, $r^{*}_0$ represents the position of the maximum of the effective potential $V(r^*)$ and the value $V_0$ is precisely the evaluation of the effective potential at the maximum, i.e., $V(r^{*}_0) \equiv V_0$.

\subsection{Wave equation for massive scalar perturbations 
}

Let us start by analyzing the dynamics of a test massive scalar field, represented by $\Phi$, propagating in a four-dimensional spacetime within a prescribed gravitational background. We also assume that the scalar field is real. 
Being $S[g_{\mu \nu}, \Phi]$ the action, by vary it we obtain 
the Klein-Gordon equation of a massive scalar field $\Phi$ in curved spacetime, namely:
\begin{equation}
\frac{1}{\sqrt{-g}}\partial_\mu \left(\sqrt{-g}g^{\mu \nu}\partial_\nu\Phi\right)
- \mu^2 \Phi =0 \, ,
\label{KGg-M}
\end{equation}
where $\mu$ represents the mass of the scalar field.

The solution can be decomposed into a radial and an angular part as
\begin{equation}\label{tr}
 \Phi(t,r,\theta,\phi)=\frac{1}{r}\,e^{-i \omega t}Y_{\ell}(\theta, \phi)\Psi_s(r) \, ,
\end{equation}
where $\omega$ is the angular frequency of the scalar field.

Replacing this last ansatz in the Klein-Gordon equation and using the tortoise coordinates, we arrive at the following equation
\begin{equation}
\frac{d^2\Psi_s}{dr_*^2}+\left(\omega^2-V_{s}(r_*)\right)\Psi_s=0,
\label{wave-equation-b}
\end{equation}
where
\begin{equation}
V(r) = f(r) \left(\frac{\ell (\ell +1)}{r^2} + \frac{f'(r)}{r} + \mu^2\right)
\label{potential-mu}
\end{equation}




\begin{figure} [H]
\begin{center}
\includegraphics[width=0.497\linewidth]{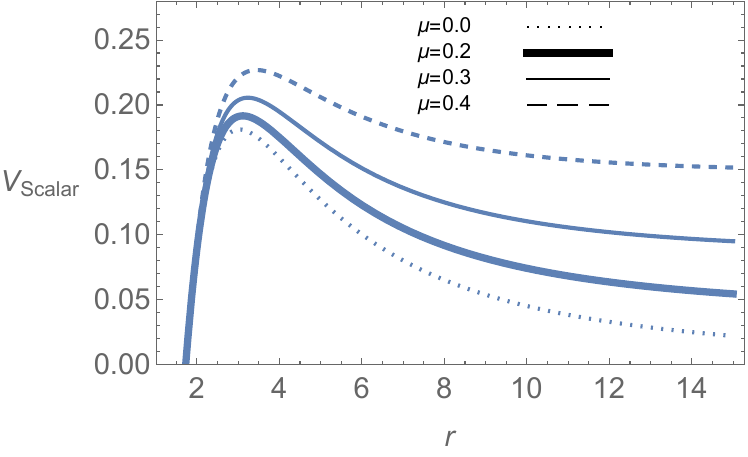} 
\includegraphics[width=0.497\linewidth]{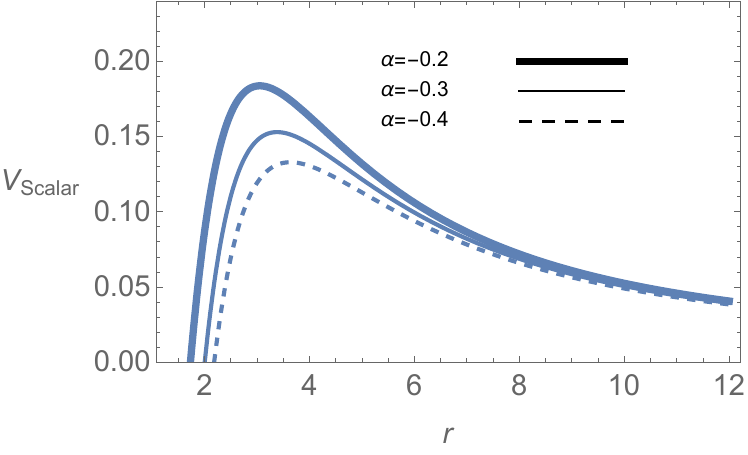}
\caption{
{\bf{Left panel:}} The figure shows $V_{Scalar}(r)$ versus $r$ for varying $\mu$. Here $M = 1, \alpha=-0.2$, g=1 and $\ell = 2$.
{\bf{Right panel:}} The figure shows $V_{Scalar}(r)$ versus $r$ for varying $\alpha$. Here $M = 1, \mu = 0.1$ and $g = 1$ and $\ell=2$.
}
\label{potscalar}
 \end{center}
 \end{figure}

\begin{figure} [H]
\begin{center}
\includegraphics[width=0.497\linewidth]{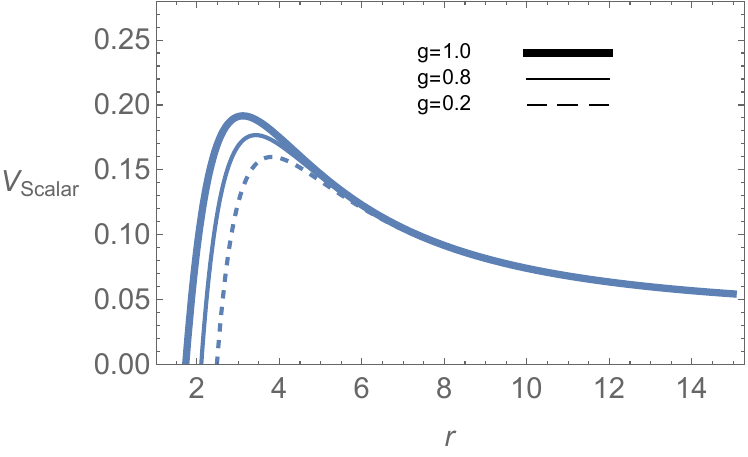} 

\caption{The figure shows $V_{Scalar}(r)$ versus $r$ for varying $g$. Here $M = 1, \alpha=-0.2, \mu = 0.2$ and $\ell = 2$.}
\label{potscalarg}
 \end{center}
 \end{figure}

In Fig.\eqref{potscalar} (left) the scalar potential is plotted against the radial coordinate $r$ for varying $\mu$, for fixing $M=1,\alpha=-0.2, g=1$ and $\ell=2$. 
As $\mu$ increases, so does the potential, and the maximum is slightly to the right of the massless case. We noticed an interesting behavior of $V(r)$ for large r as follows: when expanded for large $r$, $V(r)$ is,
\begin{equation}
    V_{\text{eff}} \rightarrow \mu^2 - \left( \frac{2 \mu^2 M}{r} + \frac{\mu^2 \alpha \ln(|\alpha|)}{r} \right) + \frac{\ell(\ell+1)}{r^2} + \frac{\mu^2 \alpha \ln(r)}{r} + \mathcal{O}(r^{-3}).
\end{equation}

One can observe that the coupling term of $\mu^2$ with the BH mass $M$ and the  dark matter parameter $\alpha$ generates a Newtonian-like attraction at $\mathcal{O}(1/r)$. Of course, for it to be an attraction, $M$ has to be larger than $\alpha \ln(|\alpha|)$ since $\alpha <0$. The angular momentum $\ell$ creates a potential barrier at $\mathcal{O}(r^{-2})$. Since $\ln(r)/r \rightarrow 0$ for large $r$, one could see that $V(r) \rightarrow \mu^2$ for very large $r$. Since it is clear that the potential differs for the massive case compared to the massless case, it is expected to modify the quasinormal frequencies.

In Fig.\eqref{potscalar} (right) the scalar potential is plotted against the radial coordinate $r$ for varying $\alpha$, for fixing $M=1,\mu=0.1, g=1$ and $\ell=2$.  As $\alpha$ decreases, so does the potential. For a fixed value of $\mu$, the asymptotic value seems similar for different values of $\alpha$.  
In Fig.\eqref{potscalarg} the scalar potential is plotted against the radial coordinate $r$ for varying $g$, for fixing $M=1,\alpha=-0.2, \mu=0.2$ and $\ell=2$. 
Similarly to the previous cases, the height of the potential increases as $g$ increases. The maximum of the potential is also shifted to the left as $g$ increases.

\begin{figure} [H]
\begin{center}
\includegraphics[width=0.497\linewidth]{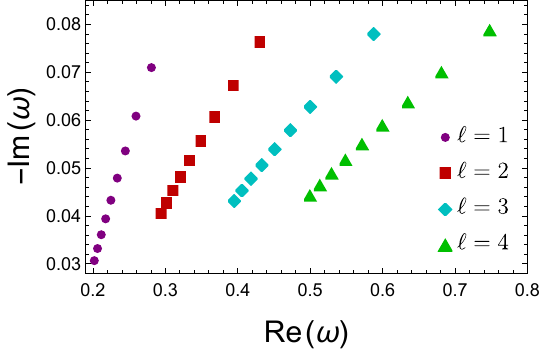} 
\includegraphics[width=0.497\linewidth]{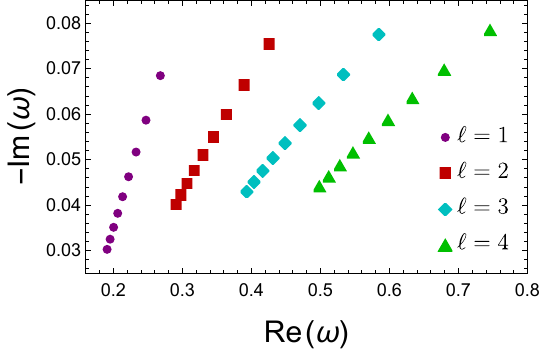} 
\caption{
{\bf{Left panel:}} Quasinormal modes for massive $(\mu = 0.2)$ scalar perturbations utilizing the Pöschl-Teller fitting approach assuming $M=1, \ g=0.1, \ n=0 $ varying $\alpha$ in the range $-0.9 \leq \alpha \leq -0.1$ with four different values of $\ell$ (see figure for details).
{\bf{Right panel:}} Quasinormal modes for massive $(\mu = 0.2)$ scalar perturbations utilizing the WKB approximation assuming $M=1, \ g=0.1, \ n=0 $ varying $\alpha$ in the range $-0.9 \leq \alpha \leq -0.1$ with four different values of $\ell$ (see figure for details).
}
\label{Fig3}
 \end{center}
 \end{figure}

\begin{figure} [H]
\begin{center}
\includegraphics[width=0.497\linewidth]{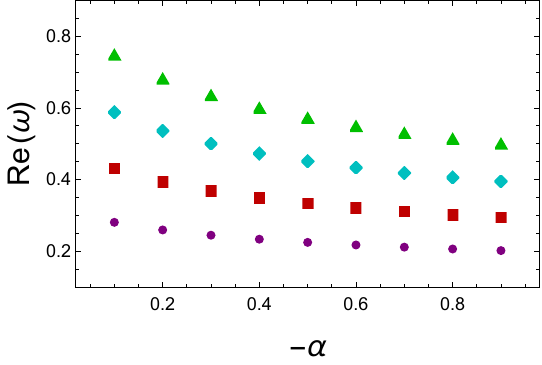} 
\includegraphics[width=0.497\linewidth]{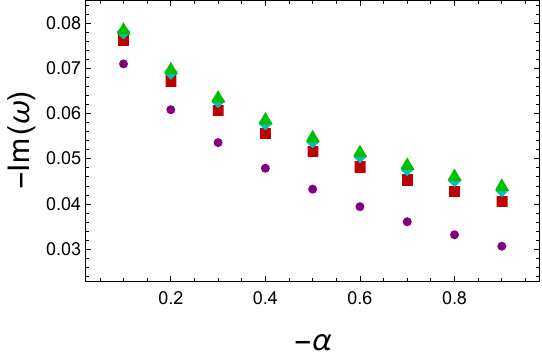}
\\
\includegraphics[width=0.497\linewidth]{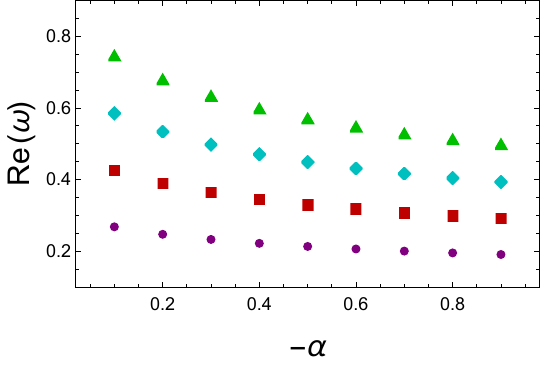} 
\includegraphics[width=0.497\linewidth]{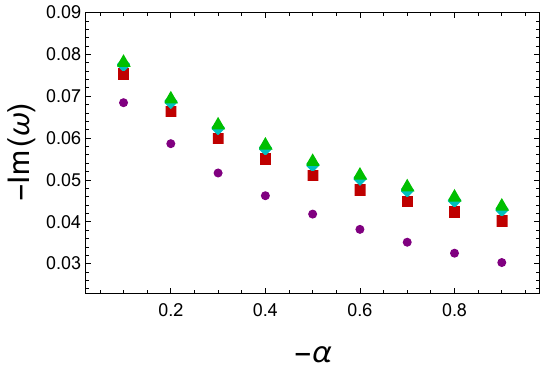}
\caption{
{\bf{First row:}} The left panel shows the real part of the quasinormal frequency $\omega_R$, while the right panel shows the imaginary part of the quasinormal frequency $\omega_I$ both cases against the parameter $\alpha$, 
for massive $(\mu = 0.2)$ scalar perturbations
assuming $M=1, \ g=0.1, \ n=0 $ with four different values of $\ell$ (color code as in the previous figure) using  the Pöschl-Teller fitting approach.
{\bf{Second row:}} The left panel shows the real part of the quasinormal frequency $\omega_R$, while the right panel shows the imaginary part of the quasinormal frequency $\omega_I$ both cases against the parameter $\alpha$, 
for massive $(\mu = 0.2)$ scalar perturbations
assuming $M=1, \ g=0.1, \ n=0 $ with four different values of $\ell$ (color code as in the previous figure) using the WKB approximation.
}
\label{Fig4}
\end{center}
\end{figure}

We have calculated the quasinormal modes for massless and massive scalar fields. The massless case has already been published \cite{Sun:2023slzl} and is just added for comparison. We have  computed the QNMs for massive scalar perturbations using two complementary methods: i) WBK approximation and ii) Pöschl-Teller fitting approach. We have considered several cases, which are summarized in Tables \eqref{table:First set}, \eqref{table:Second set}, \eqref{table:Third set} and \eqref{table:Fourth set}, and we have made a graphical representation in Figures \eqref{Fig3} and \eqref{Fig4}. Since the imaginary part is negative, all modes are found to be stable against scalar perturbations.

For the massive scalar field, when the multipole number $\ell$ increases, Re($\omega$) increases, and -Im($\omega$) increases: hence, for high multipole numbers, the damping is high and the oscillating frequency is high. In comparison, for a massless scalar field perturbation (as given in Table I), when the multipole number $\ell$ increases, Re($\omega$) increases and -Im($\omega$) decreases. Having a mass for the scalar field does induce more damping for the field. When the overtone number $n$ increases, Re($\omega$) decreases and Im($\omega$) increases: hence, for high overtones, the damping is high and the oscillation is low. This is in line with the massless field QNM frequencies. When the dark matter parameter $\alpha$ increases,  both Re($\omega$)  and --Im($\omega$) increases. Hence for high $\alpha$, oscillation frequency is high and damping is also high. This is very similar to the behavior of QNM frequencies for the massless scalar field computed in our paper. However, in reference \cite{Sun:2023slzl} the behavior of QNM frequencies for massless scalar field are slightly different: Re($\omega$) behaves similar to our observation, but as for -Im($\omega$), there is a range of $\alpha$ where -Im($\omega$) decreases.

\begin{figure} [H]
\begin{center}
\includegraphics[width=0.497\linewidth]{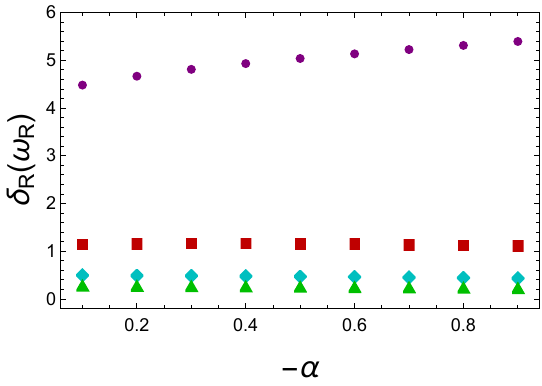} 
\includegraphics[width=0.497\linewidth]{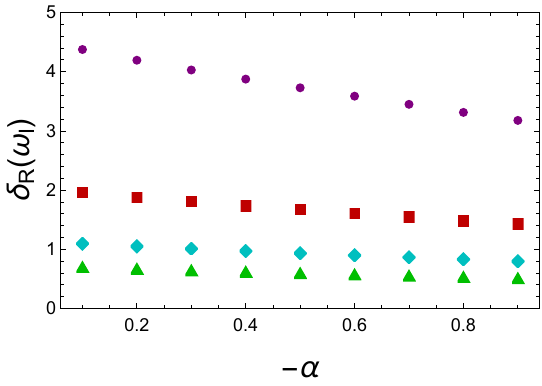}
\caption{
The figure shows the percent error of the real and imaginary quasinormal frequencies against the parameter $-\alpha$.
{\bf{Left panel:}} It shows the percent error of the real part of the quasinormal frequency $\delta_R(\omega_R)$ versus $-\alpha$, for different values of the angular number $\ell$ (color code as previous figures)
for massive $(\mu = 0.2)$ scalar perturbations
assuming $M=1, \ g=0.1, \ n=0 $. The percent error compares the modes obtained using the two complementary methods mentioned in the manuscript.
{\bf{Right panel:}} It shows the percent error of the imaginary part of the quasinormal frequency $\delta_R(\omega_I)$ versus $-\alpha$, for different values of the angular number $\ell$ (color code as previous figures)
for massive $(\mu = 0.2)$ scalar perturbations
assuming $M=1, \ g=0.1, \ n=0 $. The percent error compares the modes obtained using the two complementary methods mentioned in the manuscript.
}
\label{FigNew1}
\end{center}
\end{figure}

In Fig.\eqref{FigNew1} we have included  the percent error between the WKB method and the Pöschl-Teller method.
We have used the standard definition of the percent error, i.e., 
\begin{align}
    \delta_R(\omega_R) &\equiv \Bigg| \frac{\omega_R^{PT} - \ \omega_R^{WKB}}{\omega_R^{PT}}\Bigg|\times100\%
    \\
      \delta_R(\omega_I) &\equiv \Bigg| \frac{\omega_I^{PT} - \ \omega_I^{WKB}}{\omega_I^{PT}}\Bigg|\times100\%
\end{align}
According to the figures, the real part has an error margin of around $5\%$ for the lower value of $\ell$, and this margin is drastically reduced to $1\%$ or lower as the parameter increases. This behaviour is consistent, given that the WKB approximation is known to work better for large values of $\ell$. Similarly, the figure for the imaginary part has a percentage error around $3-4\%$ and starts to decrease as $-\alpha$ increases. Furthermore, as $\ell$ increases, the percentage error reduces to $2\%$ or less.

\begin{table}[ph!]
\centering
\caption{Quasinormal frequencies for massless $(\mu=0.0)$ scalar perturbations (varying $\ell$, $n$ and $\alpha$) fixing $M=1,g=0.1$ for the model considered in this work using the WKB approximation.
}
{
\begin{tabular}{c|c|cccc} 
\toprule
$\alpha$  &$n$  &  $\ell=1$ & $\ell=2$ & $\ell=3$ & $\ell=4$ 
\\ 
\colrule
\hline
-0.1 & 0 & 0.247428 - 0.0812135 I & 0.410636 - 0.0802953 I & 0.573818 - 0.0800706 I & 0.737122 - 0.0799817 I \\
-0.1 & 1 &                        & 0.394699 - 0.2451120 I & 0.562051 - 0.2424120 I & 0.727852 - 0.2412850 I \\
-0.1 & 2 &  				  	  &                        & 0.541633 - 0.4094670 I & 0.710954 - 0.4058460 I \\
-0.1 & 3 &                        &                        &                        & 0.688501 - 0.5740170 I \\
\botrule
\hline
-0.2 & 0 & 0.224921 - 0.0723728 I & 0.373267 - 0.0716020 I & 0.521624 - 0.0714134 I & 0.670094 - 0.0713388 I \\
-0.2 & 1 &                        & 0.359517 - 0.2184040 I & 0.511476 - 0.2161150 I & 0.662100 - 0.2151590 I \\
-0.2 & 2 &  				  	  &                        & 0.493814 - 0.3648700 I & 0.647497 - 0.3617710 I \\
-0.2 & 3 &                        &                        &                        & 0.628060 - 0.5115150 I \\
\botrule
\hline
-0.3 & 0 & 0.209229 - 0.0660777 I & 0.347216 - 0.0654130 I & 0.485242 - 0.0652504 I & 0.623373 - 0.0651861 I \\
-0.3 & 1 &                        & 0.335048 - 0.1993850 I & 0.476264 - 0.1973920 I & 0.616301 - 0.1965580 I \\
-0.3 & 2 &  				  	  &                        & 0.460599 - 0.3331110 I & 0.603361 - 0.3303880 I \\
-0.3 & 3 &                        &                        &                        & 0.586108 - 0.4670040 I \\
\botrule
\hline
-0.4 & 0 & 0.197304 - 0.0611985 I & 0.327425 - 0.0606165 I & 0.457605 - 0.0604742 I & 0.587884 - 0.0604179 I \\
-0.4 & 1 &                        & 0.316498 - 0.1846440 I & 0.449545 - 0.1828810 I & 0.581536 - 0.1821430 I \\
-0.4 & 2 &  				  	  &                        & 0.435449 - 0.3084930 I & 0.569900 - 0.3060640 I \\
-0.4 & 3 &                        &                        &                        & 0.554365 - 0.4325000 I \\
\botrule
\hline
-0.5 & 0 & 0.187819 - 0.0572328 I & 0.311689 - 0.0567183 I & 0.435631 - 0.0565925 I & 0.559669 - 0.0565428 I \\
-0.5 & 1 &                        & 0.301783 - 0.1726630 I & 0.428327 - 0.1710870 I & 0.553916 - 0.1704270 I \\
-0.5 & 2 &  				  	  &                        & 0.415522 - 0.2884830 I & 0.543354 - 0.2862940 I \\
-0.5 & 3 &                        &                        &                        & 0.529233 - 0.4044520 I \\
\botrule
\hline
-0.6 & 0 & 0.180058 - 0.0539057 I & 0.298817 - 0.0534481 I & 0.417662 - 0.0533362 I & 0.536596 - 0.0532920 I \\
-0.6 & 1 &                        & 0.289778 - 0.1626110 I & 0.410999 - 0.1611930 I & 0.531349 - 0.1605980 I \\
-0.6 & 2 &  				  	  &                        & 0.399292 - 0.2716930 I & 0.521699 - 0.2697070 I \\
-0.6 & 3 &                        &                        &                        & 0.508781 - 0.3809160 I \\
\botrule
\hline
-0.7 & 0 & 0.173588 - 0.0510481 I & 0.288094 - 0.0506395 I & 0.402693 - 0.0505397 I & 0.517379 - 0.0505002 I \\
-0.7 & 1 &                        & 0.279808 - 0.1539780 I & 0.396587 - 0.1526950 I & 0.512571 - 0.1521570 I \\
-0.7 & 2 &  				  	  &                        & 0.385837 - 0.2572700 I & 0.503715 - 0.2554600 I \\
-0.7 & 3 &                        &                        &                        & 0.491845 - 0.3606980 I \\
\botrule
\hline
-0.8 & 0 & 0.168128 - 0.0485475 I & 0.279051 - 0.0481822 I & 0.390074 - 0.0480930 I & 0.501179 - 0.0480577 I \\
-0.8 & 1 &                        & 0.271433 - 0.1464230 I & 0.384462 - 0.1452600 I & 0.496760 - 0.1447720 I \\
-0.8 & 2 &  				  	  &                        & 0.374560 - 0.2446480 I & 0.488610 - 0.2429930 I \\
-0.8 & 3 &                        &                        &                        & 0.477669 - 0.3430030 I \\
\botrule
\hline
-0.9 & 0 & 0.163486 - 0.0463248 I & 0.271372 - 0.0459982 I & 0.379360 - 0.0459185 I & 0.487426 - 0.0458869 I \\
-0.9 & 1 &                        & 0.264355 - 0.1397070 I & 0.374193 - 0.1386510 I & 0.483358 - 0.1382070 I \\
-0.9 & 2 &  				  	  &                        & 0.365056 - 0.2334270 I & 0.475844 - 0.2319110 I \\
-0.9 & 3 &                        &                        &                        & 0.465743 - 0.3272700 I \\
\botrule
\hline
\end{tabular} 
\label{table:First set}
}
\end{table}
\begin{table}[ph!]
\centering
\caption{Quasinormal frequencies for massive $(\mu = 0.2)$ scalar perturbations (varying $\ell$, $n$ and $\alpha$) fixing $M=1,g=0.1$ for the model considered in this work using the WKB approximation.
}
{
\begin{tabular}{c|c|cccc} 
\toprule
$\alpha$  &$n$  &  $\ell=1$ & $\ell=2$ & $\ell=3$ & $\ell=4$ 
\\ 
\colrule
\hline
-0.1 & 0 & 0.268153 - 0.0684593 I & 0.425180 - 0.0753569 I & 0.584642 - 0.0774946 I & 0.745683 - 0.0784084 I \\
-0.1 & 1 &                        & 0.402088 - 0.2344950 I & 0.569935 - 0.2358620 I & 0.734955 - 0.2370140 I \\
-0.1 & 2 &  				  	  &                        & 0.545313 - 0.4017220 I & 0.715673 - 0.4000350 I \\
-0.1 & 3 &                        &                        &                        & 0.690715 - 0.5680330 I \\
\botrule
\hline
-0.2 & 0 & 0.247355 - 0.0586715 I & 0.388934 - 0.0663470 I & 0.533262 - 0.0686799 I & 0.679291 - 0.0696712 I \\
-0.2 & 1 &                        & 0.367589 - 0.2070510 I & 0.520020 - 0.2091380 I & 0.669768 - 0.2106200 I \\
-0.2 & 2 &  				  	  &                        & 0.497890 - 0.3565790 I & 0.652650 - 0.3555700 I \\
-0.2 & 3 &                        &                        &                        & 0.630544 - 0.5050990 I \\
\botrule
\hline
-0.3 & 0 & 0.232973 - 0.0516713 I & 0.363732 - 0.0599359 I & 0.497489 - 0.0624090 I & 0.633045 - 0.0634546 I \\
-0.3 & 1 &                        & 0.343669 - 0.1875020 I & 0.485322 - 0.1901160 I & 0.624402 - 0.1918360 I \\
-0.3 & 2 &  				  	  &                        & 0.465006 - 0.3244230 I & 0.608860 - 0.3239110 I \\
-0.3 & 3 &                        &                        &                        & 0.588824 - 0.4602700 I \\
\botrule
\hline
-0.4 & 0 & 0.222098 - 0.0462442 I & 0.344616 - 0.0549777 I & 0.470331 - 0.0575568 I & 0.597926 - 0.0586420 I \\
-0.4 & 1 &                        & 0.325582 - 0.1723630 I & 0.459022 - 0.1753870 I & 0.589983 - 0.1772890 I \\
-0.4 & 2 &  				  	  &                        & 0.440147 - 0.2995040 I & 0.575691 - 0.2993820 I \\
-0.4 & 3 &                        &                        &                        & 0.557290 - 0.4255220 I \\
\botrule
\hline
-0.5 & 0 & 0.213462 - 0.0418534 I & 0.329421 - 0.0509637 I & 0.448738 - 0.0536231 I & 0.570005 - 0.0547373 I \\
-0.5 & 1 &                        & 0.311265 - 0.1600850 I & 0.438153 - 0.1634380 I & 0.562645 - 0.1654830 I \\
-0.5 & 2 &  				  	  &                        & 0.420481 - 0.2792640 I & 0.549395 - 0.2794630 I \\
-0.5 & 3 &                        &                        &                        & 0.532351 - 0.3972860 I \\
\botrule
\hline
-0.6 & 0 & 0.206382 - 0.0382082 I & 0.316985 - 0.0476155 I & 0.431069 - 0.0503347 I & 0.547162 - 0.0514689 I \\
-0.6 & 1 &                        & 0.299608 - 0.1498200 I & 0.421114 - 0.1534390 I & 0.540307 - 0.1555970 I \\
-0.6 & 2 &  				  	  &                        & 0.404489 - 0.2623040 I & 0.527957 - 0.2627710 I \\
-0.6 & 3 &                        &                        &                        & 0.512079 - 0.3736080 I \\
\botrule
\hline
-0.7 & 0 & 0.200443 - 0.0351338 I & 0.306601 - 0.0447623 I & 0.416331 - 0.0475235 I & 0.528118 - 0.0486703 I \\
-0.7 & 1 &                        & 0.289942 - 0.1410460 I & 0.406943 - 0.1448820 I & 0.521712 - 0.1471280 I \\
-0.7 & 2 &  				  	  &                        & 0.391252 - 0.2477640 I & 0.510160 - 0.2484600 I \\
-0.7 & 3 &                        &                        &                        & 0.495312 - 0.3532870 I \\
\botrule
\hline
-0.8 & 0 & 0.195375 - 0.0325181 I & 0.297813 - 0.0422910 I & 0.403878 - 0.0450780 I & 0.512042 - 0.0462306 I \\
-0.8 & 1 &                        & 0.281832 - 0.1334190 I & 0.395013 - 0.1374280 I & 0.506044 - 0.1397400 I \\
-0.8 & 2 &  				  	  &                        & 0.380177 - 0.2350730 I & 0.495216 - 0.2359660 I \\
-0.8 & 3 &                        &                        &                        & 0.481297 - 0.3355260 I \\
\botrule
\hline
-0.9 & 0 & 0.190994 - 0.0302867 I & 0.290308 - 0.0401223 I & 0.393271 - 0.0429198 I & 0.498366 - 0.0440718 I \\
-0.9 & 1 &                        & 0.274984 - 0.1266940 I & 0.384896 - 0.1308400 I & 0.492746 - 0.1332000 I \\
-0.9 & 2 &  				  	  &                        & 0.370860 - 0.2238270 I & 0.482588 - 0.2248910 I \\
-0.9 & 3 &                        &                        &                        & 0.469522 - 0.3197620 I \\
\botrule
\hline
\end{tabular} 
\label{table:Second set}
}
\end{table}


\begin{table}[ph!]
\centering
\caption{Quasinormal frequencies for massless $(\mu=0.0)$ scalar perturbations (varying $\ell$ and $n$ and $\alpha$) fixing $M=1,g=0.1$ for the model considered in this work using the fitting approximation. }
{
\begin{tabular}{c|c|cccc} 
\toprule
$\alpha$  &$n$  &  $\ell=1$ & $\ell=2$ & $\ell=3$ & $\ell=4$ 
\\ 
\colrule
\hline
-0.1 & 0 & 0.253333 - 0.0832982 I & 0.413957 - 0.0811590 I & 0.576130 - 0.0805254 I & 0.738900 - 0.0802602 I \\
-0.1 & 1 &                        & 0.413957 - 0.2434770 I & 0.576130 - 0.2415760 I & 0.738900 - 0.2407810 I \\
-0.1 & 2 &  				  	  &                        & 0.576130 - 0.4026270 I & 0.738900 - 0.4013010 I \\
-0.1 & 3 &                        &                        &                        & 0.738900 - 0.5618210 I \\
\botrule
\hline
-0.2 & 0 & 0.230000 - 0.0741478 I & 0.376134 - 0.0723361 I & 0.523623 - 0.0718000 I & 0.671632 - 0.0715756 I \\
-0.2 & 1 &                        & 0.376134 - 0.2170080 I & 0.523623 - 0.2154000 I & 0.671632 - 0.2147270 I \\
-0.2 & 2 &  				  	  &                        & 0.523623 - 0.3590000 I & 0.671632 - 0.3578780 I \\
-0.2 & 3 &                        &                        &                        & 0.671632 - 0.5010290 I \\
\botrule
\hline
-0.3 & 0 & 0.213713 - 0.0676299 I & 0.349756 - 0.0660542 I & 0.487016 - 0.0655881 I & 0.624739 - 0.0653930 I \\
-0.3 & 1 &                        & 0.349756 - 0.1981620 I & 0.487016 - 0.1967640 I & 0.624739 - 0.1961790 I \\
-0.3 & 2 &  				  	  &                        & 0.487016 - 0.3279410 I & 0.624739 - 0.3269650 I \\
-0.3 & 3 &                        &                        &                        & 0.624739 - 0.4577510 I \\
\botrule
\hline
-0.4 & 0 & 0.201325 - 0.0625774 I & 0.329710 - 0.0611856 I & 0.459202 - 0.0607740 I & 0.589115 - 0.0606016 I \\
-0.4 & 1 &                        & 0.329710 - 0.1835579 I & 0.459202 - 0.1823220 I & 0.589115 - 0.1818050 I \\
-0.4 & 2 &  				  	  &                        & 0.459202 - 0.3038700 I & 0.589115 - 0.3030080 I \\
-0.4 & 3 &                        &                        &                        & 0.589115 - 0.4242110 I \\
\botrule
\hline
-0.5 & 0 & 0.191461 - 0.0584706 I & 0.313765 - 0.0572288 I & 0.437085 - 0.0568616 I & 0.560789 - 0.0567076 I \\
-0.5 & 1 &                        & 0.313765 - 0.1716860 I & 0.437085 - 0.1705850 I & 0.560789 - 0.1701230 I \\
-0.5 & 2 &  				  	  &                        & 0.437085 - 0.2843080 I & 0.560789 - 0.2835380 I \\
-0.5 & 3 &                        &                        &                        & 0.560789 - 0.3969530 I \\
\botrule
\hline
-0.6 & 0 & 0.183381 - 0.0550251 I & 0.300718 - 0.0539095 I & 0.418994 - 0.0535795 I & 0.537624 - 0.0534410 I \\
-0.6 & 1 &                        & 0.300718 - 0.1617280 I & 0.418994 - 0.1607380 I & 0.537624 - 0.1603230 I \\
-0.6 & 2 &  				  	  &                        & 0.418994 - 0.2678970 I & 0.537624 - 0.2672050 I \\
-0.6 & 3 &                        &                        &                        & 0.537624 - 0.3740870 I \\
\botrule
\hline
-0.7 & 0 & 0.176635 - 0.0520660 I & 0.289843 - 0.0510589 I & 0.403921 - 0.0507608 I & 0.518326 - 0.0506357 I \\
-0.7 & 1 &                        & 0.289843 - 0.1531770 I & 0.403921 - 0.1522820 I & 0.518326 - 0.1519070 I \\
-0.7 & 2 &  				  	  &                        & 0.403921 - 0.2538040 I & 0.518326 - 0.2531790 I \\
-0.7 & 3 &                        &                        &                        & 0.518326 - 0.3544500 I \\
\botrule
\hline
-0.8 & 0 & 0.170934 - 0.0494771 I & 0.280666 - 0.0485650 I & 0.391209 - 0.0482948 I & 0.502055 - 0.0481814 I \\
-0.8 & 1 &                        & 0.280666 - 0.1456950 I & 0.391209 - 0.1448850 I & 0.502055 - 0.1445440 I \\
-0.8 & 2 &  				  	  &                        & 0.391209 - 0.2414740 I & 0.502055 - 0.2409070 I \\
-0.8 & 3 &                        &                        &                        & 0.502055 - 0.3372700 I \\
\botrule
\hline
-0.9 & 0 & 0.166077 - 0.0471765 I & 0.272867 - 0.0463487 I & 0.380412 - 0.0461033 I & 0.488239 - 0.0460002 I \\
-0.9 & 1 &                        & 0.272867 - 0.1390460 I & 0.380412 - 0.1383100 I & 0.488239 - 0.1380010 I \\
-0.9 & 2 &  				  	  &                        & 0.380412 - 0.2305170 I & 0.488239 - 0.2300010 I \\
-0.9 & 3 &                        &                        &                        & 0.488239 - 0.3220010 I \\
\botrule
\hline
\end{tabular} 
\label{table:Third set}
}
\end{table}
\begin{table}[ph!]
\centering
\caption{Quasinormal frequencies for massive $(\mu = 0.2)$ scalar perturbations (varying $\ell$, $n$ and $\alpha$) fixing $M=1,g=0.1$ for the model considered in this work using the fitting approximation.
}
{
\begin{tabular}{c|c|cccc} 
\toprule
$\alpha$  &$n$  &  $\ell=1$ & $\ell=2$ & $\ell=3$ & $\ell=4$ 
\\ 
\colrule
\hline
-0.1 & 0 & 0.280729 - 0.0709648 I & 0.430088 - 0.0762787 I & 0.587552 - 0.0779642 I & 0.747747 - 0.0786922 I \\
-0.1 & 1 &                        & 0.430088 - 0.2288360 I & 0.587552 - 0.2338930 I & 0.747747 - 0.2360770 I \\
-0.1 & 2 &  				  	  &                        & 0.587552 - 0.3898210 I & 0.747747 - 0.3934610 I \\
-0.1 & 3 &                        &                        &                        & 0.747747 - 0.5508460 I \\
\botrule
\hline
-0.2 & 0 & 0.259460 - 0.0608371 I & 0.393463 - 0.0671387 I & 0.535888 - 0.0690813 I & 0.681129 - 0.0699134 I \\
-0.2 & 1 &                        & 0.393463 - 0.2014160 I & 0.535888 - 0.2072440 I & 0.681129 - 0.2097400 I \\
-0.2 & 2 &  				  	  &                        & 0.535888 - 0.3454070 I & 0.681129 - 0.3495670 I \\
-0.2 & 3 &                        &                        &                        & 0.681129 - 0.4893940 I \\
\botrule
\hline
-0.3 & 0 & 0.244743 - 0.0535647 I & 0.367983 - 0.0606331 I & 0.499908 - 0.0627614 I & 0.634718 - 0.0636668 I \\
-0.3 & 1 &                        & 0.367983 - 0.1818990 I & 0.499908 - 0.1882840 I & 0.634718 - 0.1910000 I \\
-0.3 & 2 &  				  	  &                        & 0.499908 - 0.3138070 I & 0.634718 - 0.3183340 I \\
-0.3 & 3 &                        &                        &                        & 0.634718 - 0.4456680 I \\
\botrule
\hline
-0.4 & 0 & 0.233616 - 0.0478958 I & 0.348642 - 0.0556006 I & 0.472584 - 0.0578708 I & 0.599470 - 0.0588309 I \\
-0.4 & 1 &                        & 0.348642 - 0.1668020 I & 0.472584 - 0.1736120 I & 0.599470 - 0.1764930 I \\
-0.4 & 2 &  				  	  &                        & 0.472584 - 0.2893540 I & 0.599470 - 0.2941540 I \\
-0.4 & 3 &                        &                        &                        & 0.599470 - 0.4118160 I \\
\botrule
\hline
-0.5 & 0 & 0.224784 - 0.0432744 I & 0.333257 - 0.0515252 I & 0.450854 - 0.0539058 I & 0.571441 - 0.0549071 I \\
-0.5 & 1 &                        & 0.333257 - 0.1545750 I & 0.450854 - 0.1617170 I & 0.571441 - 0.1647210 I \\
-0.5 & 2 &  				  	  &                        & 0.450854 - 0.2695290 I & 0.571441 - 0.2745350 I \\
-0.5 & 3 &                        &                        &                        & 0.571441 - 0.3843500 I \\
\botrule
\hline
-0.6 & 0 & 0.217550 - 0.0393969 I & 0.320652 - 0.0481246 I & 0.433064 - 0.0505909 I & 0.548506 - 0.0516228 I \\
-0.6 & 1 &                        & 0.320652 - 0.1443740 I & 0.433064 - 0.1517730 I & 0.548506 - 0.1548680 I \\
-0.6 & 2 &  				  	  &                        & 0.433064 - 0.2529540 I & 0.548506 - 0.2581140 I \\
-0.6 & 3 &                        &                        &                        & 0.548506 - 0.3613590 I \\
\botrule
\hline
-0.7 & 0 & 0.211491 - 0.0360778 I & 0.310116 - 0.0452258 I & 0.418219 - 0.0477568 I & 0.529380 - 0.0488103 I \\
-0.7 & 1 &                        & 0.310116 - 0.1356770 I & 0.418219 - 0.1432700 I & 0.529380 - 0.1464310 I \\
-0.7 & 2 &  				  	  &                        & 0.418219 - 0.2387840 I & 0.529380 - 0.2440520 I \\
-0.7 & 3 &                        &                        &                        & 0.529380 - 0.3416720 I \\
\botrule
\hline
-0.8 & 0 & 0.206332 - 0.0331953 I & 0.301185 - 0.0427140 I & 0.405668 - 0.0452913 I & 0.513230 - 0.0463585 I \\
-0.8 & 1 &                        & 0.301185 - 0.1281420 I & 0.405668 - 0.1358740 I & 0.513230 - 0.1390760 I \\
-0.8 & 2 &  				  	  &                        & 0.405668 - 0.2264560 I & 0.513230 - 0.2317930 I \\
-0.8 & 3 &                        &                        &                        & 0.513230 - 0.3245100 I \\
\botrule
\hline
-0.9 & 0 & 0.201882 - 0.0306657 I & 0.293544 - 0.0405089 I & 0.394970 - 0.0431153 I & 0.499486 - 0.0441891 I \\
-0.9 & 1 &                        & 0.293544 - 0.1215270 I & 0.394970 - 0.1293460 I & 0.499486 - 0.1325670 I \\
-0.9 & 2 &  				  	  &                        & 0.394970 - 0.2155760 I & 0.499486 - 0.2209450 I \\
-0.9 & 3 &                        &                        &                        & 0.499486 - 0.3093240 I \\
\botrule
\hline
\end{tabular} 
\label{table:Fourth set}
}
\end{table}

\begin{figure} [H]
\begin{center}
\includegraphics[width=0.497\linewidth]{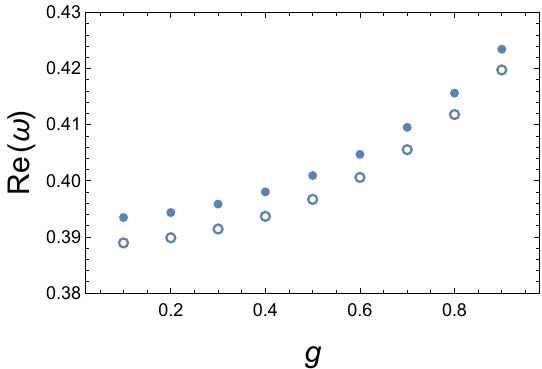} 
\includegraphics[width=0.497\linewidth]{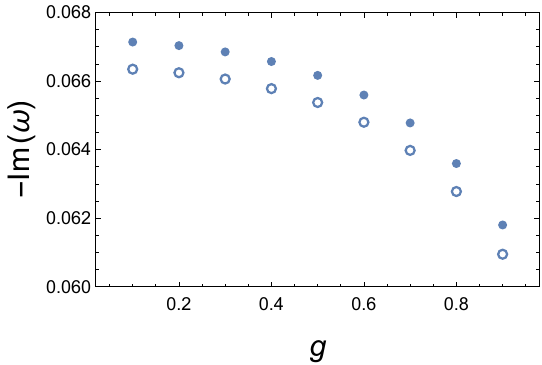} 
\caption{
{\bf{Left panel:}}
Real quasinormal frequencies $\omega_R$ against the parameter $g$ for massive $(\mu = 0.2)$ scalar perturbations utilizing the Pöschl-Teller fitting approach (solid blue points) and the WKB method (empty blue points) assuming $M=1, \ n=0, \ \ell =2, \ \alpha = -0.2 $ varying $g$ in the range $0.1 \leq g \leq 0.9$. 
{\bf{Right panel:}} 
Imaginary quasinormal frequencies $\omega_I$ against the parameter $g$ for massive $(\mu = 0.2)$ scalar perturbations utilizing the Pöschl-Teller fitting approach (solid blue points) and the WKB method (empty blue points) assuming $M=1, \ n=0, \ \ell =2, \ \alpha = -0.2 $ varying $g$ in the range $0.1 \leq g \leq 0.9$. 
}
\label{Fig6New}
 \end{center}
 \end{figure}

In Fig.\eqref{Fig6New}, Re($\omega$) and -Im($\omega$) are plotted by varying the non linear parameter $g$ for the massive scalar field. Here the lowest order QNM modes with $n=0$ are considered to observe how the frequencies behave when one vary the non-linear parameter. When $g$ increases, Re($\omega$) increases leading to higher oscillations; on the other hand -Im($\omega$) decreases leading to less damping.

\begin{figure} [H]
\begin{center}
\includegraphics[width=0.497\linewidth]{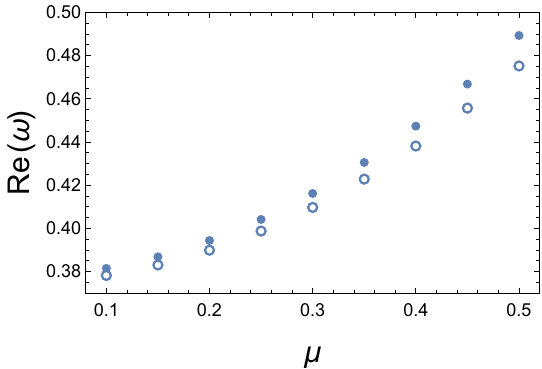} 
\includegraphics[width=0.497\linewidth]{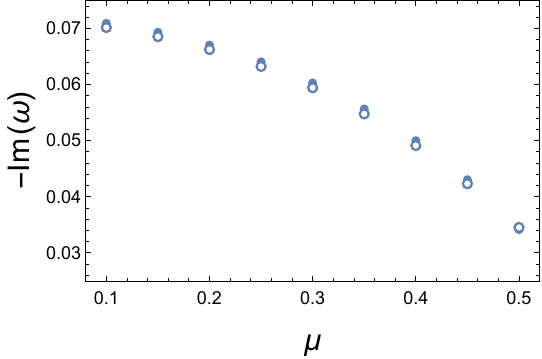} 
\caption{
{\bf{Left panel:}}
Real quasinormal frequencies $\omega_R$ against the parameter $\mu$ for scalar perturbations utilizing the Pöschl-Teller fitting approach (solid blue points) and the WKB method (empty blue points) assuming $M=1, \ n=0, \ \ell =2, \ \alpha = -0.2, \ g=0.2 $ varying $\mu$ in the range $0.1 \leq \mu \leq 0.5$. 
{\bf{Right panel:}} 
Imaginary quasinormal frequencies $\omega_R$ against the parameter $\mu$ for scalar perturbations utilizing the Pöschl-Teller fitting approach (solid blue points) and the WKB method (empty blue points) assuming $M=1, \ n=0, \ \ell =2, \ \alpha = -0.2, \ g=0.2 $ varying $\mu$ in the range $0.1 \leq \mu \leq 0.5$.
}
\label{Fig7New}
 \end{center}
 \end{figure}

In Fig.\eqref{Fig7New}, Re($\omega$) and -Im($\omega$) are plotted by varying the mass of the scalar field,  $\mu$ for the massive scalar. Here the lowest order QNM modes with $n=0$ are considered to observe how the frequencies behave when one vary the mass $\mu$. When $\mu$ increases, Re($\omega$) increases leading to higher oscillations; on the other hand -Im($\omega$) decreases leading to less damping.

\subsection{Wave equations for the spin $\frac{1}{2}$ Dirac field perturbations}


Dirac fields could exist as Hawking radiation around a BH. Hence it is important to study scattering properties, and quasinormal modes of Dirac fields. While there are many works focused on scalar  field perturbations around BHs, studies focusing on Dirac fields are not that many. We will mention few works related to Dirac field perturbations around BHs here: Dirac perturbations of the Lifshitz BH were studied by Catalan et al. \cite{Catalan:2014ccgv}, Kerr-Newmann de-Sitter BH was studied by Konoplya \cite{Konoplya:2007kozh}, regular BHs were studied by Li and Ma \cite{Li:2013lima}, Born-Infeld BHs were studied by Fernando \cite{Fernando:2010fer}, and Bardeen de Sitter BHs were studied by Wahlang et al. \cite{Wahlang:2017wjc}.

Following the work of Unruh \cite{Unruh:1973unr}, one can derive the equations of motion for the massless Dirac field with the $\gamma$ matrices as follows: the massless Dirac field around the BH is given by,

\begin{equation} \label{dirac}
\gamma^{\mu} ( \partial_{\mu} - \Gamma_{\mu} ) \xi =0
\end{equation}
Here, $\gamma^{\mu}$ matrices are defined by
$$ \gamma^t = \frac{\gamma^0}{ \sqrt{f}}; \hspace{0.5 cm} \gamma^r = \gamma^1 \sqrt{f} $$
\ben
\gamma^{\theta} = \frac{\gamma^2}{r}; \hspace{0.5 cm}  \gamma^{\phi} = \frac{ \gamma^3}{ r \sin \theta}
\en

Here $\gamma^a$ matrices appearing in  eq.$\refb{dirac}$ are  defined by,
\begin{eqnarray}
\gamma^{0} =  \left(\begin{array}{cc} i & 0 \\ 0 & - i
\end{array} \right) ; \;\;\;\; 
\gamma^{a} =  \left(\begin{array}{cc}  0  & i  \sigma^a \\  - i  \sigma^a & 0
\end{array} \right)
\end{eqnarray}
Here, $ \sigma^{a}$  are the well known  Pauli matrices  given by,
\begin{eqnarray}
\sigma^{1} =  \left(\begin{array}{cc} 1 & 0 \\ 0 & -1
\end{array} \right) ; \;\;\;\; 
\sigma^{2} =  \left(\begin{array}{cc}  0 & - i \\ i & 0
\end{array} \right);  \;\;\;\; 
\sigma^{3} = \left(\begin{array}{cc} 0 & 1 \\ 1 & 0
\end{array} \right)
\end{eqnarray}
The $\gamma^a$ matrices satisfy the anti-commuting relations,
\begin{equation}
\{ \gamma^a, \gamma^b \} = 2 \eta^{a b} I
\end{equation}
The spin connections $\Gamma_{\mu}$ in eq.$\refb{dirac}$ are defined as,

\begin{equation}
\Gamma_{\mu} = - \frac{1}{8} [ \gamma^a , \gamma^b ] e^{\nu}_{a} e_{b \nu; \mu}
\end{equation}
with,
 \begin{equation}
e_{ b \nu ; \mu} = \partial_{\mu} e_{b \nu} - \Gamma_{\mu \nu}^{\beta} e_{b \beta}
\end{equation}
$\Gamma_{ \nu \mu} ^{\beta} $ are the Christoffel symbols. $e_{\nu}^{a}$ is the tetrads and $e^{\mu}_{a}$ is the inverse of the tetrads: the metric tensor $g_{\alpha \beta}$ of the space-time considered and the tetrads $e^{a}_{\alpha}$ are related by,
\begin{equation}
g_{\alpha \beta} = \eta_{a b} e^{a}_{\alpha} e^{b}_{\beta}
\end{equation}
Here, $\eta_{a b} = ( -1,1,1,1)$. Tetrads for the given geometry  are given by,
\begin{equation} \label{tetrad}
e^a_{\mu} = diag \left( \sqrt{f}, \frac{1}{\sqrt{f}}, r, r \sin \theta \right)
\end{equation}
With the given $\gamma^a$ matrices and $e^a_{\mu}$ in eq.\refb{tetrad},  spin connections for the space-time is given by,

$$ \Gamma_t = \frac{ f'}{ 4 } \gamma^ 0 \gamma^ 1$$
$$ \Gamma_r = 0$$
$$ \Gamma_{\theta} = \frac{ \sqrt{f}}{2 } \gamma^1 \gamma^2$$
\begin{equation}
\Gamma_{\varphi} = \frac{\cos \theta }{2} \gamma^2 \gamma^3 + \frac{\sqrt{f}}{2} \sin \theta \gamma^1 \gamma^3
\end{equation}

To facilitate computations, the function $\xi$ can be  redefined as,
\begin{equation}
\xi = \frac{\Psi}{ f^{1/4} }
\end{equation}
Then the Dirac equation $\refb{dirac}$ simplifies to,
\begin{equation} \label{dirac2}
\frac{\gamma^0}{\sqrt{f}} \frac{\partial \Psi}{\partial t} + \sqrt{f} \gamma^1 
\left( \frac{\partial}{\partial r} + \frac{1}{r} \right) \Psi +
\frac{\gamma^2}{r} \left( \frac{\partial}{\partial \theta} + \frac{ \cot \theta} {2} \right) \Psi + \frac{ \gamma^3} { r \sin \theta} 
\frac{ \partial \Psi} { \partial \varphi} = 0
\end{equation}
Since the Dirac field considered here is massless and spin $\frac{1}{2}$, the solutions to the eq.$\refb{dirac2}$ is circular polarized: please see the papers by Brill and Wheeler \cite{Brill:1957baw}.   The spinors considered have right handed circular polarization and the allowable spin states satisfy the identity,
\begin{equation} \label{circular}
( 1 - i \gamma_5 ) \Psi = 0
\end{equation}
Here $\gamma_5 = \gamma^0 \gamma^1 \gamma^2 \gamma^3$. 
Then, eq.$\refb{circular}$ gives a simplified set of components for $\Psi$ as,

\begin{eqnarray} \label{new11}
\Psi =  \left(\begin{array}{l} \Phi( t,r,\theta, \varphi) \\ 
\Phi( t,r,\theta, \varphi) \\ 
\end{array} \right) 
\end{eqnarray}
where,

\begin{eqnarray} \label{new12}
\Phi = \left(\begin{array}{l}   \frac{i K(r)}{r} S_1(\theta, \varphi) \\ 
 \frac{G(r)}{r} S_2(\theta, \varphi) 
\end{array} \right) e^{- i \omega t}
\end{eqnarray}

Substituting  eq.$\refb{new11}$  and eq.$\refb{new12}$ to the Dirac equation$\refb{dirac2}$ yields,

\begin{equation} \label{wave1}
\left( \frac{ i \omega r}{ \sqrt{f} } K -  r \sqrt{f} \frac{ dK }{ dr}\right) \frac{1}{G} + \left( \frac{\cot \theta }{2} S_2 + \frac{i}{\sin \theta} \frac{ \partial S_2} {\partial \varphi} + \frac{\partial S_2} {\partial \theta} \right) \frac{1}{ S_1}=0
\end{equation}

\begin{equation} \label{wave2}
\left( \frac{  i \omega r}{ \sqrt{f} } G +  r \sqrt{f} \frac{ dG}{ dr} \right) \frac{1}{K} + \left( \frac{\cot \theta}{2} S_1 - \frac{i}{\sin \theta} \frac{ \partial S_1} {\partial \varphi} + \frac{\partial S_1} {\partial \theta} \right) \frac{1}{ S_2}=0
\end{equation}
We can write the above equation as four separate differential equations:

\ben \label{wave1}
\left( \frac{ i \omega r}{ \sqrt{f} } K -  r \sqrt{f} \frac{ dK }{ dr}\right) \frac{1}{G} = \lambda
\en
\ben \label{wave2}
\left( \frac{  i \omega r}{ \sqrt{f} } G +  r \sqrt{f} \frac{ dG}{ dr} \right) \frac{1}{K} = - \lambda
\en
\ben \label{ang1}
\left( \frac{\cot \theta}{2}  + \frac{i}{\sin \theta} \frac{ \partial } {\partial \varphi} + \frac{\partial} {\partial \theta} \right) S_2 = - \lambda S_1
\en
\ben \label{ang2}
\left( \frac{\cot \theta}{2}  - \frac{i}{\sin \theta} \frac{ \partial } {\partial \varphi} + \frac{\partial} {\partial \theta} \right) S_1 = \lambda S_2
\en

Solutions to the angular equations $\refb{ang1}$ and $\refb{ang2}$, are given by spin weighted spherical harmonics $_{\frac{1}{2}}Y_{\ell m} $ and $_{-\frac{1}{2}}Y_{\ell m} $. Specifically, $S_1$ = $_{\frac{1}{2}}Y_{\ell m} $ and $S_2$ = $_{-\frac{1}{2}}Y_{\ell m} $. Here, $\lambda = \ell + \frac{1}{2}$. A lengthy description of spin weighted spherical harmonics can be found in \cite{Castillo:2007gft} and \cite{Goldberg:1967jng}.  Note that $ -\ell <m < \ell$ and $\ell$ takes the values $\ell = 1/2, 3/2,...$. Hence $\ell$ can be written in terms of $\lambda$ which is a positive  integer as, $\ell = \lambda - \frac{1}{2}$. Here $\lambda$ will be called the multipole number in the rest of the paper.

Solving the radial part, we have two relevant equations only. Thus, the eq.$\refb{wave1}$ and eq.$\refb{wave2}$ can be simplified to be,
\begin{align} 
\label{fun1}
\frac{dG}{dr^*} + i \omega G + W(r)  K &= 0 
\\
\label{fun2}
\frac{dK}{dr^*} - i \omega K + W(r)  G &= 0
\end{align}
%
%
Here, the function $W(r)$ is given by,
\ben
W(r) = \frac{\lambda \sqrt{f}}{r}, \; \; \; \; \; \lambda=  (1,2,3,...)
\en
and $r_*$ is the  ``tortoise'' coordinate given by,
\ben \label{tortoise}
\frac{dr_{*}}{dr} = \frac{1}{f}
\en
Two new functions $R_{\pm}$ can be  defined as,
\ben
R_{\pm} = K \pm G
\en
and  eq.$\refb{fun1}$ and eq.$\refb{fun2}$  can be decoupled as,
\begin{equation} \label{final}
\frac{d^2 R_{\pm}}{dr^{*2}} + ( \omega^2 - V_{Dirac}^{\pm} ) R_{\pm} = 0
\end{equation}
Here, $V_{Dirac}^{\pm}$ are related to $W(r)$ as,
\ben \label{potential}
V_{Dirac}^{\pm} = W^2 \pm f \left( \frac{d W}{dr} \right) 
\en

The two potentials $V_{Dirac}^{+}$ and $V_{Dirac}^{-}$ as given above are related and they will produce the same physical consequences and identical QNM spectra, as discussed by Anderson and Price \cite{anderson:1991anp}. Therefore, we will use only $V_{Dirac}^-$ for all computations in the rest of the paper and will be referred to as just $V_{Dirac}(r)$.

\begin{figure} [H]
\begin{center}
\includegraphics[width=0.497\linewidth]{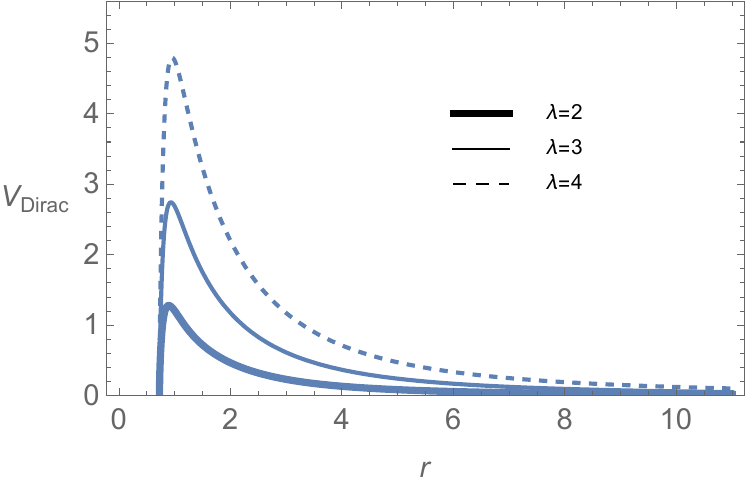} 
\includegraphics[width=0.497\linewidth]{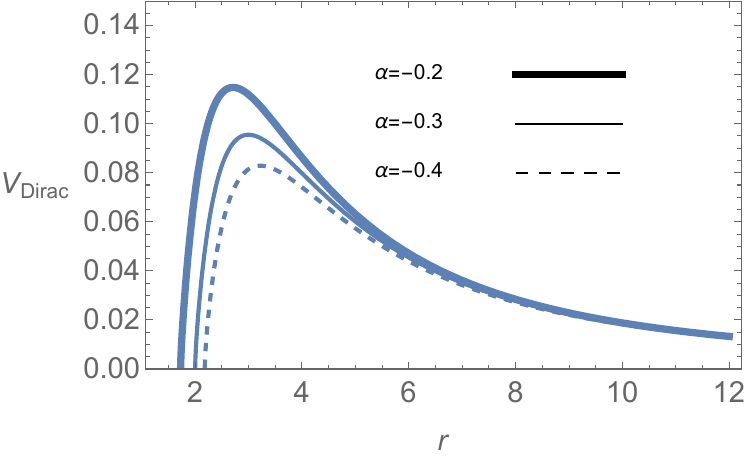}
\caption{
{\bf{Left panel:}} The figure shows $V_{Dirac}(r)$ versus $r$ for varying $\lambda$. Here $M = 1, \alpha=-1$ and $g = 1$.
{\bf{Right panel:}} The figure shows $V_{Dirac}(r)$ versus $r$ for varying $\alpha$. Here $M = 1, \lambda=2$ and $g = 1$.
}
\label{potlambda}
 \end{center}
 \end{figure}

\begin{figure} [H]
\begin{center}
\includegraphics[width=0.497\linewidth]{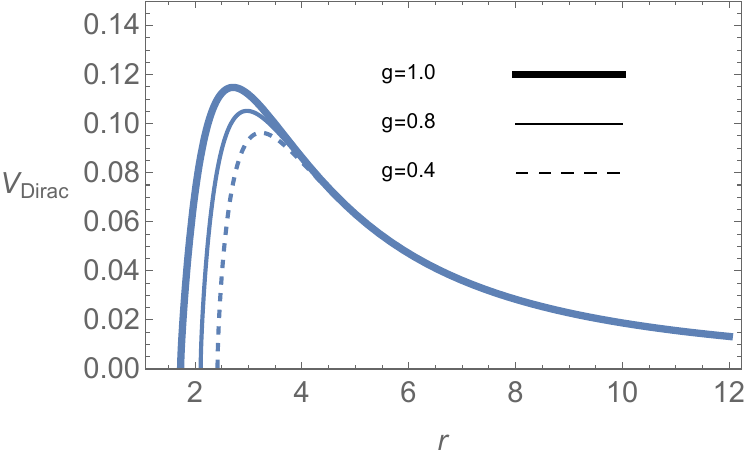} 
\caption{The figure shows $V_{Dirac}(r)$ versus $r$ for varying $g$. Here $M = 1, \alpha=-0.2$ and $\lambda = 2$.}
\label{potgvaluedirac}
 \end{center}
 \end{figure}
In Fig.$\refb{potlambda}$ the potential is plotted vs $r$ for varying $\lambda$ and $\alpha$. When $\lambda$ increases, the potential also increases. This behavior is similar to the massless scalar wave potential where when $\ell$ increases the potential increases \cite{Sun:2023slzl}. When $\alpha$ increases, the potential increases. This behavior is similar to the massless scalar field studied in \cite{Sun:2023slzl}. On the other hand, in Fig.$\refb{potgvaluedirac}$, when $g$ increases, the height of the potential increases similar the massless scalar field where the height increases with $g$ \cite{Sun:2023slzl}. 

\begin{table}[ph!]
\centering
\caption{Quasinormal frequencies for massless Dirac perturbations (varying $\lambda$, $n$ and $\alpha$) fixing $M=1,g=0.1$ for the model considered in this work using the WKB approximation.}
{
\begin{tabular}{c|c|cccc} 
\toprule
$\alpha$  &$n$  &  $\lambda=2$ & $\lambda=3$ & $\lambda=4$ & $\lambda=5$ 
\\ 
\colrule
\hline
     &    0 & 0.323603 - 0.0856526 I & 0.488191 - 0.0823897 I & 0.652350 - 0.0812438 I & 0.816285 - 0.0807247 I \\
     &    1 & 0.315600 - 0.2634740 I & 0.478658 - 0.2520920 I & 0.643782 - 0.2469750 I & 0.808861 - 0.2443970 I \\
-0.1 &    2 &  				  	     & 0.467560 - 0.4301260 I & 0.631366 - 0.4197200 I & 0.796828 - 0.4134790 I \\
     &    3 &                        &                        & 0.619840 - 0.5985340 I & 0.783847 - 0.5884580 I \\
     &    4 &                        &                        &                        & 0.772379 - 0.7678220 I \\
\botrule
\hline
     &    0 & 0.294343 - 0.0761651 I & 0.443931 - 0.0733889 I & 0.593145 - 0.0724149 I & 0.742167 - 0.0719734 I \\
     &    1 & 0.287196 - 0.2341810 I & 0.435604 - 0.2243760 I & 0.585707 - 0.2200040 I & 0.735737 - 0.2178070 I \\
-0.2 &    2 &  				  	     & 0.425731 - 0.3826310 I & 0.574802 - 0.3736290 I & 0.725237 - 0.3682680 I \\
     &    3 &                        &                        & 0.564524 - 0.5326040 I & 0.713780 - 0.5238500 I \\
     &    4 &                        &                        &                        & 0.703534 - 0.6833310 I \\
\botrule
\hline
     &    0 & 0.273960 - 0.0694036 I & 0.413090 - 0.0669787 I & 0.551886 - 0.0661285 I & 0.690510 - 0.0657429 I \\
     &    1 & 0.267439 - 0.2132870 I & 0.405638 - 0.2046270 I & 0.545266 - 0.2007940 I & 0.684802 - 0.1988730 I \\
-0.3 &    2 &  				  	     & 0.396655 - 0.3487690 I & 0.535459 - 0.3407860 I & 0.675413 - 0.3360620 I \\
     &    3 &                        &                        & 0.526085 - 0.4856010 I & 0.665064 - 0.4778080 I \\
     &    4 &                        &                        &                        & 0.655701 - 0.6230970 I \\
\botrule
\hline
     &    0 & 0.258485 - 0.0641603 I & 0.389669 - 0.0620098 I & 0.520551 - 0.0612561 I & 0.651278 - 0.0609140 I \\
     &    1 & 0.252464 - 0.1970720 I & 0.382909 - 0.1893140 I & 0.514575 - 0.1859020 I & 0.646136 - 0.1841960 I \\
-0.4 &    2 &  				  	     & 0.374630 - 0.3224960 I & 0.505635 - 0.3153160 I & 0.637624 - 0.3110920 I \\
     &    3 &                        &                        & 0.496976 - 0.4491340 I & 0.628149 - 0.4420990 I \\
     &    4 &                        &                        &                        & 0.619483 - 0.5763640 I \\
\botrule
\hline
     &    0 & 0.246190 - 0.0598969 I & 0.371056 - 0.0579710 I & 0.495644 - 0.0572961 I & 0.620093 - 0.0569895 I \\
     &    1 & 0.240586 - 0.1838770 I & 0.364867 - 0.1768630 I & 0.490200 - 0.1737970 I & 0.615418 - 0.1722670 I \\
-0.5 &    2 &  				  	     & 0.357173 - 0.3011210 I & 0.481977 - 0.2946040 I & 0.607630 - 0.2907920 I \\
     &    3 &                        &                        & 0.473912 - 0.4194650 I & 0.598880 - 0.4130570 I \\
     &    4 &                        &                        &                        & 0.590793 - 0.5383430 I \\
\botrule
\hline
     &    0 & 0.236143 - 0.0563185 I & 0.355840 - 0.0545824 I & 0.475283 - 0.0539738 I & 0.594597 - 0.0536971 I \\
     &    1 & 0.230899 - 0.1727930 I & 0.350140 - 0.1664130 I & 0.470291 - 0.1636400 I & 0.590318 - 0.1622590 I \\
-0.6 &    2 &  				  	     & 0.342949 - 0.2831710 I & 0.462683 - 0.2772180 I & 0.583149 - 0.2737560 I \\
     &    3 &                        &                        & 0.455126 - 0.3945490 I & 0.575021 - 0.3886760 I \\
     &    4 &                        &                        &                        & 0.567430 - 0.5064110 I \\
\botrule
\hline
     &    0 & 0.227781 - 0.0532439 I & 0.343172 - 0.0516717 I & 0.458328 - 0.0511204 I & 0.573365 - 0.0508694 I \\
     &    1 & 0.222857 - 0.1632610 I & 0.337900 - 0.1574340 I & 0.453731 - 0.1549150 I & 0.569433 - 0.1536620 I \\
-0.7 &    2 &  				  	     & 0.331152 - 0.2677360 I & 0.446663 - 0.2622770 I & 0.562805 - 0.2591190 I \\
     &    3 &                        &                        & 0.439555 - 0.3731260 I & 0.555223 - 0.3677210 I \\
     &    4 &                        &                        &                        & 0.548070 - 0.4789550 I \\
\botrule
\hline
     &    0 & 0.220740 - 0.0505522 I & 0.332499 - 0.0491246 I & 0.444041 - 0.0486236 I & 0.555473 - 0.0483953 I \\
     &    1 & 0.216105 - 0.1549090 I & 0.327610 - 0.1495740 I & 0.439796 - 0.1472790 I & 0.551848 - 0.1461390 I \\
-0.8 &    2 &  				  	     & 0.321260 - 0.2542160 I & 0.433211 - 0.2491960 I & 0.545703 - 0.2463090 I \\
     &    3 &                        &                        & 0.426506 - 0.3543590 I & 0.538611 - 0.3493730 I \\
     &    4 &                        &                        &                        & 0.531851 - 0.4549030 I \\
\botrule
\hline
     &    0 & 0.214770 - 0.0481586 I & 0.323446 - 0.0468605 I & 0.431919 - 0.0464045 I & 0.540290 - 0.0461964 I \\
     &    1 & 0.210401 - 0.1474760 I & 0.318903 - 0.1425840 I & 0.427992 - 0.1404910 I & 0.536944 - 0.1394530 I \\
-0.9 &    2 &  				  	     & 0.312919 - 0.2421850 I & 0.421847 - 0.2375620 I & 0.531237 - 0.2349180 I \\
     &    3 &                        &                        & 0.415512 - 0.3376580 I & 0.524592 - 0.3330510 I \\
     &    4 &                        &                        &                        & 0.518192 - 0.4334970 I \\
\botrule
\hline
\end{tabular} 
\label{table:Fifth set}
}
\end{table}

\begin{table}[ph!]
\centering
\caption{Quasinormal frequencies for massless Dirac perturbations (varying $\lambda$, $n$ and $\alpha$) fixing $M=1,g=0.1$ for the model considered in this work using the Pöschl-Teller fitting approach.
}
{
\begin{tabular}{c|c|cccc} 
\toprule
$\alpha$  &$n$  &  $\lambda=2$ & $\lambda=3$ & $\lambda=4$ & $\lambda=5$ 
\\ 
\colrule
\hline
     &    0 & 0.327641 - 0.0820623 I & 0.491146 - 0.0808075 I & 0.654575 - 0.0803641 I & 0.818050 - 0.0801665 I \\
     &    1 & 0.327641 - 0.2461870 I & 0.491146 - 0.2424230 I & 0.654575 - 0.2410920 I & 0.818050 - 0.2405000 I \\
-0.1 &    2 &  				  	     & 0.491146 - 0.4040380 I & 0.654575 - 0.4018210 I & 0.818050 - 0.4008330 I \\
     &    3 &                        &                        & 0.654575 - 0.5625490 I & 0.818050 - 0.5611660 I \\
     &    4 &                        &                        &                        & 0.818050 - 0.7214990 I \\
\botrule
\hline
     &    0 & 0.297847 - 0.0730996 I & 0.446486 - 0.0720373 I & 0.595069 - 0.0716631 I & 0.743691 - 0.0714962 I \\
     &    1 & 0.297847 - 0.2192990 I & 0.446486 - 0.2161120 I & 0.595069 - 0.2149890 I & 0.743691 - 0.2144890 I \\
-0.2 &    2 &  				  	     & 0.446486 - 0.3601870 I & 0.595069 - 0.3583150 I & 0.743691 - 0.3574810 I \\
     &    3 &                        &                        & 0.595069 - 0.5016410 I & 0.743691 - 0.5004740 I \\
     &    4 &                        &                        &                        & 0.743691 - 0.6434660 I \\
\botrule
\hline
     &    0 & 0.277078 - 0.0667158 I & 0.415356 - 0.0657930 I & 0.553591 - 0.0654686 I & 0.691863 - 0.0653239 I \\
     &    1 & 0.277078 - 0.2001470 I & 0.415356 - 0.1973790 I & 0.553591 - 0.1964060 I & 0.691863 - 0.1959720 I \\
-0.3 &    2 &  				  	     & 0.415356 - 0.3289650 I & 0.553591 - 0.3273430 I & 0.691863 - 0.3266190 I \\
     &    3 &                        &                        & 0.553591 - 0.4582800 I & 0.691863 - 0.4572670 I \\
     &    4 &                        &                        &                        & 0.691863 - 0.5879150 I \\
\botrule
\hline
     &    0 & 0.261300 - 0.0617669 I & 0.391709 - 0.0609534 I & 0.522085 - 0.0606679 I & 0.652495 - 0.0605404 I \\
     &    1 & 0.261300 - 0.1853010 I & 0.391709 - 0.1828600 I & 0.522085 - 0.1820040 I & 0.652495 - 0.1816210 I \\
-0.4 &    2 &  				  	     & 0.391709 - 0.3047670 I & 0.522085 - 0.3033390 I & 0.652495 - 0.3027020 I \\
     &    3 &                        &                        & 0.522085 - 0.4246750 I & 0.652495 - 0.4237830 I \\
     &    4 &                        &                        &                        & 0.652495 - 0.5448640 I \\
\botrule
\hline
     &    0 & 0.248754 - 0.0577441 I & 0.372909 - 0.0570202 I & 0.497039 - 0.0567663 I & 0.621199 - 0.0566529 I \\
     &    1 & 0.248754 - 0.1732320 I & 0.372909 - 0.1710610 I & 0.497039 - 0.1702990 I & 0.621199 - 0.1699590 I \\
-0.5 &    2 &  				  	     & 0.372909 - 0.2851010 I & 0.497039 - 0.2838320 I & 0.621199 - 0.2832650 I \\
     &    3 &                        &                        & 0.497039 - 0.3973640 I & 0.621199 - 0.3965700 I \\
     &    4 &                        &                        &                        & 0.621199 - 0.5098760 I \\
\botrule
\hline
     &    0 & 0.238494 - 0.0543685 I & 0.357536 - 0.0537205 I & 0.476559 - 0.0534933 I & 0.595609 - 0.0533917 I \\
     &    1 & 0.238494 - 0.1631050 I & 0.357536 - 0.1611610 I & 0.476559 - 0.1604800 I & 0.595609 - 0.1601750 I \\
-0.6 &    2 &  				  	     & 0.357536 - 0.2686020 I & 0.476559 - 0.2674670 I & 0.595609 - 0.2669590 I \\
     &    3 &                        &                        & 0.476559 - 0.3744530 I & 0.595609 - 0.3737420 I \\
     &    4 &                        &                        &                        & 0.595609 - 0.4805250 I \\
\botrule
\hline
     &    0 & 0.229947 - 0.0514688 I & 0.344731 - 0.0508866 I & 0.459501 - 0.0506825 I & 0.574296 - 0.0505910 I \\
     &    1 & 0.229947 - 0.1544070 I & 0.344731 - 0.1526600 I & 0.459501 - 0.1520470 I & 0.574296 - 0.1517730 I \\
-0.7 &    2 &  				  	     & 0.344731 - 0.2544330 I & 0.459501 - 0.2534120 I & 0.574296 - 0.2529550 I \\
     &    3 &                        &                        & 0.459501 - 0.3547770 I & 0.574296 - 0.3541370 I \\
     &    4 &                        &                        &                        & 0.574296 - 0.4553190 I \\
\botrule
\hline
     &    0 & 0.222741 - 0.0489313 I & 0.333938 - 0.0484071 I & 0.445124 - 0.0482232 I & 0.556332 - 0.0481407 I \\
     &    1 & 0.222741 - 0.1467940 I & 0.333938 - 0.1452210 I & 0.445124 - 0.1446700 I & 0.556332 - 0.1444220 I \\
-0.8 &    2 &  				  	     & 0.333938 - 0.2420360 I & 0.445124 - 0.2411160 I & 0.556332 - 0.2407030 I \\
     &    3 &                        &                        & 0.445124 - 0.3375630 I & 0.556332 - 0.3369850 I \\
     &    4 &                        &                        &                        & 0.556332 - 0.4332660 I \\
\botrule
\hline
     &    0 & 0.216622 - 0.0466756 I & 0.324775 - 0.0462035 I & 0.432920 - 0.0460376 I & 0.541085 - 0.0459630 I \\
     &    1 & 0.216622 - 0.1400270 I & 0.324775 - 0.1386100 I & 0.432920 - 0.1381130 I & 0.541085 - 0.1378890 I \\
-0.9 &    2 &  				  	     & 0.324775 - 0.2310170 I & 0.432920 - 0.2301880 I & 0.541085 - 0.2298150 I \\
     &    3 &                        &                        & 0.432920 - 0.3222630 I & 0.541085 - 0.3217410 I \\
     &    4 &                        &                        &                        & 0.541085 - 0.4136670 I \\
\botrule
\hline
\end{tabular} 
\label{table:Sixth set}
}
\end{table}

\begin{figure} [H]
\begin{center}
\includegraphics[width=0.497\linewidth]{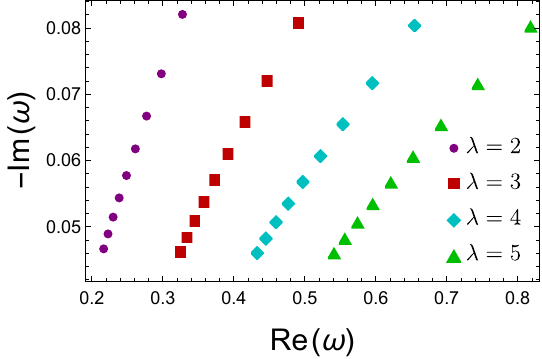} 
\includegraphics[width=0.497\linewidth]{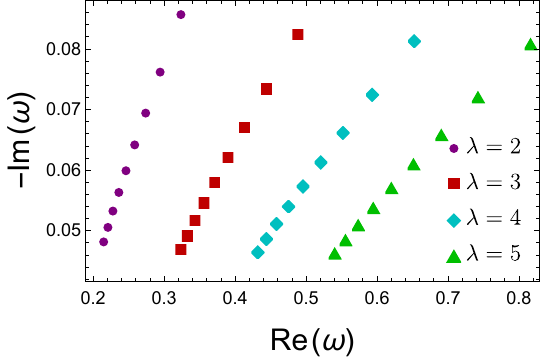} 
\caption{
{\bf{Left panel:}} Quasinormal modes computed utilizing the Pöschl-Teller fitting approach assuming $M=1, \ g=0.1, \ n=0 $ varying $\alpha$ in the range $-0.9 \leq \alpha \leq -0.1$ with four different values of $\lambda$ (see figure for details).
{\bf{Right panel:}} Quasinormal modes computed utilizing the WKB approximation assuming $M=1, \ g=0.1, \ n=0 $ varying $\alpha$ in the range $-0.9 \leq \alpha \leq -0.1$ with four different values of $\lambda$ (see figure for details).
}
\label{Fig7}
 \end{center}
 \end{figure}



\begin{figure} [H]
\begin{center}
\includegraphics[width=0.497\linewidth]{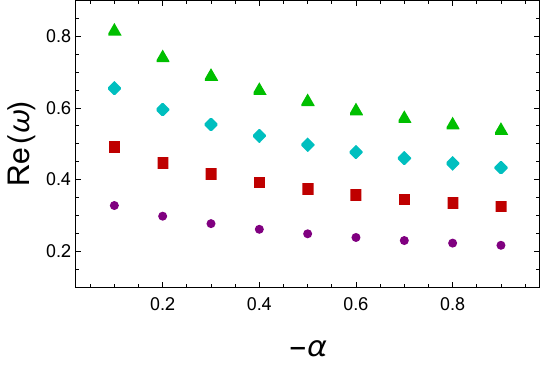} 
\includegraphics[width=0.497\linewidth]{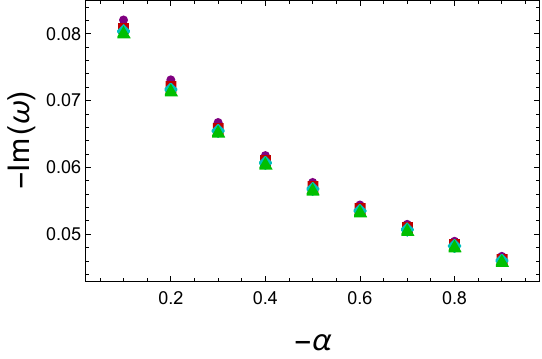}
\\
\includegraphics[width=0.497\linewidth]{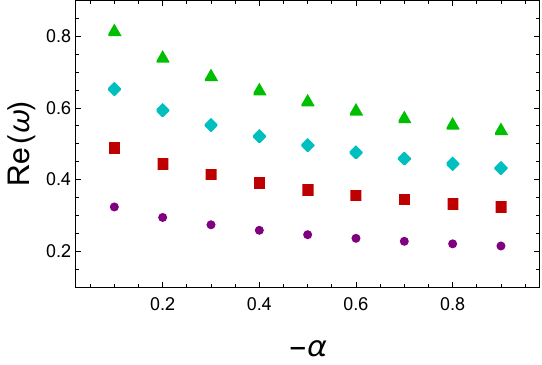} 
\includegraphics[width=0.497\linewidth]{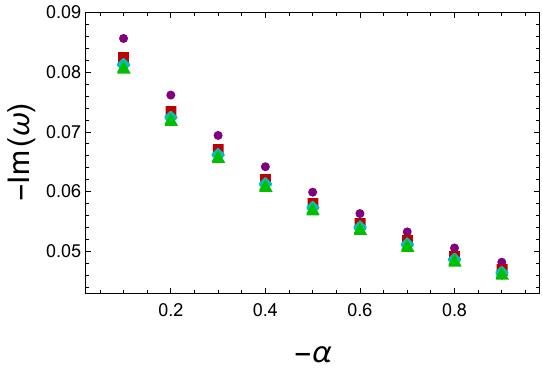}
\caption{
{\bf{First row:}} The left panel shows the real part of the quasinormal frequency $\omega_R$, while the right panel shows the imaginary part of the quasinormal frequency $\omega_I$ both cases against the parameter $\alpha$, assuming $M=1, \ g=0.1, \ n=0 $ with four different values of $\lambda$ (color code as in the previous figure) using the Pöschl-Teller fitting approach.
{\bf{Second row:}} The left panel shows the real part of the quasinormal frequency $\omega_R$, while the right panel shows the imaginary part of the quasinormal frequency $\omega_I$ both cases against the parameter $\alpha$, assuming $M=1, \ g=0.1, \ n=0 $ with four different values of $\lambda$ (color code as in the previous figure) using the WKB approximation.
}
\label{Fig8}
 \end{center}
 \end{figure}


\begin{figure} [H]
\begin{center}
\includegraphics[width=0.497\linewidth]{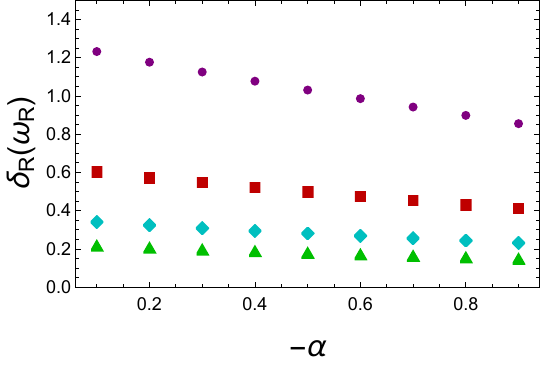} 
\includegraphics[width=0.497\linewidth]{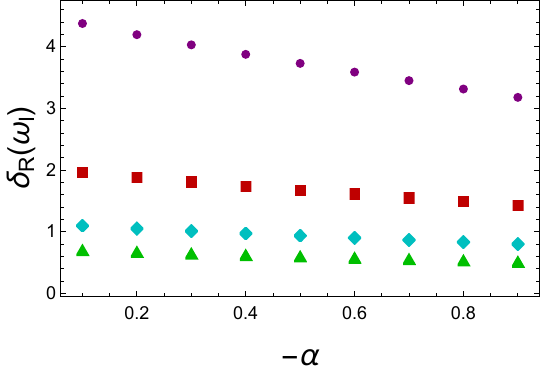}
\caption{
The figure shows the percent error of the real and imaginary quasinormal frequencies against the parameter $-\alpha$.
{\bf{Left panel:}} It shows the percent error of the real part of the quasinormal frequency $\delta_R(\omega_R)$ versus $-\alpha$, for different values of the value  $\xi$ (color code as previous figures)
for massless Dirac perturbations
assuming $M=1, \ g=0.1, \ n=0 $. The percent error compares the modes obtained using the two complementary methods mentioned in the manuscript.
{\bf{Right panel:}} It shows the percent error of the imaginary part of the quasinormal frequency $\delta_R(\omega_I)$ versus $-\alpha$, for different values of the parameter $\xi$ (color code as previous figures)
for massless Dirac perturbations
assuming $M=1, \ g=0.1, \ n=0 $. The percent error compares the modes obtained using the two complementary methods mentioned in the manuscript.
}
\label{FigNew2}
\end{center}
\end{figure}


\begin{figure} [H]
\begin{center}
\includegraphics[width=0.497\linewidth]{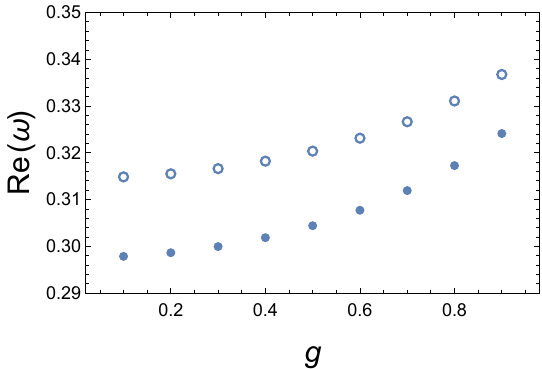} 
\includegraphics[width=0.497\linewidth]{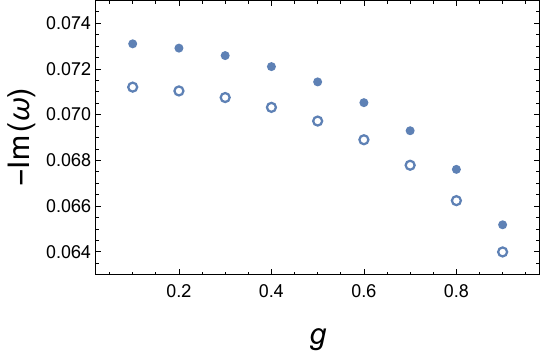} 
\caption{
{\bf{Left panel:}}
Real quasinormal frequencies $\omega_R$ against the parameter $g$ for Dirac perturbations utilizing the Pöschl-Teller fitting approach (solid blue points) and the WKB method (empty blue points) assuming $M=1, \ n=0, \ \lambda =2, \ \alpha = -0.2 $ varying $g$ in the range $0.1 \leq g \leq 0.9$. 
{\bf{Right panel:}} 
Imaginary quasinormal frequencies $\omega_I$ against the parameter $g$ for Dirac perturbations utilizing the Pöschl-Teller fitting approach (solid blue points) and the WKB method (empty blue points) assuming $M=1, \ n=0, \ \lambda =2, \ \alpha = -0.2 $ varying $g$ in the range $0.1 \leq g \leq 0.9$. 
}
\label{Fig13New}
 \end{center}
 \end{figure}

We have calculated the quasinormal modes for massless Dirac fields. Similar to the  scalar field case, we have  computed the QNMs using the i) WBK approximations and ii) Pöschl-Teller fitting approach. We have collected our numerical results in Tables \eqref{table:Fifth set} and \eqref{table:Sixth set} and we added Figures \eqref{Fig7} and \eqref{Fig8}. Since the imaginary part is negative, all modes are found to be stable against scalar perturbations.

Let us make remarks on our observation of QNM frequencies here: When the multipole number $\lambda$ increased (while  keeping $n$ and $\alpha$ fixed), we observed that Re $\omega$ increased and -Im $\omega$ decreased. Hence, oscillations are lower and damping is higher for large $\lambda$. When $\alpha$ and $\lambda$ are fixed to study QNM frequencies for overtones, we observed that when $n$ is increased, Re $\omega$ decreased and -Im $\omega$ increased. When the dark matter parameter $\alpha$ is increased, both Re $\omega$ and -Im $\omega$ increased. This is very similar to the behavior of the QNM frequencies for the massless scalar field.

In Fig.\eqref{FigNew2} we have included  the percent error between the WKB method and the Pöschl-Teller method using the same definition as for the scalar case.
 %
%
%
The figures show that: i) the real part has an error margin of around $1.2\%$ for the lower value of $\xi$, and this margin is drastically reduced to $0.6\%$ or lower as the parameter increases. This behaviour is consistent, given that the WKB approximation is known to work better for large values of $\xi$. ii) similarly, the figure for the imaginary part has a percentage error of around  $4.3\%$ and starts to decrease as $-\alpha$ increases. Furthermore, as $\lambda$ increases, the percentage error reduces to $2\%$ or less.

In Fig\eqref{Fig13New}, we have plotted the QNM frequencies by varying the non-linear parameter $g$ for the Dirac field. The behavior is similar to the massive scalar field; Re $\omega$ increases with $g$ and -Im $\omega$ decreases with $g$.

\section{Absorption cross-sections} \label{sec4}
As explained in the introduction, the GBF of the BH is given by $|T(\omega)|^2$ which is the probability of the wave tunneling through the potential created by the field in consideration. Here, we employ the WKB approximation developed by Schutz and Will \cite{Schutz:1985km}, and Iyer and Will \cite{Iyer:1986np} to compute the GBFs and partial absorption cross sections. The WKB approximation has high accuracy when $\omega^2 \approx V_0$ where $V_0$ is the potential at $r=r_0$: $r_0$ is the $r$ value when the potential is maximum. In the WKB approximation the reflection coefficient $R(\omega)$ is given by
\begin{equation} 
R(\omega) = (1+e^{-2\pi i b})^{-1/2}
\, 
\label{Refl}
\end{equation}
and 
\begin{equation}
|T(\omega)|^2 = 1 - |R(\omega)|^2 
\, 
\label{}
\end{equation}
where $T(\omega)$ is the transmission coefficient. The value $b$ is eq.$\refb{Refl}$ is given by 

\begin{equation}
b = i \frac{\omega^2 - V_0}{\sqrt{- 2 V''_0}} -\Lambda_2 - \Lambda_3
\, .
\label{}
\end{equation}
In the above equation, $\Lambda_2, \Lambda_3$ are given by 

\begin{equation}
\Lambda_2 =\frac{1}{ \sqrt{- 2 V''_0}} \left[\frac{1}{8} \left(\frac{V_0^{(4)}}{V''_0}\right) \left( 
\frac{1}{4} + b^2 \right) -  \frac{1}{288} \left(\frac{V_0^{(3)}}{V''_0}\right)   \left(7 + 60 b^2 \right) \right]
\, .
\label{}
\end{equation}
\begin{eqnarray}
\Lambda_3 &=& \frac{1}{2 V''_0} \Bigg[\frac{5}{6912} \left(\frac{V_0^{(3)}}{V''_0}\right)^4 \left(77 + 188 b^2 \right) -  \frac{1}{384} \left(\frac{(V_0^{(3)})^2 V_0^{(4)}}{(V''_0)^3}\right)   \left(51 + 100 b^2 \right) \nonumber \\ &&
+ \frac{1}{2304} \left(\frac{V_0^{(4)}}{V''_0}\right)^2 \left(67 + 68 b^2 \right) 
+  \frac{1}{288} \left(\frac{(V_0^{(3)})^2 V_0^{(5)}}{(V''_0)^2}\right)   \left(19 + 28 b^2 \right) 
- \frac{1}{288} \left(\frac{V_0^{(6)}}{V''_0}\right) \left(5 +4 b^2 \right) 
\Bigg]
\, .
\label{}
\end{eqnarray}
Here $V_0^{(n)} = d^nV_0/d r^n$.

The absorption cross section is given by the expression
\begin{equation}
\sigma = \sum_{\ell = 0}^\infty \sigma_{\ell}
\, ,
\label{}
\end{equation}
where the partial cross section can be computed by
\begin{equation}
\sigma_{\ell} = \frac{\pi}{\omega^2} (2 \ell + 1)  |T(\omega)|^2
\, .
\label{}
\end{equation}
The WKB method to compute $R(\omega), T(\omega)$ and GBFs have been used in many papers including \cite{Tosh:2016tssa}.

\subsection{Massive scalar field Case}

\begin{figure*}[t]
\begin{center}
\includegraphics[width=1\linewidth]{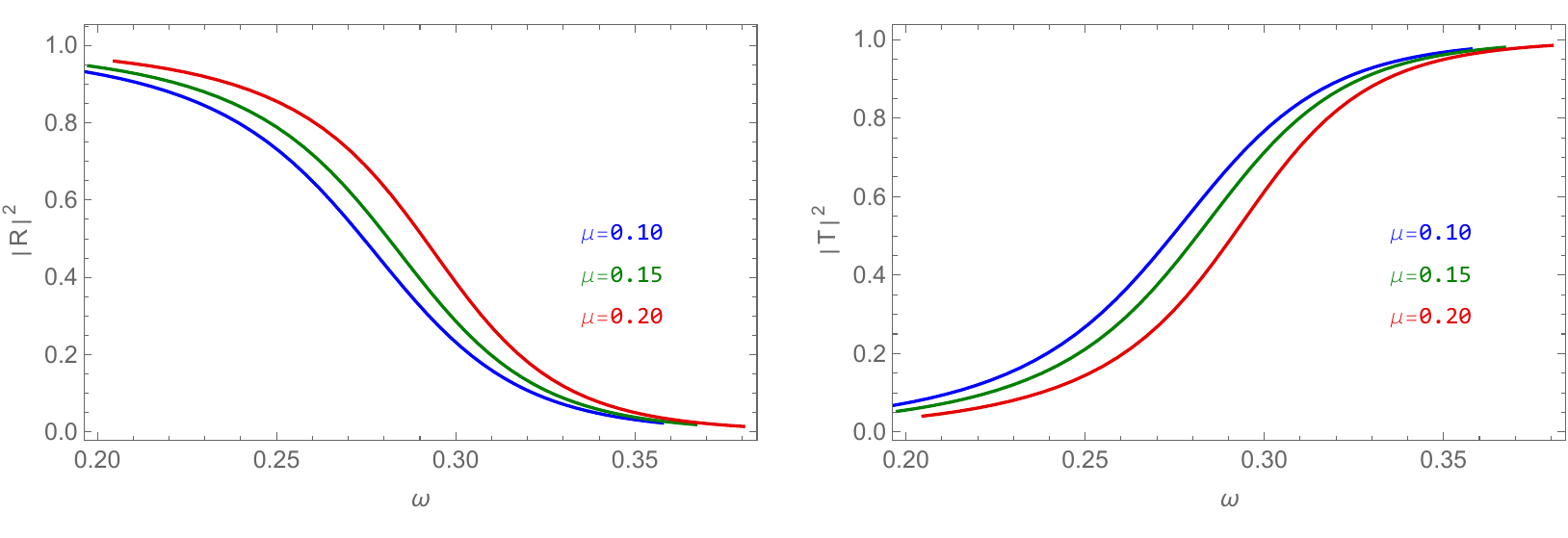}
\caption{The $|R|^2$ and $|T|^2$ coefficients for the massive scalar case are displayed as a function of $\omega$ for various values of $\mu$, with $M=1$, $g = 0.1$, $\ell = 1$ and $\alpha = -0.1$ held constant.}
\label{Mu-R-T}
\end{center}
\end{figure*}

\begin{figure*}[t]
\begin{center}
\includegraphics[width=0.8\linewidth]{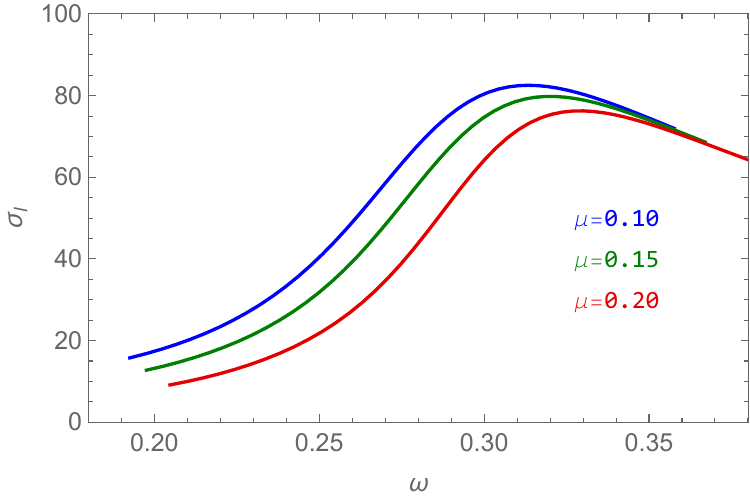}
\caption{The figure depicts the partial absorption cross section as a function of $\omega$ for the massive scalar case for various values of $\mu$, while keeping $M=1$, $g=0.1$, $\ell = 1$ and $\alpha=-0.1$ constant.}
\label{Mu-sigma}
\end{center}
\end{figure*}

In Fig.\eqref{Mu-R-T} , $|R|^2$ and $|T|^2$ are plotted for various values of the mass $\mu$ of the massive scalar field. It is observed that when $\mu$ is increased $|R|^2$ is increased and $|T|^2$ is decreased. Hence the GBF is decreased for high $\mu$. The behavior of $\sigma_{\ell}$ as a function of $\omega$ is shown in  Fig.\eqref{Mu-sigma} for fixed values of $M$, $g$, $\ell$ and varying $\mu$. It is observed that the partial absorption cross section $\sigma_{\ell}$ decreased when $\mu$ increases. Hence, the mass of the scalar field suppresses the absorption of the wave by the BH. By observing the effective potential in Fig.\eqref{potscalar}, one can see that the height of the potential increases with $\mu$, which also implies less absorption of the wave.

\begin{figure*}[t]
\begin{center}
\includegraphics[width=1\linewidth]{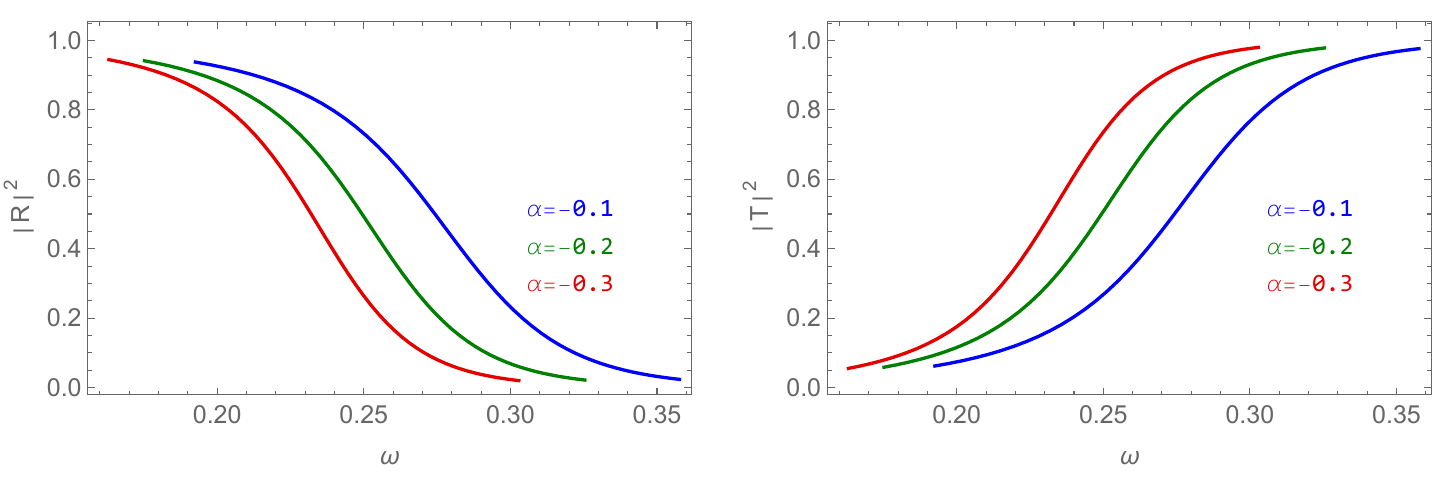}
\caption{The figures show $|R|^2$ and $|T|^2$ as a function of $\omega$ for the values $\alpha$ depicted in the graphs for the massive scalar case, where $M=1$, $g = 0.1$, $ \ell = 1$ and $\mu = 0.1$ remain constant.}
\label{alpha-R-T}
\end{center}
\end{figure*}

\begin{figure*}[t]
\begin{center}
\includegraphics[width=0.8\linewidth]{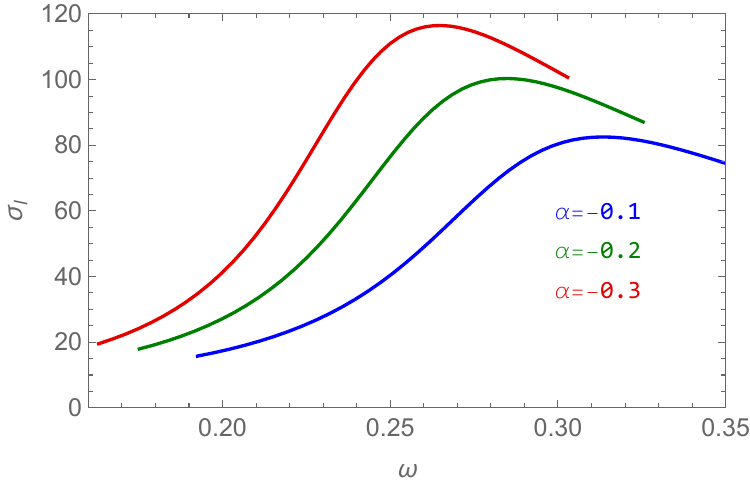}
\caption{The figure depicts the partial absorption cross section as a function of $\omega$ for the values $\alpha$ displayed in the graph for the massive case, with $M=1$, $g=0.1$, $\ell = 1$ and $\mu=0.1$ constant.}
\label{alpha-sigma}
\end{center}
\end{figure*}

Meanwhile, Figs.\eqref{alpha-R-T} and \eqref{alpha-sigma} illustrate the effect of the alpha parameter on $|R|^2$, $|T|^2$ and $\sigma_\ell$. It is observed from the figures that when the dark matter parameter $\alpha$ is increased, the reflection coefficient increases and the partial absorption cross section decreases. It means that the dark matter parameter supresses the absorption of the wave by the BH. By observing the effective potential in Fig.\eqref{potscalar}, this seems obvious, given that the height of the potential increases when $\alpha$ increases.

\subsection{Massless Dirac field case}
\begin{figure}[h!]
\centering
\includegraphics[scale=0.7]{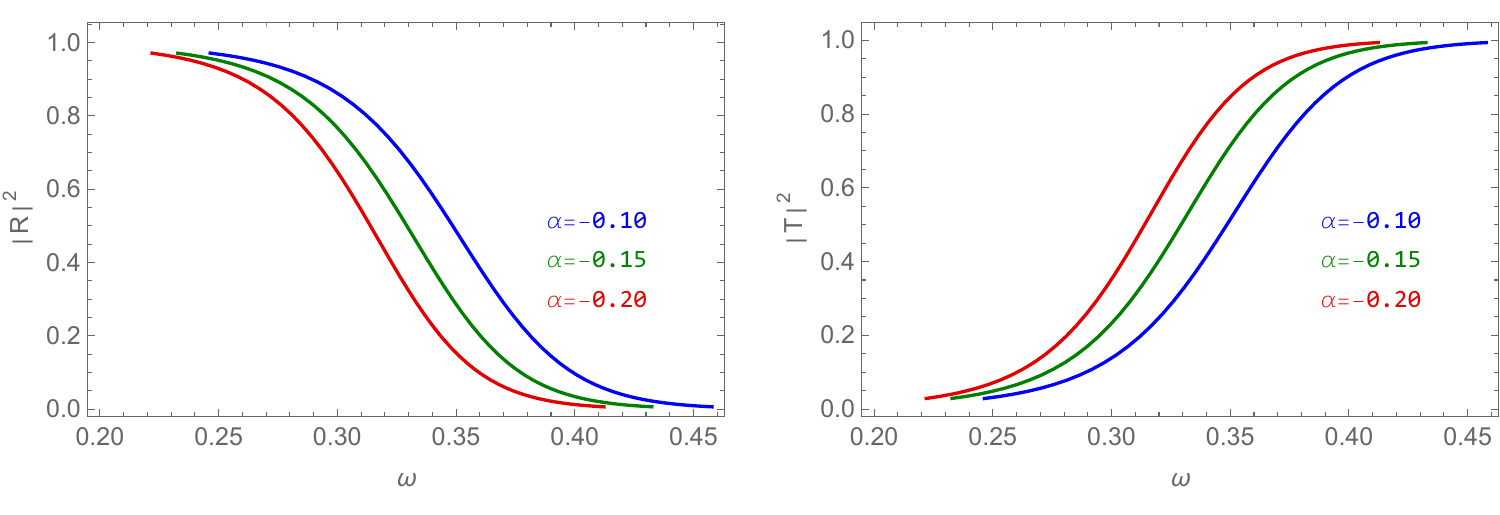}
\captionsetup{width=0.9\linewidth}
\caption{The $|R|^2$ and $|T|^2$ coefficients for the massless Dirac field case are plotted as a function of $\omega$ for the specified values of $\alpha$, keeping $M=1$, $\ell = 3/2$, and $g = 0.6$ constant.}
\label{Dirac-R-T}
\end{figure}

\begin{figure*}[t]
\begin{center}
\includegraphics[width=0.8\linewidth]{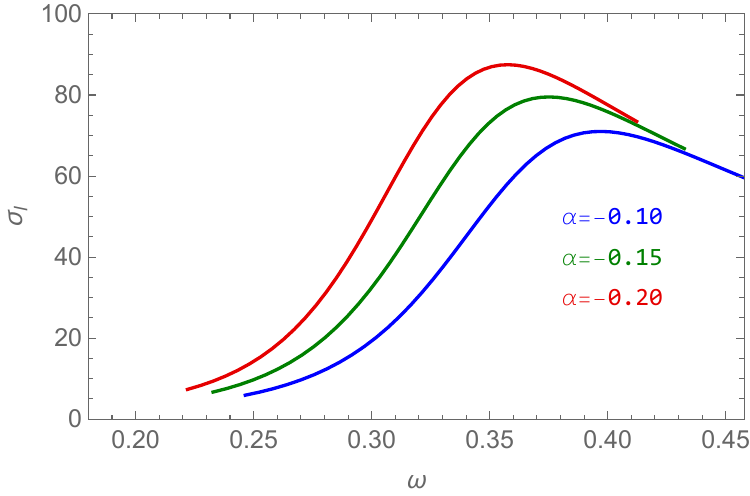}
\caption{The figure displays the partial absorption cross section as a function of $\omega$ for the massless Dirac case for different values of $\alpha$, where $M=1$, $\ell = 3/2$, and $g =0.6$ are kept fixed.}
\label{Dirac-sigma}
\end{center}
\end{figure*}

In Fig.\eqref{Dirac-R-T}, $|R|^2$ and $|T|^2$ are plotted against $r$ for various values of the dark matter parameter, $\alpha$. It is observed that $|R|^2$ increases and $|T|^2$ decreases with $\alpha$. Hence, the GBF decreases with $\alpha$. The partial absorption cross section, $\sigma_{\ell}$ decreases with $\alpha$ as in Fig.\eqref{Dirac-sigma} implying the dark matter supresses the absorption of the Dirac field by the BH. When observed the potential for the Dirac field in Fig.\eqref{potlambda}, it is confirmed since the height of the potential increased  when $\alpha$ increased suppressing  the absorption of the wave.

\section{Final remarks} \label{sec5}

In this paper we have studied massive scalar field and Dirac field perturbations of a regular Bardeen BH immersed in perfect fluid dark matter in a four-dimensional space-time.
First, we gave a descriptive introduction and summary of the relevant equations and basic ingredients regarding first-order perturbation theory. Second, we  analyze the effective potentials for both fields. Third, we compute the corresponding QNMs using the WKB approximation and  Pöschl-Teller approximation. Fourth, we computed GBFs (or absorption cross sections) for both the massive scalar field and the Dirac field.

Let us discuss the effective potentials for both fields. For the massive scalar field,  when $\mu$ increased, the potential increased. When $\alpha$ increased (with keeping $M$, $\mu$, $n$, $\ell$ fixed), the potential increased. For the Dirac field, when the multipole number, $\lambda$, is increased, the potential increased. When $g$ is increased, the potential increased.

 We utilized WKB approach and Pöschl-Teller approach to compute QNM frequencies. We computed QNM frequencies for massless scalar field also to compare with the massive scalar field values. As for QNMs of the massive scalar field and the massless Dirac field, all the modes are supposed to be stable since the imaginary part of the QNM frequencies is negative.

As for the massive scalar field, when $\ell$ increased both Re $\omega$ and -Im $\omega$ increased. Also we noted that  having a mass for the scalar field does induce more damping for the field. When the overtone number $n$ increases, Re $\omega$ decreases and -Im $\omega$ increases: hence, for high overtones, the damping is high and the oscillation is low. This is in line with the massless field QNM frequencies. When $\alpha$ is increased both Re $\omega$ and -Im $\omega$ increases. We observed a similar behavior for the massless case we computed in our paper. However, in \cite{Sun:2023slzl} Re $\omega$ behaves similarly to our calculations but, -Im $\omega$ decreased for a range of values of $\alpha$.

\begin{figure} [H]
\begin{center}
\includegraphics[width=0.497\linewidth]{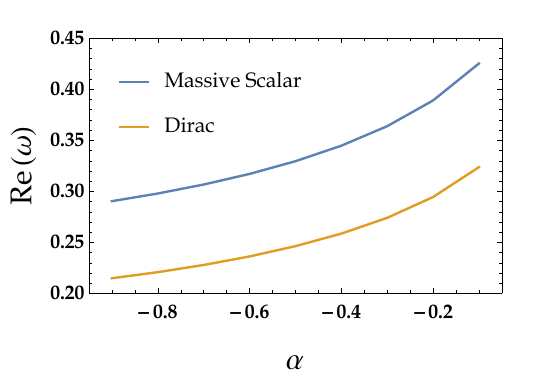} 
\includegraphics[width=0.497\linewidth]{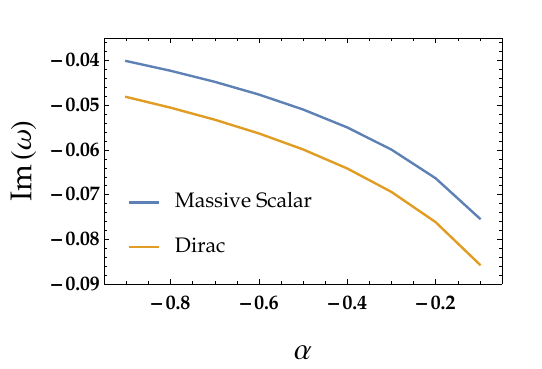}
\caption{
{\bf{First row:}} 
The left panel shows the real part of the quasinormal frequency $\omega_R$, while the right panel shows the imaginary part of the quasinormal frequency $\omega_I$ both cases against the parameter $\alpha$, assuming $M=1, \ g=0.1, \ n=0 $.
We have considered the case $\ell=2$ for the massive scalar case, and $\lambda=2$ for the Dirac field, 
using  the WKB approximation.}

\label{Fig15}
 \end{center}
 \end{figure}

Finally, for the sake of comparison, Fig.\eqref{Fig15} shows the real and imaginary parts of the QNM frequencies for the massive scalar case (blue line) and the Dirac case (orange line).
From the left-hand side, we can see that:
(i) in both cases, the real part of the QNM frequencies increases with  $\alpha$;
(ii) the real part of the QNM frequencies for the massive case is always larger than those for the Dirac case.
From the right-hand side, we observe that:
(i) in both cases, the imaginary part of the quasinormal frequencies decreases when $\alpha$ increases;
(ii) The imaginary part of the quasinormal frequencies for the massive case is always larger than those for the Dirac case.

Now, let us  discuss the absorption cross sections and the GBFs. For the massive scalar field, the GBF $|T|^2$ increased and the partial absorption cross section $\sigma_{\ell}$ decreased when $\mu$ increased. Hence $\mu$ suppresses the absorption of the wave by the BH. When the dark matter parameter $\alpha$ is increased, $|T|^2$ and $\sigma_{\ell}$ are decreased. Hence $\alpha$ suppresses  the absorption of the wave. As for the massless Dirac field, when $\alpha$ increases both $|T|^2$ and $\sigma_{\ell}$  decreases suppressing the absorption of the wave.

In extending this work, it would be interesting to study how the inclusion of a negative cosmological constant to the PFDM BH affects the QNM frequencies and absorption cross sections. Given that AdS BHs have attracted a lot of attention due to the AdS/CFT correspondence, this certainly would be an interesting problem to pursue.

Now that the massless, massive scalar field and massless Dirac field perturbations are completed, a natural extension would be to study electromagnetic perturbations: in this regard, one cannot use the usual effective potential in the literature utilized for the Maxwell electromagnetic  perturbations given by $V_{EM} = \ell(\ell+1)f(r)/r^2$. Instead, an exhaustive approach has to be used along the lines of the paper \cite{Li:2015lly} since the electromagnetic fields are non-linear.
Another avenue to proceed would be to study the perturbations due to charged scalar field of this BH along the lines of the work by Huang and Liu has studying charged scalar perturbation around a magnetic regular BH \cite{Huang:2016hl}. 

\section*{Acknowledgments}

We thank the anonymous reviewer for useful comments and suggestions. A. R. acknowledges financial support from the Generalitat Valenciana through PROMETEO PROJECT CIPROM/2022/13.
A. R. is funded by the María Zambrano contract ZAMBRANO 21-25 (Spain) (with funding from NextGenerationEU).
The author L. B. is supported by DIUFRO through the project: DI24-0087.


\bibliographystyle{apsrev4-1}
\bibliography{Library.bib}

\begin{thebibliography}{124}%
\makeatletter
\providecommand \@ifxundefined [1]{%
 \@ifx{#1\undefined}
}%
\providecommand \@ifnum [1]{%
 \ifnum #1\expandafter \@firstoftwo
 \else \expandafter \@secondoftwo
 \fi
}%
\providecommand \@ifx [1]{%
 \ifx #1\expandafter \@firstoftwo
 \else \expandafter \@secondoftwo
 \fi
}%
\providecommand \natexlab [1]{#1}%
\providecommand \enquote  [1]{``#1''}%
\providecommand \bibnamefont  [1]{#1}%
\providecommand \bibfnamefont [1]{#1}%
\providecommand \citenamefont [1]{#1}%
\providecommand \href@noop [0]{\@secondoftwo}%
\providecommand \href [0]{\begingroup \@sanitize@url \@href}%
\providecommand \@href[1]{\@@startlink{#1}\@@href}%
\providecommand \@@href[1]{\endgroup#1\@@endlink}%
\providecommand \@sanitize@url [0]{\catcode `\\12\catcode `\$12\catcode `\&12\catcode `\#12\catcode `\^12\catcode `\_12\catcode `\%12\relax}%
\providecommand \@@startlink[1]{}%
\providecommand \@@endlink[0]{}%
\providecommand \url  [0]{\begingroup\@sanitize@url \@url }%
\providecommand \@url [1]{\endgroup\@href {#1}{\urlprefix }}%
\providecommand \urlprefix  [0]{URL }%
\providecommand \Eprint [0]{\href }%
\providecommand \doibase [0]{http://dx.doi.org/}%
\providecommand \selectlanguage [0]{\@gobble}%
\providecommand \bibinfo  [0]{\@secondoftwo}%
\providecommand \bibfield  [0]{\@secondoftwo}%
\providecommand \translation [1]{[#1]}%
\providecommand \BibitemOpen [0]{}%
\providecommand \bibitemStop [0]{}%
\providecommand \bibitemNoStop [0]{.\EOS\space}%
\providecommand \EOS [0]{\spacefactor3000\relax}%
\providecommand \BibitemShut  [1]{\csname bibitem#1\endcsname}%
\let\auto@bib@innerbib\@empty
\bibitem [{\citenamefont {Sun}\ \emph {et~al.}(2023)\citenamefont {Sun}, \citenamefont {Li}, \citenamefont {Zhang},\ and\ \citenamefont {Li}}]{Sun:2023slzl}%
  \BibitemOpen
  \bibfield  {author} {\bibinfo {author} {\bibfnamefont {Q.}~\bibnamefont {Sun}}, \bibinfo {author} {\bibfnamefont {Q.}~\bibnamefont {Li}}, \bibinfo {author} {\bibfnamefont {Y.}~\bibnamefont {Zhang}}, \ and\ \bibinfo {author} {\bibfnamefont {Q.-Q.}\ \bibnamefont {Li}},\ }\href@noop {} {\bibfield  {journal} {\bibinfo  {journal} {Mod.Phys. Lett. A.}\ }\textbf {\bibinfo {volume} {38}},\ \bibinfo {pages} {2350102} (\bibinfo {year} {2023})},\ \Eprint {http://arxiv.org/abs/2302.10758} {arXiv:2302.10758 [gr-qc]} \BibitemShut {NoStop}%
\bibitem [{\citenamefont {Einstein}(1915)}]{Einstein:1915ca}%
  \BibitemOpen
  \bibfield  {author} {\bibinfo {author} {\bibfnamefont {A.}~\bibnamefont {Einstein}},\ }\href@noop {} {\bibfield  {journal} {\bibinfo  {journal} {Sitzungsber. Preuss. Akad. Wiss. Berlin (Math. Phys. )}\ }\textbf {\bibinfo {volume} {1915}},\ \bibinfo {pages} {844} (\bibinfo {year} {1915})}\BibitemShut {NoStop}%
\bibitem [{\citenamefont {Abbott}\ \emph {et~al.}(2016)\citenamefont {Abbott} \emph {et~al.}}]{LIGOScientific:2016aoc}%
  \BibitemOpen
  \bibfield  {author} {\bibinfo {author} {\bibfnamefont {B.~P.}\ \bibnamefont {Abbott}} \emph {et~al.} (\bibinfo {collaboration} {LIGO Scientific, Virgo}),\ }\href {\doibase 10.1103/PhysRevLett.116.061102} {\bibfield  {journal} {\bibinfo  {journal} {Phys. Rev. Lett.}\ }\textbf {\bibinfo {volume} {116}},\ \bibinfo {pages} {061102} (\bibinfo {year} {2016})},\ \Eprint {http://arxiv.org/abs/1602.03837} {arXiv:1602.03837 [gr-qc]} \BibitemShut {NoStop}%
\bibitem [{\citenamefont {Hawking}(1975)}]{Hawking:1975vcx}%
  \BibitemOpen
  \bibfield  {author} {\bibinfo {author} {\bibfnamefont {S.~W.}\ \bibnamefont {Hawking}},\ }\href {\doibase 10.1007/BF02345020} {\bibfield  {journal} {\bibinfo  {journal} {Commun. Math. Phys.}\ }\textbf {\bibinfo {volume} {43}},\ \bibinfo {pages} {199} (\bibinfo {year} {1975})},\ \bibinfo {note} {[Erratum: Commun.Math.Phys. 46, 206 (1976)]}\BibitemShut {NoStop}%
\bibitem [{\citenamefont {Bekenstein}(1973)}]{Bekenstein:1973ur}%
  \BibitemOpen
  \bibfield  {author} {\bibinfo {author} {\bibfnamefont {J.~D.}\ \bibnamefont {Bekenstein}},\ }\href {\doibase 10.1103/PhysRevD.7.2333} {\bibfield  {journal} {\bibinfo  {journal} {Phys. Rev. D}\ }\textbf {\bibinfo {volume} {7}},\ \bibinfo {pages} {2333} (\bibinfo {year} {1973})}\BibitemShut {NoStop}%
\bibitem [{\citenamefont {Bardeen}\ \emph {et~al.}(1973)\citenamefont {Bardeen}, \citenamefont {Carter},\ and\ \citenamefont {Hawking}}]{Bardeen:1973gs}%
  \BibitemOpen
  \bibfield  {author} {\bibinfo {author} {\bibfnamefont {J.~M.}\ \bibnamefont {Bardeen}}, \bibinfo {author} {\bibfnamefont {B.}~\bibnamefont {Carter}}, \ and\ \bibinfo {author} {\bibfnamefont {S.~W.}\ \bibnamefont {Hawking}},\ }\href {\doibase 10.1007/BF01645742} {\bibfield  {journal} {\bibinfo  {journal} {Commun. Math. Phys.}\ }\textbf {\bibinfo {volume} {31}},\ \bibinfo {pages} {161} (\bibinfo {year} {1973})}\BibitemShut {NoStop}%
\bibitem [{\citenamefont {Hawking}\ and\ \citenamefont {Page}(1983)}]{Hawking:1982dh}%
  \BibitemOpen
  \bibfield  {author} {\bibinfo {author} {\bibfnamefont {S.~W.}\ \bibnamefont {Hawking}}\ and\ \bibinfo {author} {\bibfnamefont {D.~N.}\ \bibnamefont {Page}},\ }\href {\doibase 10.1007/BF01208266} {\bibfield  {journal} {\bibinfo  {journal} {Commun. Math. Phys.}\ }\textbf {\bibinfo {volume} {87}},\ \bibinfo {pages} {577} (\bibinfo {year} {1983})}\BibitemShut {NoStop}%
\bibitem [{\citenamefont {Regge}\ and\ \citenamefont {Wheeler}(1957)}]{Regge:1957td}%
  \BibitemOpen
  \bibfield  {author} {\bibinfo {author} {\bibfnamefont {T.}~\bibnamefont {Regge}}\ and\ \bibinfo {author} {\bibfnamefont {J.~A.}\ \bibnamefont {Wheeler}},\ }\href {\doibase 10.1103/PhysRev.108.1063} {\bibfield  {journal} {\bibinfo  {journal} {Phys. Rev.}\ }\textbf {\bibinfo {volume} {108}},\ \bibinfo {pages} {1063} (\bibinfo {year} {1957})}\BibitemShut {NoStop}%
\bibitem [{\citenamefont {Zerilli}(1970{\natexlab{a}})}]{Zerilli:1970se}%
  \BibitemOpen
  \bibfield  {author} {\bibinfo {author} {\bibfnamefont {F.~J.}\ \bibnamefont {Zerilli}},\ }\href {\doibase 10.1103/PhysRevLett.24.737} {\bibfield  {journal} {\bibinfo  {journal} {Phys. Rev. Lett.}\ }\textbf {\bibinfo {volume} {24}},\ \bibinfo {pages} {737} (\bibinfo {year} {1970}{\natexlab{a}})}\BibitemShut {NoStop}%
\bibitem [{\citenamefont {Zerilli}(1970{\natexlab{b}})}]{Zerilli:1970wzz}%
  \BibitemOpen
  \bibfield  {author} {\bibinfo {author} {\bibfnamefont {F.~J.}\ \bibnamefont {Zerilli}},\ }\href {\doibase 10.1103/PhysRevD.2.2141} {\bibfield  {journal} {\bibinfo  {journal} {Phys. Rev. D}\ }\textbf {\bibinfo {volume} {2}},\ \bibinfo {pages} {2141} (\bibinfo {year} {1970}{\natexlab{b}})}\BibitemShut {NoStop}%
\bibitem [{\citenamefont {Zerilli}(1974)}]{Zerilli:1974ai}%
  \BibitemOpen
  \bibfield  {author} {\bibinfo {author} {\bibfnamefont {F.~J.}\ \bibnamefont {Zerilli}},\ }\href {\doibase 10.1103/PhysRevD.9.860} {\bibfield  {journal} {\bibinfo  {journal} {Phys. Rev. D}\ }\textbf {\bibinfo {volume} {9}},\ \bibinfo {pages} {860} (\bibinfo {year} {1974})}\BibitemShut {NoStop}%
\bibitem [{\citenamefont {Moncrief}(1975)}]{Moncrief:1975sb}%
  \BibitemOpen
  \bibfield  {author} {\bibinfo {author} {\bibfnamefont {V.}~\bibnamefont {Moncrief}},\ }\href {\doibase 10.1103/PhysRevD.12.1526} {\bibfield  {journal} {\bibinfo  {journal} {Phys. Rev. D}\ }\textbf {\bibinfo {volume} {12}},\ \bibinfo {pages} {1526} (\bibinfo {year} {1975})}\BibitemShut {NoStop}%
\bibitem [{\citenamefont {Teukolsky}(1972)}]{Teukolsky:1972my}%
  \BibitemOpen
  \bibfield  {author} {\bibinfo {author} {\bibfnamefont {S.~A.}\ \bibnamefont {Teukolsky}},\ }\href {\doibase 10.1103/PhysRevLett.29.1114} {\bibfield  {journal} {\bibinfo  {journal} {Phys. Rev. Lett.}\ }\textbf {\bibinfo {volume} {29}},\ \bibinfo {pages} {1114} (\bibinfo {year} {1972})}\BibitemShut {NoStop}%
\bibitem [{\citenamefont {Chandrasekhar}(1985)}]{Chandrasekhar:1985kt}%
  \BibitemOpen
  \bibfield  {author} {\bibinfo {author} {\bibfnamefont {S.}~\bibnamefont {Chandrasekhar}},\ }\href@noop {} {\emph {\bibinfo {title} {{The mathematical theory of black holes}}}}\ (\bibinfo {year} {1985})\BibitemShut {NoStop}%
\bibitem [{\citenamefont {Li}\ and\ \citenamefont {Yang}(2015)}]{Li:2015ly}%
  \BibitemOpen
  \bibfield  {author} {\bibinfo {author} {\bibfnamefont {J.}~\bibnamefont {Li}}\ and\ \bibinfo {author} {\bibfnamefont {N.}~\bibnamefont {Yang}},\ }\href@noop {} {\bibfield  {journal} {\bibinfo  {journal} {Eur.Phys.J.Plus}\ }\textbf {\bibinfo {volume} {75}},\ \bibinfo {pages} {131} (\bibinfo {year} {2015})},\ \Eprint {http://arxiv.org/abs/2014.5988.} {arXiv:2014.5988. [gr-qc]} \BibitemShut {NoStop}%
\bibitem [{\citenamefont {Amado}\ and\ \citenamefont {Gwak}(2024)}]{Amado:2024ag}%
  \BibitemOpen
  \bibfield  {author} {\bibinfo {author} {\bibfnamefont {J.~B.}\ \bibnamefont {Amado}}\ and\ \bibinfo {author} {\bibfnamefont {B.}~\bibnamefont {Gwak}},\ }\href@noop {} {\bibfield  {journal} {\bibinfo  {journal} {JHEP}\ ,\ \bibinfo {pages} {189}} (\bibinfo {year} {2024})},\ \Eprint {http://arxiv.org/abs/2309.11355} {arXiv:2309.11355 [gr-qc]} \BibitemShut {NoStop}%
\bibitem [{\citenamefont {Priyadarshinee}(2024)}]{Priya:2024sp}%
  \BibitemOpen
  \bibfield  {author} {\bibinfo {author} {\bibfnamefont {S.}~\bibnamefont {Priyadarshinee}},\ }\href@noop {} {\bibfield  {journal} {\bibinfo  {journal} {Eur.Phys.J.Plus}\ }\textbf {\bibinfo {volume} {3}},\ \bibinfo {pages} {258} (\bibinfo {year} {2024})},\ \Eprint {http://arxiv.org/abs/2308.05719} {arXiv:2308.05719 [gr-qc]} \BibitemShut {NoStop}%
\bibitem [{\citenamefont {Zhang}\ and\ \citenamefont {Wang}(2024)}]{Zhang:2024zw}%
  \BibitemOpen
  \bibfield  {author} {\bibinfo {author} {\bibfnamefont {C.}~\bibnamefont {Zhang}}\ and\ \bibinfo {author} {\bibfnamefont {A.}~\bibnamefont {Wang}},\ }\href@noop {} {\bibfield  {journal} {\bibinfo  {journal} {JCAP}\ }\textbf {\bibinfo {volume} {10}},\ \bibinfo {pages} {070} (\bibinfo {year} {2024})},\ \Eprint {http://arxiv.org/abs/2407.19654} {arXiv:2407.19654 [gr-qc]} \BibitemShut {NoStop}%
\bibitem [{\citenamefont {Livine}\ \emph {et~al.}(2024)\citenamefont {Livine}, \citenamefont {Montagnon}, \citenamefont {Oshita},\ and\ \citenamefont {Roussille}}]{Livine:2024lm}%
  \BibitemOpen
  \bibfield  {author} {\bibinfo {author} {\bibfnamefont {E.~R.}\ \bibnamefont {Livine}}, \bibinfo {author} {\bibfnamefont {C.}~\bibnamefont {Montagnon}}, \bibinfo {author} {\bibfnamefont {N.}~\bibnamefont {Oshita}}, \ and\ \bibinfo {author} {\bibfnamefont {H.}~\bibnamefont {Roussille}},\ }\href@noop {} {\bibfield  {journal} {\bibinfo  {journal} {JCAP}\ }\textbf {\bibinfo {volume} {10}},\ \bibinfo {pages} {037} (\bibinfo {year} {2024})},\ \Eprint {http://arxiv.org/abs/2405.12671} {arXiv:2405.12671 [gr-qc]} \BibitemShut {NoStop}%
\bibitem [{\citenamefont {Chung}\ and\ \citenamefont {Yunes}(2024{\natexlab{a}})}]{Chung:2024cy}%
  \BibitemOpen
  \bibfield  {author} {\bibinfo {author} {\bibfnamefont {A.~K.-W.}\ \bibnamefont {Chung}}\ and\ \bibinfo {author} {\bibfnamefont {N.}~\bibnamefont {Yunes}},\ }\href@noop {} {\bibfield  {journal} {\bibinfo  {journal} {Phys.Rev.Lett}\ }\textbf {\bibinfo {volume} {133}},\ \bibinfo {pages} {181401} (\bibinfo {year} {2024}{\natexlab{a}})},\ \Eprint {http://arxiv.org/abs/2405.12280} {arXiv:2405.12280 [gr-qc]} \BibitemShut {NoStop}%
\bibitem [{\citenamefont {Chung}\ and\ \citenamefont {Yunes}(2024{\natexlab{b}})}]{Chung:2024cy1}%
  \BibitemOpen
  \bibfield  {author} {\bibinfo {author} {\bibfnamefont {A.~K.-W.}\ \bibnamefont {Chung}}\ and\ \bibinfo {author} {\bibfnamefont {N.}~\bibnamefont {Yunes}},\ }\href@noop {} {\bibfield  {journal} {\bibinfo  {journal} {Phys.Rev.D}\ }\textbf {\bibinfo {volume} {110}},\ \bibinfo {pages} {064019} (\bibinfo {year} {2024}{\natexlab{b}})},\ \Eprint {http://arxiv.org/abs/2406.11986} {arXiv:2406.11986 [gr-qc]} \BibitemShut {NoStop}%
\bibitem [{\citenamefont {Lagos}\ \emph {et~al.}(2025)\citenamefont {Lagos}, \citenamefont {Andrade}, \citenamefont {Rafecas-Ventosa},\ and\ \citenamefont {Hui}}]{Lagos:2025la}%
  \BibitemOpen
  \bibfield  {author} {\bibinfo {author} {\bibfnamefont {M.}~\bibnamefont {Lagos}}, \bibinfo {author} {\bibfnamefont {T.}~\bibnamefont {Andrade}}, \bibinfo {author} {\bibfnamefont {J.}~\bibnamefont {Rafecas-Ventosa}}, \ and\ \bibinfo {author} {\bibfnamefont {L.}~\bibnamefont {Hui}},\ }\href@noop {} {\bibfield  {journal} {\bibinfo  {journal} {Phys.Rev.D}\ }\textbf {\bibinfo {volume} {111}},\ \bibinfo {pages} {024018} (\bibinfo {year} {2025})},\ \Eprint {http://arxiv.org/abs/2411.02264} {arXiv:2411.02264 [gr-qc]} \BibitemShut {NoStop}%
\bibitem [{\citenamefont {Rinc\'on}\ and\ \citenamefont {Panotopoulos}(2018{\natexlab{a}})}]{Rincon:2018sgd}%
  \BibitemOpen
  \bibfield  {author} {\bibinfo {author} {\bibfnamefont {A.}~\bibnamefont {Rinc\'on}}\ and\ \bibinfo {author} {\bibfnamefont {G.}~\bibnamefont {Panotopoulos}},\ }\href {\doibase 10.1103/PhysRevD.97.024027} {\bibfield  {journal} {\bibinfo  {journal} {Phys. Rev. D}\ }\textbf {\bibinfo {volume} {97}},\ \bibinfo {pages} {024027} (\bibinfo {year} {2018}{\natexlab{a}})},\ \Eprint {http://arxiv.org/abs/1801.03248} {arXiv:1801.03248 [hep-th]} \BibitemShut {NoStop}%
\bibitem [{\citenamefont {Panotopoulos}\ and\ \citenamefont {Rinc\'on}(2017{\natexlab{a}})}]{Panotopoulos:2017hns}%
  \BibitemOpen
  \bibfield  {author} {\bibinfo {author} {\bibfnamefont {G.}~\bibnamefont {Panotopoulos}}\ and\ \bibinfo {author} {\bibfnamefont {A.}~\bibnamefont {Rinc\'on}},\ }\href {\doibase 10.1142/S0218271818500347} {\bibfield  {journal} {\bibinfo  {journal} {Int. J. Mod. Phys. D}\ }\textbf {\bibinfo {volume} {27}},\ \bibinfo {pages} {1850034} (\bibinfo {year} {2017}{\natexlab{a}})},\ \Eprint {http://arxiv.org/abs/1711.04146} {arXiv:1711.04146 [hep-th]} \BibitemShut {NoStop}%
\bibitem [{\citenamefont {Panotopoulos}\ and\ \citenamefont {Rinc\'on}(2019)}]{Panotopoulos:2019qjk}%
  \BibitemOpen
  \bibfield  {author} {\bibinfo {author} {\bibfnamefont {G.}~\bibnamefont {Panotopoulos}}\ and\ \bibinfo {author} {\bibfnamefont {A.}~\bibnamefont {Rinc\'on}},\ }\href {\doibase 10.1140/epjp/i2019-12686-x} {\bibfield  {journal} {\bibinfo  {journal} {Eur. Phys. J. Plus}\ }\textbf {\bibinfo {volume} {134}},\ \bibinfo {pages} {300} (\bibinfo {year} {2019})},\ \Eprint {http://arxiv.org/abs/1904.10847} {arXiv:1904.10847 [gr-qc]} \BibitemShut {NoStop}%
\bibitem [{\citenamefont {Panotopoulos}\ and\ \citenamefont {Rinc\'on}(2021)}]{Panotopoulos:2020mii}%
  \BibitemOpen
  \bibfield  {author} {\bibinfo {author} {\bibfnamefont {G.}~\bibnamefont {Panotopoulos}}\ and\ \bibinfo {author} {\bibfnamefont {A.}~\bibnamefont {Rinc\'on}},\ }\href {\doibase 10.1016/j.dark.2020.100743} {\bibfield  {journal} {\bibinfo  {journal} {Phys. Dark Univ.}\ }\textbf {\bibinfo {volume} {31}},\ \bibinfo {pages} {100743} (\bibinfo {year} {2021})},\ \Eprint {http://arxiv.org/abs/2011.02860} {arXiv:2011.02860 [gr-qc]} \BibitemShut {NoStop}%
\bibitem [{\citenamefont {Rinc\'on}\ and\ \citenamefont {Santos}(2020)}]{Rincon:2020cos}%
  \BibitemOpen
  \bibfield  {author} {\bibinfo {author} {\bibfnamefont {A.}~\bibnamefont {Rinc\'on}}\ and\ \bibinfo {author} {\bibfnamefont {V.}~\bibnamefont {Santos}},\ }\href {\doibase 10.1140/epjc/s10052-020-08445-2} {\bibfield  {journal} {\bibinfo  {journal} {Eur. Phys. J. C}\ }\textbf {\bibinfo {volume} {80}},\ \bibinfo {pages} {910} (\bibinfo {year} {2020})},\ \Eprint {http://arxiv.org/abs/2009.04386} {arXiv:2009.04386 [gr-qc]} \BibitemShut {NoStop}%
\bibitem [{\citenamefont {Rinc\'on}\ \emph {et~al.}(2024)\citenamefont {Rinc\'on}, \citenamefont {\"Ovg\"un},\ and\ \citenamefont {Pantig}}]{Rincon:2024won}%
  \BibitemOpen
  \bibfield  {author} {\bibinfo {author} {\bibfnamefont {A.}~\bibnamefont {Rinc\'on}}, \bibinfo {author} {\bibfnamefont {A.}~\bibnamefont {\"Ovg\"un}}, \ and\ \bibinfo {author} {\bibfnamefont {R.~C.}\ \bibnamefont {Pantig}},\ }\href {\doibase 10.1016/j.dark.2024.101623} {\bibfield  {journal} {\bibinfo  {journal} {Phys. Dark Univ.}\ }\textbf {\bibinfo {volume} {46}},\ \bibinfo {pages} {101623} (\bibinfo {year} {2024})},\ \Eprint {http://arxiv.org/abs/2409.10930} {arXiv:2409.10930 [gr-qc]} \BibitemShut {NoStop}%
\bibitem [{\citenamefont {Balart}\ \emph {et~al.}(2023)\citenamefont {Balart}, \citenamefont {Belmar-Herrera}, \citenamefont {Panotopoulos},\ and\ \citenamefont {Rinc\'on}}]{Balart:2023swp}%
  \BibitemOpen
  \bibfield  {author} {\bibinfo {author} {\bibfnamefont {L.}~\bibnamefont {Balart}}, \bibinfo {author} {\bibfnamefont {S.}~\bibnamefont {Belmar-Herrera}}, \bibinfo {author} {\bibfnamefont {G.}~\bibnamefont {Panotopoulos}}, \ and\ \bibinfo {author} {\bibfnamefont {A.}~\bibnamefont {Rinc\'on}},\ }\href {\doibase 10.1016/j.aop.2023.169329} {\bibfield  {journal} {\bibinfo  {journal} {Annals Phys.}\ }\textbf {\bibinfo {volume} {454}},\ \bibinfo {pages} {169329} (\bibinfo {year} {2023})},\ \Eprint {http://arxiv.org/abs/2304.09777} {arXiv:2304.09777 [gr-qc]} \BibitemShut {NoStop}%
\bibitem [{\citenamefont {Skvortsova}(2025)}]{Skvortsova:2024eqi}%
  \BibitemOpen
  \bibfield  {author} {\bibinfo {author} {\bibfnamefont {M.}~\bibnamefont {Skvortsova}},\ }\href {\doibase 10.1209/0295-5075/adaee2} {\bibfield  {journal} {\bibinfo  {journal} {EPL}\ }\textbf {\bibinfo {volume} {149}},\ \bibinfo {pages} {59001} (\bibinfo {year} {2025})},\ \Eprint {http://arxiv.org/abs/2503.03650} {arXiv:2503.03650 [gr-qc]} \BibitemShut {NoStop}%
\bibitem [{\citenamefont {Rosato}\ \emph {et~al.}(2024)\citenamefont {Rosato}, \citenamefont {Destounis},\ and\ \citenamefont {P.}}]{Rosato:2024rdp}%
  \BibitemOpen
  \bibfield  {author} {\bibinfo {author} {\bibfnamefont {R.~F.}\ \bibnamefont {Rosato}}, \bibinfo {author} {\bibfnamefont {K.}~\bibnamefont {Destounis}}, \ and\ \bibinfo {author} {\bibfnamefont {P.}~\bibnamefont {P.}},\ }\href@noop {} {\bibfield  {journal} {\bibinfo  {journal} {Phys. Rev. D.}\ }\textbf {\bibinfo {volume} {110}},\ \bibinfo {pages} {L121501} (\bibinfo {year} {2024})},\ \Eprint {http://arxiv.org/abs/2406.01692} {arXiv:2406.01692 [gr-qc]} \BibitemShut {NoStop}%
\bibitem [{\citenamefont {Konoplya}\ and\ \citenamefont {Zinhailo1}(2019)}]{Konoplya:2019kz}%
  \BibitemOpen
  \bibfield  {author} {\bibinfo {author} {\bibfnamefont {R.~A.}\ \bibnamefont {Konoplya}}\ and\ \bibinfo {author} {\bibfnamefont {A.~F.}\ \bibnamefont {Zinhailo1}},\ }\href@noop {} {\bibfield  {journal} {\bibinfo  {journal} {Phys. Rev. D.}\ }\textbf {\bibinfo {volume} {99}},\ \bibinfo {pages} {104060} (\bibinfo {year} {2019})},\ \Eprint {http://arxiv.org/abs/1904.05341} {arXiv:1904.05341 [gr-qc]} \BibitemShut {NoStop}%
\bibitem [{\citenamefont {Jianhui~Lin}\ \emph {et~al.}(2024)\citenamefont {Jianhui~Lin}, \citenamefont {Moisés Bravo-Gaete},\ and\ \citenamefont {Zhang}}]{Lin:2024lbz}%
  \BibitemOpen
  \bibfield  {author} {\bibinfo {author} {\bibfnamefont {J.}~\bibnamefont {Jianhui~Lin}}, \bibinfo {author} {\bibfnamefont {M.}~\bibnamefont {Moisés Bravo-Gaete}}, \ and\ \bibinfo {author} {\bibfnamefont {X.}~\bibnamefont {Zhang}},\ }\href@noop {} {\bibfield  {journal} {\bibinfo  {journal} {Phys. Rev. D.}\ }\textbf {\bibinfo {volume} {109}},\ \bibinfo {pages} {104039} (\bibinfo {year} {2024})},\ \Eprint {http://arxiv.org/abs/2401.02045} {arXiv:2401.02045 [gr-qc]} \BibitemShut {NoStop}%
\bibitem [{\citenamefont {Li}\ \emph {et~al.}(2024)\citenamefont {Li}, \citenamefont {Zhang}, \citenamefont {Li},\ and\ \citenamefont {Sun}}]{li:2024lzls}%
  \BibitemOpen
  \bibfield  {author} {\bibinfo {author} {\bibfnamefont {Q.}~\bibnamefont {Li}}, \bibinfo {author} {\bibfnamefont {Y.}~\bibnamefont {Zhang}}, \bibinfo {author} {\bibfnamefont {Q.-Q.}\ \bibnamefont {Li}}, \ and\ \bibinfo {author} {\bibfnamefont {Q.}~\bibnamefont {Sun}},\ }\href@noop {} {\bibfield  {journal} {\bibinfo  {journal} {Comm.Theor.Phys.}\ }\textbf {\bibinfo {volume} {76}},\ \bibinfo {pages} {115402} (\bibinfo {year} {2024})},\ \Eprint {http://arxiv.org/abs/2307.04459} {arXiv:2307.04459 [gr-qc]} \BibitemShut {NoStop}%
\bibitem [{\citenamefont {Al-Badawi}(2023)}]{Al-Badawi:2023al}%
  \BibitemOpen
  \bibfield  {author} {\bibinfo {author} {\bibfnamefont {A.}~\bibnamefont {Al-Badawi}},\ }\href@noop {} {\bibfield  {journal} {\bibinfo  {journal} {Euro. Phys. C.}\ }\textbf {\bibinfo {volume} {83}},\ \bibinfo {pages} {380} (\bibinfo {year} {2023})},\ \Eprint {http://arxiv.org/abs/2305.07436} {arXiv:2305.07436 [gr-qc]} \BibitemShut {NoStop}%
\bibitem [{\citenamefont {Baruah}\ \emph {et~al.}(2023)\citenamefont {Baruah}, \citenamefont {Ovg¨un},\ and\ \citenamefont {Deshamukhya1}}]{Baruah:2023bod}%
  \BibitemOpen
  \bibfield  {author} {\bibinfo {author} {\bibfnamefont {A.}~\bibnamefont {Baruah}}, \bibinfo {author} {\bibfnamefont {A.}~\bibnamefont {Ovg¨un}}, \ and\ \bibinfo {author} {\bibfnamefont {A.}~\bibnamefont {Deshamukhya1}},\ }\href@noop {} {\bibfield  {journal} {\bibinfo  {journal} {Annal. Phys.}\ }\textbf {\bibinfo {volume} {455}},\ \bibinfo {pages} {169393} (\bibinfo {year} {2023})},\ \Eprint {http://arxiv.org/abs/2304.07761} {arXiv:2304.07761 [gr-qc]} \BibitemShut {NoStop}%
\bibitem [{\citenamefont {Sakalli}\ and\ \citenamefont {Kanzi}(2022)}]{Sakalli:2022sk}%
  \BibitemOpen
  \bibfield  {author} {\bibinfo {author} {\bibfnamefont {I.}~\bibnamefont {Sakalli}}\ and\ \bibinfo {author} {\bibfnamefont {S.}~\bibnamefont {Kanzi}},\ }\href@noop {} {\bibfield  {journal} {\bibinfo  {journal} {Turk. Jour. Phys.}\ }\textbf {\bibinfo {volume} {46}},\ \bibinfo {pages} {51} (\bibinfo {year} {2022})},\ \Eprint {http://arxiv.org/abs/2205.01771} {arXiv:2205.01771 [gr-qc]} \BibitemShut {NoStop}%
\bibitem [{\citenamefont {\"Ovg\"un}\ \emph {et~al.}(2023)\citenamefont {\"Ovg\"un}, \citenamefont {Pantig},\ and\ \citenamefont {Rinc\'on}}]{Ovgun:2023ego}%
  \BibitemOpen
  \bibfield  {author} {\bibinfo {author} {\bibfnamefont {A.}~\bibnamefont {\"Ovg\"un}}, \bibinfo {author} {\bibfnamefont {R.~C.}\ \bibnamefont {Pantig}}, \ and\ \bibinfo {author} {\bibfnamefont {A.}~\bibnamefont {Rinc\'on}},\ }\href {\doibase 10.1140/epjp/s13360-023-03793-w} {\bibfield  {journal} {\bibinfo  {journal} {Eur. Phys. J. Plus}\ }\textbf {\bibinfo {volume} {138}},\ \bibinfo {pages} {192} (\bibinfo {year} {2023})},\ \Eprint {http://arxiv.org/abs/2303.01696} {arXiv:2303.01696 [gr-qc]} \BibitemShut {NoStop}%
\bibitem [{\citenamefont {Panotopoulos}\ and\ \citenamefont {Rinc\'on}(2018)}]{Panotopoulos:2018pvu}%
  \BibitemOpen
  \bibfield  {author} {\bibinfo {author} {\bibfnamefont {G.}~\bibnamefont {Panotopoulos}}\ and\ \bibinfo {author} {\bibfnamefont {A.}~\bibnamefont {Rinc\'on}},\ }\href {\doibase 10.1103/PhysRevD.97.085014} {\bibfield  {journal} {\bibinfo  {journal} {Phys. Rev. D}\ }\textbf {\bibinfo {volume} {97}},\ \bibinfo {pages} {085014} (\bibinfo {year} {2018})},\ \Eprint {http://arxiv.org/abs/1804.04684} {arXiv:1804.04684 [hep-th]} \BibitemShut {NoStop}%
\bibitem [{\citenamefont {Panotopoulos}\ and\ \citenamefont {Rinc\'on}(2017{\natexlab{b}})}]{Panotopoulos:2016wuu}%
  \BibitemOpen
  \bibfield  {author} {\bibinfo {author} {\bibfnamefont {G.}~\bibnamefont {Panotopoulos}}\ and\ \bibinfo {author} {\bibfnamefont {A.}~\bibnamefont {Rinc\'on}},\ }\href {\doibase 10.1016/j.physletb.2017.07.014} {\bibfield  {journal} {\bibinfo  {journal} {Phys. Lett. B}\ }\textbf {\bibinfo {volume} {772}},\ \bibinfo {pages} {523} (\bibinfo {year} {2017}{\natexlab{b}})},\ \Eprint {http://arxiv.org/abs/1611.06233} {arXiv:1611.06233 [hep-th]} \BibitemShut {NoStop}%
\bibitem [{\citenamefont {Panotopoulos}\ and\ \citenamefont {Rinc\'on}(2017{\natexlab{c}})}]{Panotopoulos:2017yoe}%
  \BibitemOpen
  \bibfield  {author} {\bibinfo {author} {\bibfnamefont {G.}~\bibnamefont {Panotopoulos}}\ and\ \bibinfo {author} {\bibfnamefont {A.}~\bibnamefont {Rinc\'on}},\ }\href {\doibase 10.1103/PhysRevD.96.025009} {\bibfield  {journal} {\bibinfo  {journal} {Phys. Rev. D}\ }\textbf {\bibinfo {volume} {96}},\ \bibinfo {pages} {025009} (\bibinfo {year} {2017}{\natexlab{c}})},\ \Eprint {http://arxiv.org/abs/1706.07455} {arXiv:1706.07455 [hep-th]} \BibitemShut {NoStop}%
\bibitem [{\citenamefont {Lambiase}\ \emph {et~al.}(2023)\citenamefont {Lambiase}, \citenamefont {Mastrototaro}, \citenamefont {Pantig},\ and\ \citenamefont {Ovgun}}]{Lambiase:2023zeo}%
  \BibitemOpen
  \bibfield  {author} {\bibinfo {author} {\bibfnamefont {G.}~\bibnamefont {Lambiase}}, \bibinfo {author} {\bibfnamefont {L.}~\bibnamefont {Mastrototaro}}, \bibinfo {author} {\bibfnamefont {R.~C.}\ \bibnamefont {Pantig}}, \ and\ \bibinfo {author} {\bibfnamefont {A.}~\bibnamefont {Ovgun}},\ }\href {\doibase 10.1088/1475-7516/2023/12/026} {\bibfield  {journal} {\bibinfo  {journal} {JCAP}\ }\textbf {\bibinfo {volume} {12}},\ \bibinfo {pages} {026} (\bibinfo {year} {2023})},\ \Eprint {http://arxiv.org/abs/2309.13594} {arXiv:2309.13594 [gr-qc]} \BibitemShut {NoStop}%
\bibitem [{\citenamefont {Javed}\ \emph {et~al.}(2022)\citenamefont {Javed}, \citenamefont {Riaz},\ and\ \citenamefont {\"Ovg\"un}}]{Javed:2022rrs}%
  \BibitemOpen
  \bibfield  {author} {\bibinfo {author} {\bibfnamefont {W.}~\bibnamefont {Javed}}, \bibinfo {author} {\bibfnamefont {S.}~\bibnamefont {Riaz}}, \ and\ \bibinfo {author} {\bibfnamefont {A.}~\bibnamefont {\"Ovg\"un}},\ }\href {\doibase 10.3390/universe8050262} {\bibfield  {journal} {\bibinfo  {journal} {Universe}\ }\textbf {\bibinfo {volume} {8}},\ \bibinfo {pages} {262} (\bibinfo {year} {2022})},\ \Eprint {http://arxiv.org/abs/2205.02229} {arXiv:2205.02229 [gr-qc]} \BibitemShut {NoStop}%
\bibitem [{\citenamefont {Kim}\ and\ \citenamefont {Oh}(2008)}]{Kim:2007gj}%
  \BibitemOpen
  \bibfield  {author} {\bibinfo {author} {\bibfnamefont {W.}~\bibnamefont {Kim}}\ and\ \bibinfo {author} {\bibfnamefont {J.~J.}\ \bibnamefont {Oh}},\ }\href {\doibase 10.3938/jkps.52.986} {\bibfield  {journal} {\bibinfo  {journal} {J. Korean Phys. Soc.}\ }\textbf {\bibinfo {volume} {52}},\ \bibinfo {pages} {986} (\bibinfo {year} {2008})},\ \Eprint {http://arxiv.org/abs/0709.1754} {arXiv:0709.1754 [hep-th]} \BibitemShut {NoStop}%
\bibitem [{\citenamefont {L\"utf\"uo\u{g}lu}(2025{\natexlab{a}})}]{Lutfuoglu:2025ljm}%
  \BibitemOpen
  \bibfield  {author} {\bibinfo {author} {\bibfnamefont {B.~C.}\ \bibnamefont {L\"utf\"uo\u{g}lu}},\ }\href {\doibase 10.1140/epjc/s10052-025-14380-x} {\bibfield  {journal} {\bibinfo  {journal} {Eur. Phys. J. C}\ }\textbf {\bibinfo {volume} {85}},\ \bibinfo {pages} {630} (\bibinfo {year} {2025}{\natexlab{a}})},\ \Eprint {http://arxiv.org/abs/2504.18482} {arXiv:2504.18482 [gr-qc]} \BibitemShut {NoStop}%
\bibitem [{\citenamefont {L\"utf\"uo\u{g}lu}(2025{\natexlab{b}})}]{Lutfuoglu:2025hjy}%
  \BibitemOpen
  \bibfield  {author} {\bibinfo {author} {\bibfnamefont {B.~C.}\ \bibnamefont {L\"utf\"uo\u{g}lu}},\ }\href {\doibase 10.1140/epjc/s10052-025-14210-0} {\bibfield  {journal} {\bibinfo  {journal} {Eur. Phys. J. C}\ }\textbf {\bibinfo {volume} {85}},\ \bibinfo {pages} {486} (\bibinfo {year} {2025}{\natexlab{b}})},\ \Eprint {http://arxiv.org/abs/2503.16087} {arXiv:2503.16087 [gr-qc]} \BibitemShut {NoStop}%
\bibitem [{\citenamefont {Hamil}\ and\ \citenamefont {L\"utf\"uo\u{g}lu}(2025)}]{Hamil:2024njs}%
  \BibitemOpen
  \bibfield  {author} {\bibinfo {author} {\bibfnamefont {B.}~\bibnamefont {Hamil}}\ and\ \bibinfo {author} {\bibfnamefont {B.~C.}\ \bibnamefont {L\"utf\"uo\u{g}lu}},\ }\href {\doibase 10.1002/prop.202400105} {\bibfield  {journal} {\bibinfo  {journal} {Fortsch. Phys.}\ }\textbf {\bibinfo {volume} {73}},\ \bibinfo {pages} {2400105} (\bibinfo {year} {2025})},\ \Eprint {http://arxiv.org/abs/2406.02109} {arXiv:2406.02109 [gr-qc]} \BibitemShut {NoStop}%
\bibitem [{\citenamefont {Rubin}\ and\ \citenamefont {Ford}(1970)}]{Rubin:1970rf}%
  \BibitemOpen
  \bibfield  {author} {\bibinfo {author} {\bibfnamefont {V.~C.}\ \bibnamefont {Rubin}}\ and\ \bibinfo {author} {\bibfnamefont {W.~K.}\ \bibnamefont {Ford}},\ }\href@noop {} {\bibfield  {journal} {\bibinfo  {journal} {Astro. Jour.}\ }\textbf {\bibinfo {volume} {159}},\ \bibinfo {pages} {379} (\bibinfo {year} {1970})}\BibitemShut {NoStop}%
\bibitem [{\citenamefont {Akiyama}\ and\ \citenamefont {et~al. (Event Horizon Telescope~Collaboration)}(2019)}]{akiyama:2019eh}%
  \BibitemOpen
  \bibfield  {author} {\bibinfo {author} {\bibfnamefont {K.}~\bibnamefont {Akiyama}}\ and\ \bibinfo {author} {\bibnamefont {et~al. (Event Horizon Telescope~Collaboration)}},\ }\href@noop {} {\bibfield  {journal} {\bibinfo  {journal} {Astro. Jour.}\ }\textbf {\bibinfo {volume} {L1}},\ \bibinfo {pages} {875} (\bibinfo {year} {2019})},\ \Eprint {http://arxiv.org/abs/1906.11238} {arXiv:1906.11238} \BibitemShut {NoStop}%
\bibitem [{\citenamefont {Mannheim}\ and\ \citenamefont {Kazanas}(1989)}]{Mannheim:1989mk}%
  \BibitemOpen
  \bibfield  {author} {\bibinfo {author} {\bibfnamefont {P.}~\bibnamefont {Mannheim}}\ and\ \bibinfo {author} {\bibfnamefont {D.}~\bibnamefont {Kazanas}},\ }\href@noop {} {\bibfield  {journal} {\bibinfo  {journal} {Astro. Jour.}\ }\textbf {\bibinfo {volume} {342}},\ \bibinfo {pages} {635} (\bibinfo {year} {1989})}\BibitemShut {NoStop}%
\bibitem [{\citenamefont {Konoplya}\ \emph {et~al.}(2025)\citenamefont {Konoplya}, \citenamefont {Khrabustovskyi}, \citenamefont {Kříž},\ and\ \citenamefont {Zhidenko}}]{Konoplya:2025kk}%
  \BibitemOpen
  \bibfield  {author} {\bibinfo {author} {\bibfnamefont {R.~A.}\ \bibnamefont {Konoplya}}, \bibinfo {author} {\bibfnamefont {A.}~\bibnamefont {Khrabustovskyi}}, \bibinfo {author} {\bibfnamefont {J.}~\bibnamefont {Kříž}}, \ and\ \bibinfo {author} {\bibfnamefont {A.}~\bibnamefont {Zhidenko}},\ }\href@noop {} {\  (\bibinfo {year} {2025})},\ \Eprint {http://arxiv.org/abs/2501.16134} {arXiv:2501.16134} \BibitemShut {NoStop}%
\bibitem [{\citenamefont {Al-Badawi}\ and\ \citenamefont {Shaymator}(2025)}]{Al-Badawi:2024bs}%
  \BibitemOpen
  \bibfield  {author} {\bibinfo {author} {\bibfnamefont {A.}~\bibnamefont {Al-Badawi}}\ and\ \bibinfo {author} {\bibfnamefont {S.}~\bibnamefont {Shaymator}},\ }\href@noop {} {\bibfield  {journal} {\bibinfo  {journal} {Comm.Theor.Phys.}\ }\textbf {\bibinfo {volume} {77}},\ \bibinfo {pages} {035402} (\bibinfo {year} {2025})},\ \Eprint {http://arxiv.org/abs/2412.20037} {arXiv:2412.20037} \BibitemShut {NoStop}%
\bibitem [{\citenamefont {Myung}()}]{Myung:2025m}%
  \BibitemOpen
  \bibfield  {author} {\bibinfo {author} {\bibfnamefont {Y.}~\bibnamefont {Myung}},\ }\href@noop {} {\ }\Eprint {http://arxiv.org/abs/2502.13397} {arXiv:2502.13397 [gr-qc]} \BibitemShut {NoStop}%
\bibitem [{\citenamefont {Mollicone1}\ and\ \citenamefont {Destounis1}(2025)}]{Mollicone:2025md}%
  \BibitemOpen
  \bibfield  {author} {\bibinfo {author} {\bibfnamefont {A.}~\bibnamefont {Mollicone1}}\ and\ \bibinfo {author} {\bibfnamefont {K.}~\bibnamefont {Destounis1}},\ }\href@noop {} {\bibfield  {journal} {\bibinfo  {journal} {Phys. Rev. D.}\ }\textbf {\bibinfo {volume} {111}},\ \bibinfo {pages} {024017} (\bibinfo {year} {2025})},\ \Eprint {http://arxiv.org/abs/2410.11952} {arXiv:2410.11952 [gr-qc]} \BibitemShut {NoStop}%
\bibitem [{\citenamefont {Liu}\ \emph {et~al.}(2025)\citenamefont {Liu}, \citenamefont {Mu}, \citenamefont {Tao},\ and\ \citenamefont {Weng}}]{Liu:2025lmtw}%
  \BibitemOpen
  \bibfield  {author} {\bibinfo {author} {\bibfnamefont {Y.}~\bibnamefont {Liu}}, \bibinfo {author} {\bibfnamefont {B.}~\bibnamefont {Mu}}, \bibinfo {author} {\bibfnamefont {J.}~\bibnamefont {Tao}}, \ and\ \bibinfo {author} {\bibfnamefont {Y.}~\bibnamefont {Weng}},\ }\href@noop {} {\bibfield  {journal} {\bibinfo  {journal} {Nucl. Phys. B}\ }\textbf {\bibinfo {volume} {1010}},\ \bibinfo {pages} {116787} (\bibinfo {year} {2025})},\ \Eprint {http://arxiv.org/abs/2409.20333} {arXiv:2409.20333 [gr-qc]} \BibitemShut {NoStop}%
\bibitem [{\citenamefont {Hawking}\ and\ \citenamefont {Ellis}(2023)}]{Hawking:1973uf}%
  \BibitemOpen
  \bibfield  {author} {\bibinfo {author} {\bibfnamefont {S.~W.}\ \bibnamefont {Hawking}}\ and\ \bibinfo {author} {\bibfnamefont {G.~F.~R.}\ \bibnamefont {Ellis}},\ }\href {\doibase 10.1017/9781009253161} {\emph {\bibinfo {title} {{The Large Scale Structure of Space-Time}}}},\ Cambridge Monographs on Mathematical Physics\ (\bibinfo  {publisher} {Cambridge University Press},\ \bibinfo {year} {2023})\BibitemShut {NoStop}%
\bibitem [{\citenamefont {Penrose}(1965)}]{Penrose:1964wq}%
  \BibitemOpen
  \bibfield  {author} {\bibinfo {author} {\bibfnamefont {R.}~\bibnamefont {Penrose}},\ }\href {\doibase 10.1103/PhysRevLett.14.57} {\bibfield  {journal} {\bibinfo  {journal} {Phys. Rev. Lett.}\ }\textbf {\bibinfo {volume} {14}},\ \bibinfo {pages} {57} (\bibinfo {year} {1965})}\BibitemShut {NoStop}%
\bibitem [{\citenamefont {Hawking}(1965)}]{Hawking:1965mf}%
  \BibitemOpen
  \bibfield  {author} {\bibinfo {author} {\bibfnamefont {S.}~\bibnamefont {Hawking}},\ }\href {\doibase 10.1103/PhysRevLett.15.689} {\bibfield  {journal} {\bibinfo  {journal} {Phys. Rev. Lett.}\ }\textbf {\bibinfo {volume} {15}},\ \bibinfo {pages} {689} (\bibinfo {year} {1965})}\BibitemShut {NoStop}%
\bibitem [{\citenamefont {Hawking}(1966{\natexlab{a}})}]{Hawking:1966sx}%
  \BibitemOpen
  \bibfield  {author} {\bibinfo {author} {\bibfnamefont {S.}~\bibnamefont {Hawking}},\ }\href {\doibase 10.1098/rspa.1966.0221} {\bibfield  {journal} {\bibinfo  {journal} {Proc. Roy. Soc. Lond. A}\ }\textbf {\bibinfo {volume} {294}},\ \bibinfo {pages} {511} (\bibinfo {year} {1966}{\natexlab{a}})}\BibitemShut {NoStop}%
\bibitem [{\citenamefont {Hawking}(1966{\natexlab{b}})}]{Hawking:1966jv}%
  \BibitemOpen
  \bibfield  {author} {\bibinfo {author} {\bibfnamefont {S.}~\bibnamefont {Hawking}},\ }\href {\doibase 10.1098/rspa.1966.0255} {\bibfield  {journal} {\bibinfo  {journal} {Proc. Roy. Soc. Lond. A}\ }\textbf {\bibinfo {volume} {295}},\ \bibinfo {pages} {490} (\bibinfo {year} {1966}{\natexlab{b}})}\BibitemShut {NoStop}%
\bibitem [{\citenamefont {Hawking}(1967)}]{Hawking:1967ju}%
  \BibitemOpen
  \bibfield  {author} {\bibinfo {author} {\bibfnamefont {S.}~\bibnamefont {Hawking}},\ }\href {\doibase 10.1098/rspa.1967.0164} {\bibfield  {journal} {\bibinfo  {journal} {Proc. Roy. Soc. Lond. A}\ }\textbf {\bibinfo {volume} {300}},\ \bibinfo {pages} {187} (\bibinfo {year} {1967})}\BibitemShut {NoStop}%
\bibitem [{\citenamefont {Israel}(1968)}]{Israel:1967za}%
  \BibitemOpen
  \bibfield  {author} {\bibinfo {author} {\bibfnamefont {W.}~\bibnamefont {Israel}},\ }\href {\doibase 10.1007/BF01645859} {\bibfield  {journal} {\bibinfo  {journal} {Commun. Math. Phys.}\ }\textbf {\bibinfo {volume} {8}},\ \bibinfo {pages} {245} (\bibinfo {year} {1968})}\BibitemShut {NoStop}%
\bibitem [{\citenamefont {Lan}\ \emph {et~al.}(2023)\citenamefont {Lan}, \citenamefont {Yang}, \citenamefont {Guo},\ and\ \citenamefont {Miao}}]{Lan:2023cvz}%
  \BibitemOpen
  \bibfield  {author} {\bibinfo {author} {\bibfnamefont {C.}~\bibnamefont {Lan}}, \bibinfo {author} {\bibfnamefont {H.}~\bibnamefont {Yang}}, \bibinfo {author} {\bibfnamefont {Y.}~\bibnamefont {Guo}}, \ and\ \bibinfo {author} {\bibfnamefont {Y.-G.}\ \bibnamefont {Miao}},\ }\href {\doibase 10.1007/s10773-023-05454-1} {\bibfield  {journal} {\bibinfo  {journal} {Int. J. Theor. Phys.}\ }\textbf {\bibinfo {volume} {62}},\ \bibinfo {pages} {202} (\bibinfo {year} {2023})},\ \Eprint {http://arxiv.org/abs/2303.11696} {arXiv:2303.11696 [gr-qc]} \BibitemShut {NoStop}%
\bibitem [{\citenamefont {Kiselev}(2003)}]{Kiselev:2002dx}%
  \BibitemOpen
  \bibfield  {author} {\bibinfo {author} {\bibfnamefont {V.~V.}\ \bibnamefont {Kiselev}},\ }\href {\doibase 10.1088/0264-9381/20/6/310} {\bibfield  {journal} {\bibinfo  {journal} {Class. Quant. Grav.}\ }\textbf {\bibinfo {volume} {20}},\ \bibinfo {pages} {1187} (\bibinfo {year} {2003})},\ \Eprint {http://arxiv.org/abs/gr-qc/0210040} {arXiv:gr-qc/0210040} \BibitemShut {NoStop}%
\bibitem [{\citenamefont {Zhang}\ \emph {et~al.}(2021)\citenamefont {Zhang}, \citenamefont {Chen}, \citenamefont {Ma}, \citenamefont {He},\ and\ \citenamefont {Deng}}]{Zhang:2020mxi}%
  \BibitemOpen
  \bibfield  {author} {\bibinfo {author} {\bibfnamefont {H.-X.}\ \bibnamefont {Zhang}}, \bibinfo {author} {\bibfnamefont {Y.}~\bibnamefont {Chen}}, \bibinfo {author} {\bibfnamefont {T.-C.}\ \bibnamefont {Ma}}, \bibinfo {author} {\bibfnamefont {P.-Z.}\ \bibnamefont {He}}, \ and\ \bibinfo {author} {\bibfnamefont {J.-B.}\ \bibnamefont {Deng}},\ }\href {\doibase 10.1088/1674-1137/abe84c} {\bibfield  {journal} {\bibinfo  {journal} {Chin. Phys. C}\ }\textbf {\bibinfo {volume} {45}},\ \bibinfo {pages} {055103} (\bibinfo {year} {2021})},\ \Eprint {http://arxiv.org/abs/2007.09408} {arXiv:2007.09408 [gr-qc]} \BibitemShut {NoStop}%
\bibitem [{\citenamefont {Rayimbaev}\ \emph {et~al.}(2021)\citenamefont {Rayimbaev}, \citenamefont {Shaymatov},\ and\ \citenamefont {Jamil}}]{Rayimbaev:2021rs}%
  \BibitemOpen
  \bibfield  {author} {\bibinfo {author} {\bibfnamefont {J.}~\bibnamefont {Rayimbaev}}, \bibinfo {author} {\bibfnamefont {S.}~\bibnamefont {Shaymatov}}, \ and\ \bibinfo {author} {\bibfnamefont {M.}~\bibnamefont {Jamil}},\ }\href@noop {} {\bibfield  {journal} {\bibinfo  {journal} {Eur.Phys.Jour. C}\ }\textbf {\bibinfo {volume} {81}},\ \bibinfo {pages} {699} (\bibinfo {year} {2021})},\ \Eprint {http://arxiv.org/abs/2107.13436} {arXiv:2107.13436 [gr-qc]} \BibitemShut {NoStop}%
\bibitem [{\citenamefont {Hendi}\ \emph {et~al.}(2020)\citenamefont {Hendi}, \citenamefont {Nemati}, \citenamefont {Lin}, ,\ and\ \citenamefont {Jamil}}]{Hendi:2020hnlj}%
  \BibitemOpen
  \bibfield  {author} {\bibinfo {author} {\bibfnamefont {S.}~\bibnamefont {Hendi}}, \bibinfo {author} {\bibfnamefont {A.}~\bibnamefont {Nemati}}, \bibinfo {author} {\bibfnamefont {K.}~\bibnamefont {Lin}}, , \ and\ \bibinfo {author} {\bibfnamefont {M.}~\bibnamefont {Jamil}},\ }\href@noop {} {\bibfield  {journal} {\bibinfo  {journal} {Eur.Phys.Jour. C}\ }\textbf {\bibinfo {volume} {80}},\ \bibinfo {pages} {296} (\bibinfo {year} {2020})},\ \Eprint {http://arxiv.org/abs/1810.04103} {arXiv:1810.04103 [gr-qc]} \BibitemShut {NoStop}%
\bibitem [{\citenamefont {Haroon}\ \emph {et~al.}(2019)\citenamefont {Haroon}, \citenamefont {Jamil}, \citenamefont {Jusufi}, \citenamefont {Lin},\ and\ \citenamefont {Mann}}]{Haroon:2019hjjlm}%
  \BibitemOpen
  \bibfield  {author} {\bibinfo {author} {\bibfnamefont {S.}~\bibnamefont {Haroon}}, \bibinfo {author} {\bibfnamefont {M.}~\bibnamefont {Jamil}}, \bibinfo {author} {\bibfnamefont {K.}~\bibnamefont {Jusufi}}, \bibinfo {author} {\bibfnamefont {K.}~\bibnamefont {Lin}}, \ and\ \bibinfo {author} {\bibfnamefont {R.~B.}\ \bibnamefont {Mann}},\ }\href@noop {} {\bibfield  {journal} {\bibinfo  {journal} {Phys. Rev. D.}\ }\textbf {\bibinfo {volume} {99}},\ \bibinfo {pages} {044015} (\bibinfo {year} {2019})},\ \Eprint {http://arxiv.org/abs/1810.04103} {arXiv:1810.04103 [gr-qc]} \BibitemShut {NoStop}%
\bibitem [{\citenamefont {Rizwan}\ \emph {et~al.}(2019)\citenamefont {Rizwan}, \citenamefont {Jamil}, ,\ and\ \citenamefont {Jusuf}}]{Rizwan:2019rjj}%
  \BibitemOpen
  \bibfield  {author} {\bibinfo {author} {\bibfnamefont {M.}~\bibnamefont {Rizwan}}, \bibinfo {author} {\bibfnamefont {M.}~\bibnamefont {Jamil}}, , \ and\ \bibinfo {author} {\bibfnamefont {K.}~\bibnamefont {Jusuf}},\ }\href@noop {} {\bibfield  {journal} {\bibinfo  {journal} {Phys. Rev. D.}\ }\textbf {\bibinfo {volume} {99}},\ \bibinfo {pages} {024050} (\bibinfo {year} {2019})},\ \Eprint {http://arxiv.org/abs/1812.01331} {arXiv:1812.01331 [gr-qc]} \BibitemShut {NoStop}%
\bibitem [{\citenamefont {Narzilloev}\ \emph {et~al.}(2020)\citenamefont {Narzilloev}, \citenamefont {Rayimbaev}, \citenamefont {Shaymatov}, \citenamefont {Abdujabbarov}, \citenamefont {Ahmedov}, ,\ and\ \citenamefont {Bambi}}]{Narzilloev:2020nrsaab}%
  \BibitemOpen
  \bibfield  {author} {\bibinfo {author} {\bibfnamefont {B.}~\bibnamefont {Narzilloev}}, \bibinfo {author} {\bibfnamefont {J.}~\bibnamefont {Rayimbaev}}, \bibinfo {author} {\bibfnamefont {S.}~\bibnamefont {Shaymatov}}, \bibinfo {author} {\bibfnamefont {A.}~\bibnamefont {Abdujabbarov}}, \bibinfo {author} {\bibfnamefont {B.}~\bibnamefont {Ahmedov}}, , \ and\ \bibinfo {author} {\bibfnamefont {C.}~\bibnamefont {Bambi}},\ }\href@noop {} {\bibfield  {journal} {\bibinfo  {journal} {Phys. Rev. D.}\ }\textbf {\bibinfo {volume} {102}},\ \bibinfo {pages} {104062} (\bibinfo {year} {2020})},\ \Eprint {http://arxiv.org/abs/2011.06148} {arXiv:2011.06148 [gr-qc]} \BibitemShut {NoStop}%
\bibitem [{\citenamefont {Shaymatov}\ \emph {et~al.}(2021{\natexlab{a}})\citenamefont {Shaymatov}, \citenamefont {Ahmedov}, ,\ and\ \citenamefont {Jamil}}]{Shaymatov:2021saj}%
  \BibitemOpen
  \bibfield  {author} {\bibinfo {author} {\bibfnamefont {S.}~\bibnamefont {Shaymatov}}, \bibinfo {author} {\bibfnamefont {B.}~\bibnamefont {Ahmedov}}, , \ and\ \bibinfo {author} {\bibfnamefont {M.}~\bibnamefont {Jamil}},\ }\href@noop {} {\bibfield  {journal} {\bibinfo  {journal} {Eur. Phys. Jour. C.}\ }\textbf {\bibinfo {volume} {81}},\ \bibinfo {pages} {588} (\bibinfo {year} {2021}{\natexlab{a}})}\BibitemShut {NoStop}%
\bibitem [{\citenamefont {Shaymatov}\ \emph {et~al.}(2021{\natexlab{b}})\citenamefont {Shaymatov}, \citenamefont {Malafarina}, ,\ and\ \citenamefont {Ahmedov}}]{Shaymatov:2021sma}%
  \BibitemOpen
  \bibfield  {author} {\bibinfo {author} {\bibfnamefont {S.}~\bibnamefont {Shaymatov}}, \bibinfo {author} {\bibfnamefont {D.}~\bibnamefont {Malafarina}}, , \ and\ \bibinfo {author} {\bibfnamefont {B.}~\bibnamefont {Ahmedov}},\ }\href@noop {} {\bibfield  {journal} {\bibinfo  {journal} {Phys. Dark Universe}\ }\textbf {\bibinfo {volume} {34}},\ \bibinfo {pages} {100891} (\bibinfo {year} {2021}{\natexlab{b}})},\ \Eprint {http://arxiv.org/abs/2004.06811} {arXiv:2004.06811 [gr-qc]} \BibitemShut {NoStop}%
\bibitem [{\citenamefont {Ayon-Beato}\ and\ \citenamefont {Garcia}(2000)}]{Ayon-Beato:2000mjt}%
  \BibitemOpen
  \bibfield  {author} {\bibinfo {author} {\bibfnamefont {E.}~\bibnamefont {Ayon-Beato}}\ and\ \bibinfo {author} {\bibfnamefont {A.}~\bibnamefont {Garcia}},\ }\href {\doibase 10.1016/S0370-2693(00)01125-4} {\bibfield  {journal} {\bibinfo  {journal} {Phys. Lett. B}\ }\textbf {\bibinfo {volume} {493}},\ \bibinfo {pages} {149} (\bibinfo {year} {2000})},\ \Eprint {http://arxiv.org/abs/gr-qc/0009077} {arXiv:gr-qc/0009077} \BibitemShut {NoStop}%
\bibitem [{\citenamefont {Li}\ and\ \citenamefont {Yang}(2012)}]{Li:2012zx}%
  \BibitemOpen
  \bibfield  {author} {\bibinfo {author} {\bibfnamefont {M.-H.}\ \bibnamefont {Li}}\ and\ \bibinfo {author} {\bibfnamefont {K.-C.}\ \bibnamefont {Yang}},\ }\href {\doibase 10.1103/PhysRevD.86.123015} {\bibfield  {journal} {\bibinfo  {journal} {Phys. Rev. D}\ }\textbf {\bibinfo {volume} {86}},\ \bibinfo {pages} {123015} (\bibinfo {year} {2012})},\ \Eprint {http://arxiv.org/abs/1204.3178} {arXiv:1204.3178 [astro-ph.CO]} \BibitemShut {NoStop}%
\bibitem [{\citenamefont {Poschl}\ and\ \citenamefont {Teller}(1933)}]{Poschl:1933zz}%
  \BibitemOpen
  \bibfield  {author} {\bibinfo {author} {\bibfnamefont {G.}~\bibnamefont {Poschl}}\ and\ \bibinfo {author} {\bibfnamefont {E.}~\bibnamefont {Teller}},\ }\href {\doibase 10.1007/BF01331132} {\bibfield  {journal} {\bibinfo  {journal} {Z. Phys.}\ }\textbf {\bibinfo {volume} {83}},\ \bibinfo {pages} {143} (\bibinfo {year} {1933})}\BibitemShut {NoStop}%
\bibitem [{\citenamefont {Ferrari}\ and\ \citenamefont {Mashhoon}(1984)}]{Ferrari:1984zz}%
  \BibitemOpen
  \bibfield  {author} {\bibinfo {author} {\bibfnamefont {V.}~\bibnamefont {Ferrari}}\ and\ \bibinfo {author} {\bibfnamefont {B.}~\bibnamefont {Mashhoon}},\ }\href {\doibase 10.1103/PhysRevD.30.295} {\bibfield  {journal} {\bibinfo  {journal} {Phys. Rev. D}\ }\textbf {\bibinfo {volume} {30}},\ \bibinfo {pages} {295} (\bibinfo {year} {1984})}\BibitemShut {NoStop}%
\bibitem [{\citenamefont {Cardoso}\ and\ \citenamefont {Lemos}(2001)}]{Cardoso:2001hn}%
  \BibitemOpen
  \bibfield  {author} {\bibinfo {author} {\bibfnamefont {V.}~\bibnamefont {Cardoso}}\ and\ \bibinfo {author} {\bibfnamefont {J.~P.~S.}\ \bibnamefont {Lemos}},\ }\href {\doibase 10.1103/PhysRevD.63.124015} {\bibfield  {journal} {\bibinfo  {journal} {Phys. Rev. D}\ }\textbf {\bibinfo {volume} {63}},\ \bibinfo {pages} {124015} (\bibinfo {year} {2001})},\ \Eprint {http://arxiv.org/abs/gr-qc/0101052} {arXiv:gr-qc/0101052} \BibitemShut {NoStop}%
\bibitem [{\citenamefont {Cardoso}\ and\ \citenamefont {Lemos}(2003)}]{Cardoso:2003sw}%
  \BibitemOpen
  \bibfield  {author} {\bibinfo {author} {\bibfnamefont {V.}~\bibnamefont {Cardoso}}\ and\ \bibinfo {author} {\bibfnamefont {J.~P.~S.}\ \bibnamefont {Lemos}},\ }\href {\doibase 10.1103/PhysRevD.67.084020} {\bibfield  {journal} {\bibinfo  {journal} {Phys. Rev. D}\ }\textbf {\bibinfo {volume} {67}},\ \bibinfo {pages} {084020} (\bibinfo {year} {2003})},\ \Eprint {http://arxiv.org/abs/gr-qc/0301078} {arXiv:gr-qc/0301078} \BibitemShut {NoStop}%
\bibitem [{\citenamefont {Molina}(2003)}]{Molina:2003ff}%
  \BibitemOpen
  \bibfield  {author} {\bibinfo {author} {\bibfnamefont {C.}~\bibnamefont {Molina}},\ }\href {\doibase 10.1103/PhysRevD.68.064007} {\bibfield  {journal} {\bibinfo  {journal} {Phys. Rev. D}\ }\textbf {\bibinfo {volume} {68}},\ \bibinfo {pages} {064007} (\bibinfo {year} {2003})},\ \Eprint {http://arxiv.org/abs/gr-qc/0304053} {arXiv:gr-qc/0304053} \BibitemShut {NoStop}%
\bibitem [{\citenamefont {Panotopoulos}(2018)}]{Panotopoulos:2018hua}%
  \BibitemOpen
  \bibfield  {author} {\bibinfo {author} {\bibfnamefont {G.}~\bibnamefont {Panotopoulos}},\ }\href {\doibase 10.1142/S0217732318501304} {\bibfield  {journal} {\bibinfo  {journal} {Mod. Phys. Lett. A}\ }\textbf {\bibinfo {volume} {33}},\ \bibinfo {pages} {1850130} (\bibinfo {year} {2018})},\ \Eprint {http://arxiv.org/abs/1807.03278} {arXiv:1807.03278 [gr-qc]} \BibitemShut {NoStop}%
\bibitem [{\citenamefont {Hamil}\ and\ \citenamefont {L\"utf\"uo\u{g}lu}(2024)}]{Hamil:2024rsg}%
  \BibitemOpen
  \bibfield  {author} {\bibinfo {author} {\bibfnamefont {B.}~\bibnamefont {Hamil}}\ and\ \bibinfo {author} {\bibfnamefont {B.~C.}\ \bibnamefont {L\"utf\"uo\u{g}lu}},\ }\href@noop {} {\  (\bibinfo {year} {2024})},\ \Eprint {http://arxiv.org/abs/2404.06575} {arXiv:2404.06575 [gr-qc]} \BibitemShut {NoStop}%
\bibitem [{\citenamefont {Birmingham}(2001)}]{Birmingham:2001hc}%
  \BibitemOpen
  \bibfield  {author} {\bibinfo {author} {\bibfnamefont {D.}~\bibnamefont {Birmingham}},\ }\href {\doibase 10.1103/PhysRevD.64.064024} {\bibfield  {journal} {\bibinfo  {journal} {Phys. Rev. D}\ }\textbf {\bibinfo {volume} {64}},\ \bibinfo {pages} {064024} (\bibinfo {year} {2001})},\ \Eprint {http://arxiv.org/abs/hep-th/0101194} {arXiv:hep-th/0101194} \BibitemShut {NoStop}%
\bibitem [{\citenamefont {Fernando}(2004)}]{Fernando:2003ai}%
  \BibitemOpen
  \bibfield  {author} {\bibinfo {author} {\bibfnamefont {S.}~\bibnamefont {Fernando}},\ }\href {\doibase 10.1023/B:GERG.0000006694.68399.c9} {\bibfield  {journal} {\bibinfo  {journal} {Gen. Rel. Grav.}\ }\textbf {\bibinfo {volume} {36}},\ \bibinfo {pages} {71} (\bibinfo {year} {2004})},\ \Eprint {http://arxiv.org/abs/hep-th/0306214} {arXiv:hep-th/0306214} \BibitemShut {NoStop}%
\bibitem [{\citenamefont {Fernando}(2008)}]{Fernando:2008hb}%
  \BibitemOpen
  \bibfield  {author} {\bibinfo {author} {\bibfnamefont {S.}~\bibnamefont {Fernando}},\ }\href {\doibase 10.1103/PhysRevD.77.124005} {\bibfield  {journal} {\bibinfo  {journal} {Phys. Rev. D}\ }\textbf {\bibinfo {volume} {77}},\ \bibinfo {pages} {124005} (\bibinfo {year} {2008})},\ \Eprint {http://arxiv.org/abs/0802.3321} {arXiv:0802.3321 [hep-th]} \BibitemShut {NoStop}%
\bibitem [{\citenamefont {Gonzalez}\ \emph {et~al.}(2010)\citenamefont {Gonzalez}, \citenamefont {Papantonopoulos},\ and\ \citenamefont {Saavedra}}]{Gonzalez:2010vv}%
  \BibitemOpen
  \bibfield  {author} {\bibinfo {author} {\bibfnamefont {P.}~\bibnamefont {Gonzalez}}, \bibinfo {author} {\bibfnamefont {E.}~\bibnamefont {Papantonopoulos}}, \ and\ \bibinfo {author} {\bibfnamefont {J.}~\bibnamefont {Saavedra}},\ }\href {\doibase 10.1007/JHEP08(2010)050} {\bibfield  {journal} {\bibinfo  {journal} {JHEP}\ }\textbf {\bibinfo {volume} {08}},\ \bibinfo {pages} {050} (\bibinfo {year} {2010})},\ \Eprint {http://arxiv.org/abs/1003.1381} {arXiv:1003.1381 [hep-th]} \BibitemShut {NoStop}%
\bibitem [{\citenamefont {Destounis}\ \emph {et~al.}(2018)\citenamefont {Destounis}, \citenamefont {Panotopoulos},\ and\ \citenamefont {Rinc\'on}}]{Destounis:2018utr}%
  \BibitemOpen
  \bibfield  {author} {\bibinfo {author} {\bibfnamefont {K.}~\bibnamefont {Destounis}}, \bibinfo {author} {\bibfnamefont {G.}~\bibnamefont {Panotopoulos}}, \ and\ \bibinfo {author} {\bibfnamefont {A.}~\bibnamefont {Rinc\'on}},\ }\href {\doibase 10.1140/epjc/s10052-018-5576-8} {\bibfield  {journal} {\bibinfo  {journal} {Eur. Phys. J. C}\ }\textbf {\bibinfo {volume} {78}},\ \bibinfo {pages} {139} (\bibinfo {year} {2018})},\ \Eprint {http://arxiv.org/abs/1801.08955} {arXiv:1801.08955 [gr-qc]} \BibitemShut {NoStop}%
\bibitem [{\citenamefont {Ovg\"un}\ and\ \citenamefont {Jusufi}(2018)}]{Ovgun:2018gwt}%
  \BibitemOpen
  \bibfield  {author} {\bibinfo {author} {\bibfnamefont {A.}~\bibnamefont {Ovg\"un}}\ and\ \bibinfo {author} {\bibfnamefont {K.}~\bibnamefont {Jusufi}},\ }\href {\doibase 10.1016/j.aop.2018.05.013} {\bibfield  {journal} {\bibinfo  {journal} {Annals Phys.}\ }\textbf {\bibinfo {volume} {395}},\ \bibinfo {pages} {138} (\bibinfo {year} {2018})},\ \Eprint {http://arxiv.org/abs/1801.02555} {arXiv:1801.02555 [gr-qc]} \BibitemShut {NoStop}%
\bibitem [{\citenamefont {Rinc\'on}\ and\ \citenamefont {Panotopoulos}(2018{\natexlab{b}})}]{Rincon:2018ktz}%
  \BibitemOpen
  \bibfield  {author} {\bibinfo {author} {\bibfnamefont {A.}~\bibnamefont {Rinc\'on}}\ and\ \bibinfo {author} {\bibfnamefont {G.}~\bibnamefont {Panotopoulos}},\ }\href {\doibase 10.1140/epjc/s10052-018-6352-5} {\bibfield  {journal} {\bibinfo  {journal} {Eur. Phys. J. C}\ }\textbf {\bibinfo {volume} {78}},\ \bibinfo {pages} {858} (\bibinfo {year} {2018}{\natexlab{b}})},\ \Eprint {http://arxiv.org/abs/1810.08822} {arXiv:1810.08822 [gr-qc]} \BibitemShut {NoStop}%
\bibitem [{\citenamefont {Hatsuda}(2020{\natexlab{a}})}]{Hatsuda:2020sbn}%
  \BibitemOpen
  \bibfield  {author} {\bibinfo {author} {\bibfnamefont {Y.}~\bibnamefont {Hatsuda}},\ }\href {\doibase 10.1088/1361-6382/abc82e} {\bibfield  {journal} {\bibinfo  {journal} {Class. Quant. Grav.}\ }\textbf {\bibinfo {volume} {38}},\ \bibinfo {pages} {025015} (\bibinfo {year} {2020}{\natexlab{a}})},\ \Eprint {http://arxiv.org/abs/2006.08957} {arXiv:2006.08957 [gr-qc]} \BibitemShut {NoStop}%
\bibitem [{\citenamefont {Fiziev}\ and\ \citenamefont {Staicova}(2011)}]{Fiziev:2011mm}%
  \BibitemOpen
  \bibfield  {author} {\bibinfo {author} {\bibfnamefont {P.}~\bibnamefont {Fiziev}}\ and\ \bibinfo {author} {\bibfnamefont {D.}~\bibnamefont {Staicova}},\ }\href {\doibase 10.1103/PhysRevD.84.127502} {\bibfield  {journal} {\bibinfo  {journal} {Phys. Rev. D}\ }\textbf {\bibinfo {volume} {84}},\ \bibinfo {pages} {127502} (\bibinfo {year} {2011})},\ \Eprint {http://arxiv.org/abs/1109.1532} {arXiv:1109.1532 [gr-qc]} \BibitemShut {NoStop}%
\bibitem [{\citenamefont {Naderi}\ and\ \citenamefont {Rezaei-Aghdam}(2024)}]{Naderi:2024dhh}%
  \BibitemOpen
  \bibfield  {author} {\bibinfo {author} {\bibfnamefont {F.}~\bibnamefont {Naderi}}\ and\ \bibinfo {author} {\bibfnamefont {A.}~\bibnamefont {Rezaei-Aghdam}},\ }\href {\doibase 10.1140/epjc/s10052-024-13655-z} {\bibfield  {journal} {\bibinfo  {journal} {Eur. Phys. J. C}\ }\textbf {\bibinfo {volume} {84}},\ \bibinfo {pages} {1283} (\bibinfo {year} {2024})},\ \Eprint {http://arxiv.org/abs/2410.19658} {arXiv:2410.19658 [hep-th]} \BibitemShut {NoStop}%
\bibitem [{\citenamefont {Destounis}\ \emph {et~al.}(2020)\citenamefont {Destounis}, \citenamefont {Fontana},\ and\ \citenamefont {Mena}}]{Destounis:2020pjk}%
  \BibitemOpen
  \bibfield  {author} {\bibinfo {author} {\bibfnamefont {K.}~\bibnamefont {Destounis}}, \bibinfo {author} {\bibfnamefont {R.~D.~B.}\ \bibnamefont {Fontana}}, \ and\ \bibinfo {author} {\bibfnamefont {F.~C.}\ \bibnamefont {Mena}},\ }\href {\doibase 10.1103/PhysRevD.102.044005} {\bibfield  {journal} {\bibinfo  {journal} {Phys. Rev. D}\ }\textbf {\bibinfo {volume} {102}},\ \bibinfo {pages} {044005} (\bibinfo {year} {2020})},\ \Eprint {http://arxiv.org/abs/2005.03028} {arXiv:2005.03028 [gr-qc]} \BibitemShut {NoStop}%
\bibitem [{\citenamefont {Fontana}\ and\ \citenamefont {Mena}(2022)}]{Fontana:2022whx}%
  \BibitemOpen
  \bibfield  {author} {\bibinfo {author} {\bibfnamefont {R.~D.~B.}\ \bibnamefont {Fontana}}\ and\ \bibinfo {author} {\bibfnamefont {F.~C.}\ \bibnamefont {Mena}},\ }\href {\doibase 10.1007/JHEP10(2022)047} {\bibfield  {journal} {\bibinfo  {journal} {JHEP}\ }\textbf {\bibinfo {volume} {10}},\ \bibinfo {pages} {047} (\bibinfo {year} {2022})},\ \Eprint {http://arxiv.org/abs/2203.13933} {arXiv:2203.13933 [gr-qc]} \BibitemShut {NoStop}%
\bibitem [{\citenamefont {Hatsuda}\ and\ \citenamefont {Kimura}(2021)}]{Hatsuda:2021gtn}%
  \BibitemOpen
  \bibfield  {author} {\bibinfo {author} {\bibfnamefont {Y.}~\bibnamefont {Hatsuda}}\ and\ \bibinfo {author} {\bibfnamefont {M.}~\bibnamefont {Kimura}},\ }\href {\doibase 10.3390/universe7120476} {\bibfield  {journal} {\bibinfo  {journal} {Universe}\ }\textbf {\bibinfo {volume} {7}},\ \bibinfo {pages} {476} (\bibinfo {year} {2021})},\ \Eprint {http://arxiv.org/abs/2111.15197} {arXiv:2111.15197 [gr-qc]} \BibitemShut {NoStop}%
\bibitem [{\citenamefont {Leaver}(1985)}]{Leaver:1985ax}%
  \BibitemOpen
  \bibfield  {author} {\bibinfo {author} {\bibfnamefont {E.~W.}\ \bibnamefont {Leaver}},\ }\href {\doibase 10.1098/rspa.1985.0119} {\bibfield  {journal} {\bibinfo  {journal} {Proc. Roy. Soc. Lond. A}\ }\textbf {\bibinfo {volume} {402}},\ \bibinfo {pages} {285} (\bibinfo {year} {1985})}\BibitemShut {NoStop}%
\bibitem [{\citenamefont {Nollert}(1993)}]{Nollert:1993zz}%
  \BibitemOpen
  \bibfield  {author} {\bibinfo {author} {\bibfnamefont {H.-P.}\ \bibnamefont {Nollert}},\ }\href {\doibase 10.1103/PhysRevD.47.5253} {\bibfield  {journal} {\bibinfo  {journal} {Phys. Rev. D}\ }\textbf {\bibinfo {volume} {47}},\ \bibinfo {pages} {5253} (\bibinfo {year} {1993})}\BibitemShut {NoStop}%
\bibitem [{\citenamefont {Daghigh}\ \emph {et~al.}(2023)\citenamefont {Daghigh}, \citenamefont {Green},\ and\ \citenamefont {Morey}}]{Daghigh:2022uws}%
  \BibitemOpen
  \bibfield  {author} {\bibinfo {author} {\bibfnamefont {R.~G.}\ \bibnamefont {Daghigh}}, \bibinfo {author} {\bibfnamefont {M.~D.}\ \bibnamefont {Green}}, \ and\ \bibinfo {author} {\bibfnamefont {J.~C.}\ \bibnamefont {Morey}},\ }\href {\doibase 10.1103/PhysRevD.107.024023} {\bibfield  {journal} {\bibinfo  {journal} {Phys. Rev. D}\ }\textbf {\bibinfo {volume} {107}},\ \bibinfo {pages} {024023} (\bibinfo {year} {2023})},\ \Eprint {http://arxiv.org/abs/2209.09324} {arXiv:2209.09324 [gr-qc]} \BibitemShut {NoStop}%
\bibitem [{\citenamefont {Cho}\ \emph {et~al.}(2012)\citenamefont {Cho}, \citenamefont {Cornell}, \citenamefont {Doukas}, \citenamefont {Huang},\ and\ \citenamefont {Naylor}}]{Cho:2011sf}%
  \BibitemOpen
  \bibfield  {author} {\bibinfo {author} {\bibfnamefont {H.~T.}\ \bibnamefont {Cho}}, \bibinfo {author} {\bibfnamefont {A.~S.}\ \bibnamefont {Cornell}}, \bibinfo {author} {\bibfnamefont {J.}~\bibnamefont {Doukas}}, \bibinfo {author} {\bibfnamefont {T.~R.}\ \bibnamefont {Huang}}, \ and\ \bibinfo {author} {\bibfnamefont {W.}~\bibnamefont {Naylor}},\ }\href {\doibase 10.1155/2012/281705} {\bibfield  {journal} {\bibinfo  {journal} {Adv. Math. Phys.}\ }\textbf {\bibinfo {volume} {2012}},\ \bibinfo {pages} {281705} (\bibinfo {year} {2012})},\ \Eprint {http://arxiv.org/abs/1111.5024} {arXiv:1111.5024 [gr-qc]} \BibitemShut {NoStop}%
\bibitem [{\citenamefont {{Ciftci}}\ \emph {et~al.}(2003)\citenamefont {{Ciftci}}, \citenamefont {{Hall}},\ and\ \citenamefont {{Saad}}}]{2003JPhA...3611807C}%
  \BibitemOpen
  \bibfield  {author} {\bibinfo {author} {\bibfnamefont {H.}~\bibnamefont {{Ciftci}}}, \bibinfo {author} {\bibfnamefont {R.~L.}\ \bibnamefont {{Hall}}}, \ and\ \bibinfo {author} {\bibfnamefont {N.}~\bibnamefont {{Saad}}},\ }\href {\doibase 10.1088/0305-4470/36/47/008} {\bibfield  {journal} {\bibinfo  {journal} {Journal of Physics A Mathematical General}\ }\textbf {\bibinfo {volume} {36}},\ \bibinfo {pages} {11807} (\bibinfo {year} {2003})},\ \Eprint {http://arxiv.org/abs/math-ph/0309066} {arXiv:math-ph/0309066 [math-ph]} \BibitemShut {NoStop}%
\bibitem [{\citenamefont {Ciftci}\ \emph {et~al.}(2005)\citenamefont {Ciftci}, \citenamefont {Hall},\ and\ \citenamefont {Saad}}]{Ciftci:2005xn}%
  \BibitemOpen
  \bibfield  {author} {\bibinfo {author} {\bibfnamefont {H.}~\bibnamefont {Ciftci}}, \bibinfo {author} {\bibfnamefont {R.~L.}\ \bibnamefont {Hall}}, \ and\ \bibinfo {author} {\bibfnamefont {N.}~\bibnamefont {Saad}},\ }\href {\doibase 10.1016/j.physleta.2005.04.030} {\bibfield  {journal} {\bibinfo  {journal} {Phys. Lett. A}\ }\textbf {\bibinfo {volume} {340}},\ \bibinfo {pages} {388} (\bibinfo {year} {2005})},\ \Eprint {http://arxiv.org/abs/math-ph/0504056} {arXiv:math-ph/0504056} \BibitemShut {NoStop}%
\bibitem [{\citenamefont {Konoplya}\ and\ \citenamefont {Zhidenko}(2011)}]{Konoplya:2011qq}%
  \BibitemOpen
  \bibfield  {author} {\bibinfo {author} {\bibfnamefont {R.}~\bibnamefont {Konoplya}}\ and\ \bibinfo {author} {\bibfnamefont {A.}~\bibnamefont {Zhidenko}},\ }\href {\doibase 10.1103/RevModPhys.83.793} {\bibfield  {journal} {\bibinfo  {journal} {Rev. Mod. Phys.}\ }\textbf {\bibinfo {volume} {83}},\ \bibinfo {pages} {793} (\bibinfo {year} {2011})},\ \Eprint {http://arxiv.org/abs/1102.4014} {arXiv:1102.4014 [gr-qc]} \BibitemShut {NoStop}%
\bibitem [{\citenamefont {Schutz}\ and\ \citenamefont {Will}(1985)}]{Schutz:1985km}%
  \BibitemOpen
  \bibfield  {author} {\bibinfo {author} {\bibfnamefont {B.~F.}\ \bibnamefont {Schutz}}\ and\ \bibinfo {author} {\bibfnamefont {C.~M.}\ \bibnamefont {Will}},\ }\href {\doibase 10.1086/184453} {\bibfield  {journal} {\bibinfo  {journal} {Astrophys. J. Lett.}\ }\textbf {\bibinfo {volume} {291}},\ \bibinfo {pages} {L33} (\bibinfo {year} {1985})}\BibitemShut {NoStop}%
\bibitem [{\citenamefont {Iyer}\ and\ \citenamefont {Will}(1987)}]{Iyer:1986np}%
  \BibitemOpen
  \bibfield  {author} {\bibinfo {author} {\bibfnamefont {S.}~\bibnamefont {Iyer}}\ and\ \bibinfo {author} {\bibfnamefont {C.~M.}\ \bibnamefont {Will}},\ }\href {\doibase 10.1103/PhysRevD.35.3621} {\bibfield  {journal} {\bibinfo  {journal} {Phys. Rev. D}\ }\textbf {\bibinfo {volume} {35}},\ \bibinfo {pages} {3621} (\bibinfo {year} {1987})}\BibitemShut {NoStop}%
\bibitem [{\citenamefont {Iyer}(1987)}]{Iyer:1986nq}%
  \BibitemOpen
  \bibfield  {author} {\bibinfo {author} {\bibfnamefont {S.}~\bibnamefont {Iyer}},\ }\href {\doibase 10.1103/PhysRevD.35.3632} {\bibfield  {journal} {\bibinfo  {journal} {Phys. Rev. D}\ }\textbf {\bibinfo {volume} {35}},\ \bibinfo {pages} {3632} (\bibinfo {year} {1987})}\BibitemShut {NoStop}%
\bibitem [{\citenamefont {Kokkotas}\ and\ \citenamefont {Schutz}(1988)}]{Kokkotas:1988fm}%
  \BibitemOpen
  \bibfield  {author} {\bibinfo {author} {\bibfnamefont {K.~D.}\ \bibnamefont {Kokkotas}}\ and\ \bibinfo {author} {\bibfnamefont {B.~F.}\ \bibnamefont {Schutz}},\ }\href {\doibase 10.1103/PhysRevD.37.3378} {\bibfield  {journal} {\bibinfo  {journal} {Phys. Rev. D}\ }\textbf {\bibinfo {volume} {37}},\ \bibinfo {pages} {3378} (\bibinfo {year} {1988})}\BibitemShut {NoStop}%
\bibitem [{\citenamefont {Seidel}\ and\ \citenamefont {Iyer}(1990)}]{Seidel:1989bp}%
  \BibitemOpen
  \bibfield  {author} {\bibinfo {author} {\bibfnamefont {E.}~\bibnamefont {Seidel}}\ and\ \bibinfo {author} {\bibfnamefont {S.}~\bibnamefont {Iyer}},\ }\href {\doibase 10.1103/PhysRevD.41.374} {\bibfield  {journal} {\bibinfo  {journal} {Phys. Rev. D}\ }\textbf {\bibinfo {volume} {41}},\ \bibinfo {pages} {374} (\bibinfo {year} {1990})}\BibitemShut {NoStop}%
\bibitem [{\citenamefont {Konoplya}(2003)}]{Konoplya:2003ii}%
  \BibitemOpen
  \bibfield  {author} {\bibinfo {author} {\bibfnamefont {R.~A.}\ \bibnamefont {Konoplya}},\ }\href {\doibase 10.1103/PhysRevD.68.024018} {\bibfield  {journal} {\bibinfo  {journal} {Phys. Rev. D}\ }\textbf {\bibinfo {volume} {68}},\ \bibinfo {pages} {024018} (\bibinfo {year} {2003})},\ \Eprint {http://arxiv.org/abs/gr-qc/0303052} {arXiv:gr-qc/0303052} \BibitemShut {NoStop}%
\bibitem [{\citenamefont {Matyjasek}\ and\ \citenamefont {Opala}(2017)}]{Matyjasek:2017psv}%
  \BibitemOpen
  \bibfield  {author} {\bibinfo {author} {\bibfnamefont {J.}~\bibnamefont {Matyjasek}}\ and\ \bibinfo {author} {\bibfnamefont {M.}~\bibnamefont {Opala}},\ }\href {\doibase 10.1103/PhysRevD.96.024011} {\bibfield  {journal} {\bibinfo  {journal} {Phys. Rev. D}\ }\textbf {\bibinfo {volume} {96}},\ \bibinfo {pages} {024011} (\bibinfo {year} {2017})},\ \Eprint {http://arxiv.org/abs/1704.00361} {arXiv:1704.00361 [gr-qc]} \BibitemShut {NoStop}%
\bibitem [{wol()}]{wolfram}%
  \BibitemOpen
  \href {https://www.wolframalpha.com} {\enquote {\bibinfo {title} {Wolfram alpha},}\ }\BibitemShut {NoStop}%
\bibitem [{\citenamefont {Konoplya}\ \emph {et~al.}(2019)\citenamefont {Konoplya}, \citenamefont {Zhidenko},\ and\ \citenamefont {Zinhailo}}]{Konoplya:2019hlu}%
  \BibitemOpen
  \bibfield  {author} {\bibinfo {author} {\bibfnamefont {R.~A.}\ \bibnamefont {Konoplya}}, \bibinfo {author} {\bibfnamefont {A.}~\bibnamefont {Zhidenko}}, \ and\ \bibinfo {author} {\bibfnamefont {A.~F.}\ \bibnamefont {Zinhailo}},\ }\href {\doibase 10.1088/1361-6382/ab2e25} {\bibfield  {journal} {\bibinfo  {journal} {Class. Quant. Grav.}\ }\textbf {\bibinfo {volume} {36}},\ \bibinfo {pages} {155002} (\bibinfo {year} {2019})},\ \Eprint {http://arxiv.org/abs/1904.10333} {arXiv:1904.10333 [gr-qc]} \BibitemShut {NoStop}%
\bibitem [{\citenamefont {Hatsuda}(2020{\natexlab{b}})}]{Hatsuda:2019eoj}%
  \BibitemOpen
  \bibfield  {author} {\bibinfo {author} {\bibfnamefont {Y.}~\bibnamefont {Hatsuda}},\ }\href {\doibase 10.1103/PhysRevD.101.024008} {\bibfield  {journal} {\bibinfo  {journal} {Phys. Rev. D}\ }\textbf {\bibinfo {volume} {101}},\ \bibinfo {pages} {024008} (\bibinfo {year} {2020}{\natexlab{b}})},\ \Eprint {http://arxiv.org/abs/1906.07232} {arXiv:1906.07232 [gr-qc]} \BibitemShut {NoStop}%
\bibitem [{\citenamefont {Catal$\acute{a}$n}\ \emph {et~al.}(2014)\citenamefont {Catal$\acute{a}$n}, \citenamefont {Cisternas}, \citenamefont {Gonz$\acute{a}$les},\ and\ \citenamefont {V$\acute{a}$squez}}]{Catalan:2014ccgv}%
  \BibitemOpen
  \bibfield  {author} {\bibinfo {author} {\bibfnamefont {M.}~\bibnamefont {Catal$\acute{a}$n}}, \bibinfo {author} {\bibfnamefont {E.}~\bibnamefont {Cisternas}}, \bibinfo {author} {\bibfnamefont {P.~A.}\ \bibnamefont {Gonz$\acute{a}$les}}, \ and\ \bibinfo {author} {\bibfnamefont {Y.}~\bibnamefont {V$\acute{a}$squez}},\ }\href@noop {} {\bibfield  {journal} {\bibinfo  {journal} {Eur. Phys. Jour. C}\ }\textbf {\bibinfo {volume} {74}},\ \bibinfo {pages} {2813} (\bibinfo {year} {2014})}\BibitemShut {NoStop}%
\bibitem [{\citenamefont {Konoplya}\ and\ \citenamefont {Zhidenko}(2007)}]{Konoplya:2007kozh}%
  \BibitemOpen
  \bibfield  {author} {\bibinfo {author} {\bibfnamefont {R.~A.}\ \bibnamefont {Konoplya}}\ and\ \bibinfo {author} {\bibfnamefont {A.}~\bibnamefont {Zhidenko}},\ }\href@noop {} {\bibfield  {journal} {\bibinfo  {journal} {Phy. Rev. D}\ }\textbf {\bibinfo {volume} {76}},\ \bibinfo {pages} {084018} (\bibinfo {year} {2007})}\BibitemShut {NoStop}%
\bibitem [{\citenamefont {Li}\ and\ \citenamefont {Ma}(2013)}]{Li:2013lima}%
  \BibitemOpen
  \bibfield  {author} {\bibinfo {author} {\bibfnamefont {J.}~\bibnamefont {Li}}\ and\ \bibinfo {author} {\bibfnamefont {H.}~\bibnamefont {Ma}},\ }\href@noop {} {\bibfield  {journal} {\bibinfo  {journal} {Phy. Rev. D}\ }\textbf {\bibinfo {volume} {88}},\ \bibinfo {pages} {064001} (\bibinfo {year} {2013})}\BibitemShut {NoStop}%
\bibitem [{\citenamefont {Fernando}(2010)}]{Fernando:2010fer}%
  \BibitemOpen
  \bibfield  {author} {\bibinfo {author} {\bibfnamefont {S.}~\bibnamefont {Fernando}},\ }\href@noop {} {\bibfield  {journal} {\bibinfo  {journal} {Int. Jour. Mod. Phys. A}\ }\textbf {\bibinfo {volume} {25}},\ \bibinfo {pages} {669} (\bibinfo {year} {2010})}\BibitemShut {NoStop}%
\bibitem [{\citenamefont {Wahlang}\ \emph {et~al.}(2017)\citenamefont {Wahlang}, \citenamefont {Jeena},\ and\ \citenamefont {Chakrabarti}}]{Wahlang:2017wjc}%
  \BibitemOpen
  \bibfield  {author} {\bibinfo {author} {\bibfnamefont {W.}~\bibnamefont {Wahlang}}, \bibinfo {author} {\bibfnamefont {P.~A.}\ \bibnamefont {Jeena}}, \ and\ \bibinfo {author} {\bibfnamefont {S.}~\bibnamefont {Chakrabarti}},\ }\href@noop {} {\bibfield  {journal} {\bibinfo  {journal} {Int. Jour. Mod. Phys. D}\ }\textbf {\bibinfo {volume} {26}},\ \bibinfo {pages} {1750160} (\bibinfo {year} {2017})}\BibitemShut {NoStop}%
\bibitem [{\citenamefont {Unruh}(1973)}]{Unruh:1973unr}%
  \BibitemOpen
  \bibfield  {author} {\bibinfo {author} {\bibfnamefont {W.}~\bibnamefont {Unruh}},\ }\href@noop {} {\bibfield  {journal} {\bibinfo  {journal} {Phy. Rev. Lett.}\ }\textbf {\bibinfo {volume} {31}},\ \bibinfo {pages} {1265} (\bibinfo {year} {1973})}\BibitemShut {NoStop}%
\bibitem [{\citenamefont {Brill}\ and\ \citenamefont {Wheeler}(1957)}]{Brill:1957baw}%
  \BibitemOpen
  \bibfield  {author} {\bibinfo {author} {\bibfnamefont {D.~R.}\ \bibnamefont {Brill}}\ and\ \bibinfo {author} {\bibfnamefont {J.~A.}\ \bibnamefont {Wheeler}},\ }\href@noop {} {\bibfield  {journal} {\bibinfo  {journal} {Rev. Mod. Phys}\ }\textbf {\bibinfo {volume} {29}},\ \bibinfo {pages} {465} (\bibinfo {year} {1957})}\BibitemShut {NoStop}%
\bibitem [{\citenamefont {del Castillo}(2007)}]{Castillo:2007gft}%
  \BibitemOpen
  \bibfield  {author} {\bibinfo {author} {\bibfnamefont {G.~F.~T.}\ \bibnamefont {del Castillo}},\ }\href@noop {} {\bibfield  {journal} {\bibinfo  {journal} {Rev. Mexi. De. Fisc.}\ }\textbf {\bibinfo {volume} {33}},\ \bibinfo {pages} {125} (\bibinfo {year} {2007})}\BibitemShut {NoStop}%
\bibitem [{\citenamefont {Goldberg}\ \emph {et~al.}(1967)\citenamefont {Goldberg}, \citenamefont {A.~J.~Macfarlane},\ and\ \citenamefont {Sudarshan}}]{Goldberg:1967jng}%
  \BibitemOpen
  \bibfield  {author} {\bibinfo {author} {\bibfnamefont {J.~N.}\ \bibnamefont {Goldberg}}, \bibinfo {author} {\bibfnamefont {F.~R.}\ \bibnamefont {A.~J.~Macfarlane}, \bibfnamefont {E.~T.~Newman}}, \ and\ \bibinfo {author} {\bibfnamefont {E.}~\bibnamefont {Sudarshan}},\ }\href@noop {} {\bibfield  {journal} {\bibinfo  {journal} {Jour. Math. Phys.}\ }\textbf {\bibinfo {volume} {8}},\ \bibinfo {pages} {2155} (\bibinfo {year} {1967})}\BibitemShut {NoStop}%
\bibitem [{\citenamefont {Anderson}\ and\ \citenamefont {Price}(1991)}]{anderson:1991anp}%
  \BibitemOpen
  \bibfield  {author} {\bibinfo {author} {\bibfnamefont {A.}~\bibnamefont {Anderson}}\ and\ \bibinfo {author} {\bibfnamefont {R.~H.}\ \bibnamefont {Price}},\ }\href@noop {} {\bibfield  {journal} {\bibinfo  {journal} {Phy. Rev. D}\ }\textbf {\bibinfo {volume} {43}},\ \bibinfo {pages} {3147} (\bibinfo {year} {1991})}\BibitemShut {NoStop}%
\bibitem [{\citenamefont {Toshmatov}\ \emph {et~al.}(2016)\citenamefont {Toshmatov}, \citenamefont {Stuchlik}, \citenamefont {Schee},\ and\ \citenamefont {Ahmedov}}]{Tosh:2016tssa}%
  \BibitemOpen
  \bibfield  {author} {\bibinfo {author} {\bibfnamefont {B.}~\bibnamefont {Toshmatov}}, \bibinfo {author} {\bibnamefont {Stuchlik}}, \bibinfo {author} {\bibfnamefont {J.}~\bibnamefont {Schee}}, \ and\ \bibinfo {author} {\bibfnamefont {B.}~\bibnamefont {Ahmedov}},\ }\href@noop {} {\bibfield  {journal} {\bibinfo  {journal} {Phys. Rev. D.}\ }\textbf {\bibinfo {volume} {93}},\ \bibinfo {pages} {124017} (\bibinfo {year} {2016})},\ \Eprint {http://arxiv.org/abs/1605.02058} {arXiv:1605.02058 [gr-qc]} \BibitemShut {NoStop}%
\bibitem [{\citenamefont {Li}\ \emph {et~al.}(2015)\citenamefont {Li}, \citenamefont {Lin},\ and\ \citenamefont {Nan}}]{Li:2015lly}%
  \BibitemOpen
  \bibfield  {author} {\bibinfo {author} {\bibfnamefont {J.}~\bibnamefont {Li}}, \bibinfo {author} {\bibfnamefont {K.}~\bibnamefont {Lin}}, \ and\ \bibinfo {author} {\bibfnamefont {Y.}~\bibnamefont {Nan}},\ }\href@noop {} {\bibfield  {journal} {\bibinfo  {journal} {Eur. Phys. Jour. C.}\ }\textbf {\bibinfo {volume} {75}},\ \bibinfo {pages} {131} (\bibinfo {year} {2015})},\ \Eprint {http://arxiv.org/abs/1409.5988} {arXiv:1409.5988 [gr-qc]} \BibitemShut {NoStop}%
\bibitem [{\citenamefont {Huang}\ and\ \citenamefont {Liu}(2016)}]{Huang:2016hl}%
  \BibitemOpen
  \bibfield  {author} {\bibinfo {author} {\bibfnamefont {Y.}~\bibnamefont {Huang}}\ and\ \bibinfo {author} {\bibfnamefont {D.}~\bibnamefont {Liu}},\ }\href@noop {} {\bibfield  {journal} {\bibinfo  {journal} {Phys. Rev. D.}\ }\textbf {\bibinfo {volume} {93}},\ \bibinfo {pages} {104011} (\bibinfo {year} {2016})},\ \Eprint {http://arxiv.org/abs/1509.09017} {arXiv:1509.09017 [gr-qc]} \BibitemShut {NoStop}%
\end{thebibliography}%

\end{document}